\documentclass[twocolumn]{aastex7}

\newcommand\aastex{AAS\TeX}
\newcommand\latex{La\TeX}
\usepackage{color}

\begin{document}

\title{Detection of Compton scattering in the jet of 3C~84}

\author[0000-0001-9200-4006]{Ioannis Liodakis}
\affiliation{Institute of Astrophysics, Foundation for Research and Technology - Hellas, Voutes, 70013 Heraklion, Greece}
\affiliation{Max-Planck-Institut f\"{u}r Radioastronomie, Auf dem H\"{u}gel 69, D-53121 Bonn, Germany}
\email[show]{liodakis@ia.forth.gr}

\author{Sudip Chakraborty}
\affiliation{NASA Marshall Space Flight Center, Huntsville, AL 35812, USA}
\email{sudip.chakraborty@nasa.gov}

\author[0000-0003-2049-2690]{Fr\'{e}d\'{e}ric Marin}
\affiliation{Universit\'{e} de Strasbourg, CNRS, Observatoire Astronomique de Strasbourg, UMR 7550, 67000 Strasbourg, France}
\email{frederic.marin@astro.unistra.fr}

\author[0000-0003-4420-2838]{Steven R. Ehlert}
\affiliation{NASA Marshall Space Flight Center, Huntsville, AL 35812, USA}
\email{steven.r.ehlert@nasa.gov}

\author[0000-0003-1340-5675]{Thibault Barnouin}
\affiliation{Universit\'{e} de Strasbourg, CNRS, Observatoire Astronomique de Strasbourg, UMR 7550, 67000 Strasbourg, France}
\email{thibault.barnouin2@etu.unistra.fr}

\author[0000-0002-9328-2750]{Pouya M. Kouch}
\affiliation{Finnish Centre for Astronomy with ESO (FINCA), 20014 University of Turku, Finland}
\affiliation{Department of Physics and  Astronomy, Quantum, Vesilinnantie 5, FI-20014 University of Turku, Finland}
\email{pouya.kouch@utu.fi}

\author[0000-0002-1445-8683]{Kari Nilsson}
\affiliation{Finnish Centre for Astronomy with ESO (FINCA), 20014 University of Turku, Finland}
\email{kani@utu.fi}

\author[0000-0002-9155-6199]{Elina Lindfors}
\affiliation{Department of Physics and  Astronomy, Quantum, Vesilinnantie 5, FI-20014 University of Turku, Finland}
\email{elilin@utu.fi}

\author{Tapio Pursimo}
\affiliation{Nordic Optical Telescope, Apartado 474, E-38700 Santa Cruz de La Palma, Santa Cruz de Tenerife, Spain}
\affiliation{Department of Physics and Astronomy, Aarhus University, Munkegade 120, DK-8000 Aarhus C, Denmark}
\email{tpursimo@not.iac.es}

\author[0000-0001-6757-3098]{Georgios F. Paraschos}
\affiliation{Max-Planck-Institut f\"{u}r Radioastronomie, Auf dem H\"{u}gel 69, D-53121 Bonn, Germany}
\email{gfparaschos@mpifr-bonn.mpg.de}

\author[0000-0001-9815-9092]{Riccardo Middei}
\affiliation{Space Science Data Center, Agenzia Spaziale Italiana, Via del Politecnico snc, I-00133 Roma, Italy}
\affiliation{INAF Osservatorio Astronomico di Roma, Via Frascati 33, 00078 Monte Porzio Catone (RM), Italy}
\affiliation{Center for Astrophysics \textbar ~Harvard \& Smithsonian, 60 Garden Street, Cambridge, MA 02138 USA}
\email{riccardo.middei@ssdc.asi.it}

\author{Anna Trindade Falcão}
\affiliation{Harvard-Smithsonian Center for Astrophysics, 60 Garden St., Cambridge, MA 02138, USA} 
\affiliation{NASA-Goddard Space Flight Center, Code 662, Greenbelt, MD 20771, USA} 
\email{anna.trindade_falcao@cfa.harvard.edu}

\author[0000-0001-9522-5453]{Svetlana Jorstad}
\affiliation{Institute for Astrophysical Research, Boston University, 725 Commonwealth Avenue, Boston, MA 02215, USA}
\affiliation{Saint Petersburg State University, 7/9 Universitetskaya nab., St. Petersburg, 199034 Russia}
\email{jorstad@bu.edu}

\author[0000-0002-3777-6182]{Iv\'an Agudo}
\affiliation{Instituto de Astrof\'{i}sica de Andaluc\'{i}a, IAA-CSIC, Glorieta de la Astronom\'{i}a s/n, E-18008 Granada, Spain}
\email{iagudo@iaa.es}

\author{Yuri Y. Kovalev}
\affiliation{Max-Planck-Institut f\"{u}r Radioastronomie, Auf dem H\"{u}gel 69, D-53121 Bonn, Germany}
\email{ykovalev@mpifr-bonn.mpg.de}


\author[0009-0009-3051-6570]{Jacob J. Casey}
\affiliation{Department of Physics and Astronomy and Space Science Center, University of New Hampshire, Durham, NH 03824, USA}
\email{Jacob.J.Casey@unh.edu}

\author[0000-0002-5614-5028]{Laura Di Gesu}
\affiliation{ASI - Agenzia Spaziale Italiana, Via del Politecnico snc, 00133 Roma, Italy}
\email{laura.digesu@asi.it}

\author[0000-0002-3638-0637]{Philip Kaaret}
\affiliation{NASA Marshall Space Flight Center, Huntsville, AL 35812, USA}
\email{philip.kaaret@nasa.gov}

\author[0000-0001-5717-3736]{Dawoon E. Kim}
\affiliation{INAF, Istituto di Astrofisica e Planetologia Spaziali, Via Fosso del Cavaliere 100, 00133 Roma, Italy}
\email{dawoon.kim@inaf.it}

\author[0000-0001-7477-0380]{Fabian Kislat}
\affiliation{Department of Physics and Astronomy and Space Science Center, University of New Hampshire, Durham, NH 03824, USA}
\email{fabian.kislat@unh.edu}

\author[0000-0003-0411-4243]{Ajay Ratheesh}
\affiliation{Physical Research Laboratory, Thaltej, Ahmedabad, Gujarat 380009, India}
\affiliation{INAF, Istituto di Astrofisica e Planetologia Spaziali, Via Fosso del Cavaliere 100, 00133 Roma, Italy}
\email{ajay.ratheesh@inaf.it}

\author[0000-0001-7163-7015]{M. Lynne Saade}
\affiliation{Science \& Technology Institute, Universities Space Research Association, 320 Sparkman Drive, Huntsville, AL 35805, USA}
\affiliation{NASA Marshall Space Flight Center, Huntsville, AL 35812, USA}
\email{mary.l.saade@nasa.gov}

\author[0000-0002-6562-8654]{Francesco Tombesi}
\affiliation{Dipartimento di Fisica, Universit\'{a} degli Studi di Roma "Tor Vergata", Via della Ricerca Scientifica 1, 00133 Roma, Italy}
\affiliation{Istituto Nazionale di Fisica Nucleare, Sezione di Roma "Tor Vergata", Via della Ricerca Scientifica 1, 00133 Roma, Italy}
\email{francesco.tombesi@roma2.infn.it}


\author[0000-0001-7396-3332]{Alan Marscher}
\affiliation{Institute for Astrophysical Research, Boston University, 725 Commonwealth Avenue, Boston, MA 02215, USA}
\email{marscher@bu.edu}

\author{Francisco Jos\'e Aceituno}
\affiliation{Instituto de Astrof\'{i}sica de Andaluc\'{i}a, IAA-CSIC, Glorieta de la Astronom\'{i}a s/n, 18008 Granada, Spain}
\email{fja@iaa.es}

\author[0000-0003-2464-9077]{Giacomo Bonnoli}
\affiliation{INAF Osservatorio Astronomico di Brera, Via E. Bianchi 46, 23807 Merate (LC), Italy}
\affiliation{Instituto de Astrof\'{i}sica de Andaluc\'{i}a, IAA-CSIC, Glorieta de la Astronom\'{i}a s/n, 18008 Granada, Spain}
\email{giacomo.bonnoli@inaf.it}

\author[0000-0003-2036-8999]{V\'{i}ctor Casanova}
\affiliation{Instituto de Astrof\'{i}sica de Andaluc\'{i}a, IAA-CSIC, Glorieta de la Astronom\'{i}a s/n, 18008 Granada, Spain}
\email{casanova@iaa.es}

\author{Gabriel Emery}
\affiliation{Instituto de Astrof\'{i}sica de Andaluc\'{i}a, IAA-CSIC, Glorieta de la Astronom\'{i}a s/n, 18008 Granada, Spain}
\email{emery@iaa.es}

\author[0000-0002-4131-655X]{Juan Escudero Pedrosa}
\affiliation{Instituto de Astrof\'{i}sica de Andaluc\'{i}a, IAA-CSIC, Glorieta de la Astronom\'{i}a s/n, E-18008 Granada, Spain}
\affiliation{Center for Astrophysics \textbar ~Harvard \& Smithsonian, 60 Garden Street, Cambridge, MA 02138 USA}
\email{juan.escuderopedrosa@cfa.harvard.edu}

\author{Daniel Morcuende}
\affiliation{Instituto de Astrof\'{i}sica de Andaluc\'{i}a, IAA-CSIC, Glorieta de la Astronom\'{i}a s/n, 18008 Granada, Spain}
\email{daniel.morcuende@cta-observatory.org}

\author[0000-0002-4241-5875]{Jorge Otero-Santos}
\affiliation{Instituto de Astrof\'{i}sica de Andaluc\'{i}a, IAA-CSIC, Glorieta de la Astronom\'{i}a s/n, E-18008 Granada, Spain}
\affiliation{Istituto Nazionale di Fisica Nucleare, Sezione di Padova, 35131 Padova, Italy}
\email{jorge.otero@pd.infn.it}

\author{Alfredo Sota}
\affiliation{Instituto de Astrof\'{i}sica de Andaluc\'{i}a, IAA-CSIC, Glorieta de la Astronom\'{i}a s/n, 18008 Granada, Spain}
\email{sota@iaa.es}

\author{Vilppu Piirola}
\affiliation{Department of Physics and Astronomy, 20014 University of Turku, Finland}
\email{piirola@utu.fi}

\author{Rumen Bachev}
\affiliation{Institute of Astronomy and NAO, Bulgarian Academy of Sciences, 1784 Sofia, Bulgaria}
\email{bachevr@astro.bas.bg}

\author{Anton Strigachev}
\affiliation{Institute of Astronomy and NAO, Bulgarian Academy of Sciences, 1784 Sofia, Bulgaria}
\email{anton@nao-rozhen.org}

\author[0000-0002-7262-6710]{George A. Borman}
\affiliation{Crimean Astrophysical Observatory RAS, P/O Nauchny, 298409, Crimea}
\email{borman.ga@gmail.com}

\author[0000-0002-3953-6676]{Tatiana S. Grishina}
\affiliation{Saint Petersburg State University, 7/9 Universitetskaya nab., St. Petersburg, 199034 Russia}
\email{t.s.grishina@spbu.ru}

\author[0000-0002-6431-8590]{Vladimir A. Hagen-Thorn}
\affiliation{Saint Petersburg State University, 7/9 Universitetskaya nab., St. Petersburg, 199034 Russia}
\email{hth-home@yandex.ru}

\author[0000-0001-9518-337X]{Evgenia N. Kopatskaya}
\affiliation{Saint Petersburg State University, 7/9 Universitetskaya nab., St. Petersburg, 199034 Russia}
\email{enik1346@rambler.ru}

\author[0000-0002-2471-6500]{Elena G. Larionova}
\affiliation{Saint Petersburg State University, 7/9 Universitetskaya nab., St. Petersburg, 199034 Russia}
\email{sung2v@mail.ru}

\author[0000-0002-9407-7804]{Daria A. Morozova}
\affiliation{Saint Petersburg State University, 7/9 Universitetskaya nab., St. Petersburg, 199034 Russia}
\email{d.morozova@spbu.ru}

\author[0000-0003-4147-3851]{Sergey S. Savchenko}
\affiliation{Saint Petersburg State University, 7/9 Universitetskaya nab., St. Petersburg, 199034 Russia}
\affiliation{Pulkovo Observatory, St.Petersburg, 196140, Russia}
\email{s.s.savchenko@spbu.ru}

\author[0009-0002-2440-2947]{Ekaterina V. Shishkina}
\affiliation{Saint Petersburg State University, 7/9 Universitetskaya nab., St. Petersburg, 199034 Russia}
\email{e.v.shishkina99@yandex.ru}

\author[0000-0002-4218-0148]{Ivan S. Troitskiy}
\affiliation{Saint Petersburg State University, 7/9 Universitetskaya nab., St. Petersburg, 199034 Russia}
\email{i.troitsky@spbu.ru}

\author[0000-0002-9907-9876]{Yulia V. Troitskaya}
\affiliation{Saint Petersburg State University, 7/9 Universitetskaya nab., St. Petersburg, 199034 Russia}
\email{y.troitskaya@spbu.ru}

\author[0000-0002-8293-0214]{Andrey A. Vasilyev}
\affiliation{Saint Petersburg State University, 7/9 Universitetskaya nab., St. Petersburg, 199034 Russia}
\email{andrey.vasilyev@spbu.ru}

\author{Alexey V. Zhovtan}
\affiliation{Crimean Astrophysical Observatory RAS, P/O Nauchny, 298409, Crimea}
\email{astroalex2012@gmail.com}

\author[0000-0003-3025-9497]{Ioannis Myserlis}
\affiliation{Institut de Radioastronomie Millim\'{e}trique, Avenida Divina Pastora, 7, Local 20, E–18012 Granada, Spain}
\email{imyserlis@iram.es}

\author[0000-0003-0685-3621]{Mark Gurwell}
\affiliation{Center for Astrophysics $|$ Harvard \& Smithsonian, 60 Garden Street, Cambridge, MA 02138 USA}
\email{mgurwell@cfa.harvard.edu}

\author[0000-0002-3490-146X]{Garrett Keating}
\affiliation{Center for Astrophysics $|$ Harvard \& Smithsonian, 60 Garden Street, Cambridge, MA 02138 USA}
\email{garrett.keating@cfa.harvard.edu}

\author[0000-0002-1407-7944]{Ramprasad Rao}
\affiliation{Center for Astrophysics $|$ Harvard \& Smithsonian, 60 Garden Street, Cambridge, MA 02138 USA}
\email{rrao@cfa.harvard.edu}

\author[0000-0002-0112-4836]{Sincheol Kang}
\affiliation{Korea Astronomy and Space Science Institute, 776 Daedeok-daero, Yuseong-gu, Daejeon 34055, Korea}
\email{kang87@kasi.re.kr}

\author[0000-0002-6269-594X]{Sang-Sung Lee}
\affiliation{Korea Astronomy and Space Science Institute, 776 Daedeok-daero, Yuseong-gu, Daejeon 34055, Korea}
\affiliation{University of Science and Technology, Korea, 217 Gajeong-ro, Yuseong-gu, Daejeon 34113, Korea}
\email{sslee@kasi.re.kr}

\author[0000-0001-7556-8504]{Sanghyun Kim}
\affiliation{Korea Astronomy and Space Science Institute, 776 Daedeok-daero, Yuseong-gu, Daejeon 34055, Korea}
\email{sanghkim@kasi.re.kr}

\author[0009-0002-1871-5824]{Whee Yeon Cheong}
\affiliation{Korea Astronomy and Space Science Institute, 776 Daedeok-daero, Yuseong-gu, Daejeon 34055, Korea}
\email{wheeyeon@kasi.re.kr}

\author[0009-0005-7629-8450]{Hyeon-Woo Jeong}
\affiliation{Korea Astronomy and Space Science Institute, 776 Daedeok-daero, Yuseong-gu, Daejeon 34055, Korea}
\affiliation{University of Science and Technology, Korea, 217 Gajeong-ro, Yuseong-gu, Daejeon 34113, Korea}
\email{hwjeong@kasi.re.kr}

\author[0009-0003-8767-7080]{Chanwoo Song}
\affiliation{Korea Astronomy and Space Science Institute, 776 Daedeok-daero, Yuseong-gu, Daejeon 34055, Korea}
\affiliation{University of Science and Technology, Korea, 217 Gajeong-ro, Yuseong-gu, Daejeon 34113, Korea}
\email{scw317@kasi.re.kr}

\author[0009-0006-1247-0976]{Shan Li}
\affiliation{Korea Astronomy and Space Science Institute, 776 Daedeok-daero, Yuseong-gu, Daejeon 34055, Korea}
\affiliation{University of Science and Technology, Korea, 217 Gajeong-ro, Yuseong-gu, Daejeon 34113, Korea}
\email{shan_li@kasi.re.kr}

\author[0009-0001-4748-0211]{Myeong-Seok Nam}
\affiliation{Korea Astronomy and Space Science Institute, 776 Daedeok-daero, Yuseong-gu, Daejeon 34055, Korea}
\affiliation{University of Science and Technology, Korea, 217 Gajeong-ro, Yuseong-gu, Daejeon 34113, Korea}
\email{msnam@kasi.re.kr}

\author[0000-0002-9998-5238]{Diego \'{A}lvarez-Ortega}
\affiliation{Institute of Astrophysics, Foundation for Research and Technology - Hellas, Voutes, 70013 Heraklion, Greece}
\affiliation{Department of Physics, University of Crete, 70013, Heraklion, Greece}
\email{dalvarez@physics.uoc.gr}

\author[0000-0003-1117-2863]{Carolina Casadio}
\affiliation{Institute of Astrophysics, Foundation for Research and Technology - Hellas, Voutes, 70013 Heraklion, Greece}
\affiliation{Department of Physics, University of Crete, 70013, Heraklion, Greece}
\email{ccasadio@ia.forth.gr}

\author[0000-0001-7327-5441]{Emmanouil Angelakis}
\affiliation{Section of Astrophysics, Astronomy \& Mechanics, Department of Physics, National and Kapodistrian University of Athens,
Panepistimiopolis Zografos 15784, Greece}
\email{eangelakis@physics.auth.gr}

\author[0000-0002-4184-9372]{Alexander Kraus}
\affiliation{Max-Planck-Institut f\"{u}r Radioastronomie, Auf dem H\"{u}gel 69,
D-53121 Bonn, Germany}
\email{akraus@mpifr-bonn.mpg.de}

\author[0000-0003-4519-7751]{Jenni Jormanainen}
\affiliation{Department of Physics and  Astronomy, Quantum, Vesilinnantie 5, FI-20014 University of Turku, Finland}
\affiliation{Finnish Centre for Astronomy with ESO (FINCA), 20014 University of Turku, Finland}
\email{jenni.s.jormanainen@utu.fi}

\author[0000-0001-8991-7744]{Vandad Fallah Ramazani}
\affiliation{Finnish Centre for Astronomy with ESO (FINCA), 20014 University of Turku, Finland}
\affiliation{Aalto University Mets\"ahovi Radio Observatory, Mets\"ahovintie 114, FI-02540 Kylm\"al\"a, Finland}
\email{vandad.fallahramazani@cta-consortium.org}

\author[0000-0002-4945-5079 ]{Chien-Ting Chen}
\affiliation{Science and Technology Institute, Universities Space Research Association, Huntsville, AL 35805, USA}
\email{chien-ting.chen@nasa.gov}

\author[0000-0003-4925-8523]{Enrico Costa}
\affiliation{INAF Istituto di Astrofisica e Planetologia Spaziali, Via del Fosso del Cavaliere 100, 00133 Roma, Italy}
\email{enrico.costa@inaf.it}

\author{Eugene Churazov}
\affiliation{Max Planck Institute for Astrophysics, Karl-Schwarzschild-Str. 1, D-85741 Garching, Germany}
\email{churazov@mpa-garching.mpg.de}

\author[0000-0003-1074-8605]{Riccardo Ferrazzoli}
\affiliation{INAF Istituto di Astrofisica e Planetologia Spaziali, Via del Fosso del Cavaliere 100, 00133 Roma, Italy}
\email{riccardo.ferrazzoli@inaf.it}

\author{Giorgio Galanti}
\affiliation{INAF, Istituto di Astrofisica Spaziale e Fisica Cosmica di Milano, Via Alfonso Corti 12, I – 20133 Milano, Italy}
\email{gam.galanti@gmail.com}

\author[0000-0003-3701-5882]{Ildar Khabibullin}
\affiliation{Universit\"{a}ts-Sternwarte, Fakult\"{a}t fuer Physik, Ludwig-Maximilians-Universit\"{a}t Muenchen, Scheinerstr.1, 81679 Muenchen, Germany}
\affiliation{Max Planck Institute for Astrophysics, Karl-Schwarzschild-Str. 1, D-85741 Garching, Germany}
\email{ildar@MPA-Garching.MPG.DE}

\author[0000-0002-1868-8056]{Stephen L. O'Dell}
\affiliation{NASA Marshall Space Flight Center, Huntsville, AL 35812, USA}
\email{stephen.l.odell@nasa.gov}

\author[0000-0001-6897-5996]{Luigi Pacciani}
\affiliation{INAF, Istituto di Astrofisica e Planetologia Spaziali, Via Fosso del Cavaliere, 100 - I-00133 Rome, Italy}
\email{luigi.pacciani@inaf.it}

\author{Marco Roncadelli}
\affiliation{INFN, Sezione di Pavia, Via A. Bassi 6, 27100 Pavia, Italy}
\email{marco.roncadelli@pv.infn.it}

\author[0000-0002-7150-9061]{Oliver J. Roberts}
\affiliation{Science and Technology Institute, Universities Space Research Association, Huntsville, AL 35805, USA}
\email{oliver.roberts@nasa.gov}

\author[0000-0002-7781-4104]{Paolo Soffitta}
\affiliation{INAF Istituto di Astrofisica e Planetologia Spaziali, Via del Fosso del Cavaliere 100, 00133 Roma, Italy}
\email{paolo.soffitta@inaf.it}

\author[0000-0002-2954-4461]{Douglas A. Swartz}
\affiliation{Science and Technology Institute, Universities Space Research Association, Huntsville, AL 35805, USA}
\email{doug.swartz@nasa.gov}

\author[0000-0003-0256-0995]{Fabrizio Tavecchio}
\affiliation{INAF Osservatorio Astronomico di Brera, Via E. Bianchi 46, 23807 Merate (LC), Italy}
\email{fabrizio.tavecchio@inaf.it}

\author[0000-0002-5270-4240]{Martin C. Weisskopf}
\affiliation{NASA Marshall Space Flight Center, Huntsville, AL 35812, USA}
\email{martin.c.weisskopf@nasa.gov}

\author[0000-0001-7630-8085]{Irina Zhuravleva}
\affiliation{Department of Astronomy and Astrophysics, The University of Chicago, Chicago, IL 60637, USA}
\email{zhuravleva@astro.uchicago.edu}


\begin{abstract}

3C~84 is the brightest cluster galaxy in the Perseus Cluster. It is among the closest radio-loud active galaxies and among the very few that can be detected from low frequency radio up to TeV $\gamma$-rays. Here we report on the first X-ray polarization observation of 3C~84 with the Imaging X-ray Polarimetry Explorer, for a total of 2.2~Msec that coincides with a flare in $\gamma$-rays. This is the longest observation for a radio-loud active galaxy that allowed us to reach unprecedented sensitivity, leading to the detection of an X-ray polarization degree of $\rm\Pi_X=4.2\pm1.3\%$ ($\sim3.2\sigma$ confidence) at an X-ray electric vector polarization angle of $\rm \psi_X=163^{\circ}\pm9^{\circ}$, which is aligned with the radio jet direction on the sky. Optical polarization observations show fast variability about the jet axis as well. Our results strongly favor models in which X-rays are produced by Compton scattering from relativistic electrons -- specifically synchrotron self-Compton -- that takes place downstream, away from the supermassive black hole. 
\end{abstract}



\section{Introduction} 

3C~84 (also known as NGC~1275) is among the nearest and brightest radio galaxies, located in the Perseus Cluster. It is an FR-I, Seyfert-2 type galaxy \citep{Khachikian1974} that would be equivalent to a BL Lacertae-type object according to the active galactic nuclei (AGN) unification scheme \citep{Antonucci1993}. At radio wavelengths it shows a spine-sheath morphology \cite[e.g.,][]{Nagai2014,Giovannini2018,Paraschos2024} with a magnetic field of a few gauss at parsec scales \citep{Kim2019,Paraschos2021,Paraschos2023} potentially driving Kelvin-Helmoltz instabilities \citep{Paraschos2025}. Recent Event Horizon Telescope observations at 228~GHz showed regions with highly ordered magnetic fields reaching a polarization degree of $\sim17\%$ \citep{Paraschos2024-II}.

Despite showing a low jet speed and Doppler factor \cite[e.g.,][]{Liodakis2018-II,Paraschos2022,Weaver2022} its multiwavelength behavior is quite intriguing. Unlike other radio galaxies, 3C~84 has variable high-energy emission \cite[e.g.,][]{Rani2018,Imazato2021,Fukazawa2018}, shows bright $\gamma$-ray behavior at GeV energies \citep{Abdo2009}, and has even been detected in TeV $\gamma$-rays \citep{Aleksic2014, Cao2024}. The origin of the high-energy emission in radio galaxies in general, and 3C~84 in particular is still a matter of debate. If the emission is dominated by the jet, it will originate either from relativistic electrons upscattering low-energy photons to X-rays and $\gamma$-rays \cite[e.g.,][]{Poutanen1994,Mastichiadis1997,Boettcher2013}, or proton processes that include proton synchrotron and pair cascades from secondary particles created in proton-photon interactions \cite[e.g.,][]{Mannheim1993-II,Mastichiadis1996,Zhang2013,Mastichiadis2021}. On the other hand, if the emission is dominated by the accretion disk the high-energy emission will be dominated by the hot corona \cite[e.g.,][]{Haardt1991,Kara2016}. Depending on the inclination of the source the emission will be either direct corona emission or reflection from Compton-thick material surrounding the accretion disk \cite[e.g.,][]{Singh2011}. Recent X-ray observations from the Imaging X-ray Polarimetry Explorer \cite[IXPE,][]{Weisskopf2022,Soffitta2023} have confirmed this picture \citep{Gianolli2023,Ingram2023,Tagliacozzo2023,Ursini2023,Marin2024,Chakraborty2025}.

Previous studies on 3C~84 suggest that the X-ray emission is dominated by the jet \citep{Rani2018}, while $\gamma$-rays could originate from multiple emission sites \citep{Hodgson2018,Hodgson2021,Paraschos2023,Sinitsyna2025}. The high-energy emission is thought to originate from synchrotron self-Compton from relativistic electrons \cite[e.g.,][]{Tanada2018,Aleksic2014,Cao2024} that are likely accelerated in shocks \citep{Fukazawa2018}. IXPE observations of low-synchrotron \citep{Ehlert2022,Middei2023,Marshall2023,Kouch2025,Agudo2025} and high-synchrotron peak blazars \citep{Liodakis2022,Ehlert2023,DiGesu2023,Chen2024,Middei2023-II,Kouch2024,Capecchiacci2025} are also consistent with such scenarios.

Apart from 3C~84 discussed here, IXPE has observed two more radio galaxies. Centaurus A \citep{Ehlert2022} that showed low X-ray polarization ($<$8\% at 99\% confidence), and Pictor A focusing on the western hot spot \citep{Tugliani2025}. Here we report on the first X-ray polarization observation of 3C~84 aiming to understand the origin of its X-ray emission. In section \ref{sec:MWLpol} we describe the novel X-ray polarization observations and the multiwavelength campaign that accompanied them, and in section \ref{sec:disc_concl} we discuss our findings and present our conclusions. Additional details on our analysis are given in Appendices \ref{app_sec:xpol} and \ref{app_sec:mwl_obs}.

\section{Multiwavelength Polarization observations} \label{sec:MWLpol}

\subsection{X-ray polarization observations}

\begin{figure*}
        \centering
    \includegraphics[width=\linewidth]{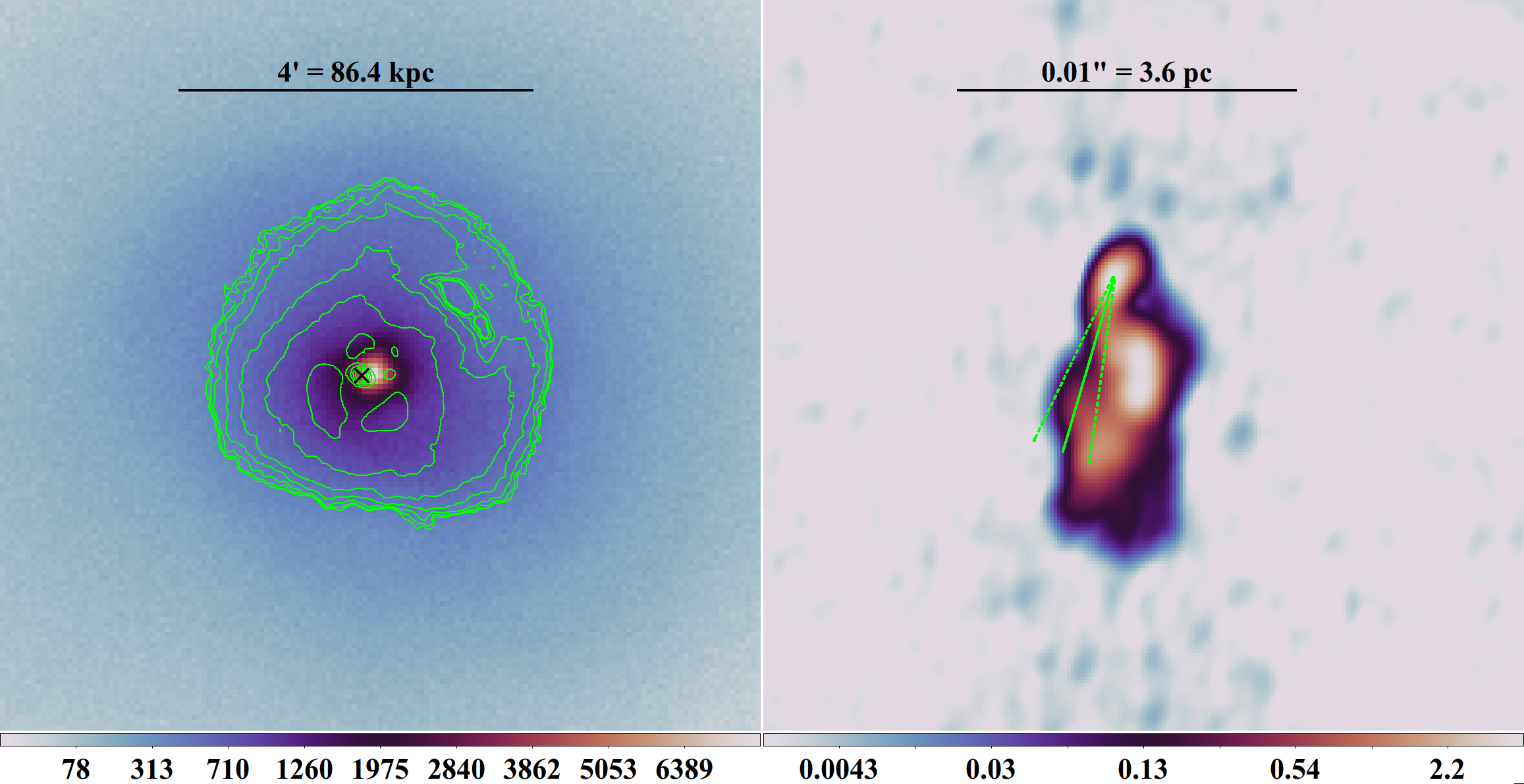}
    \caption{\textit{Left: } IXPE counts image of the Perseus Cluster overlaid with surface brightness contours from Chandra, showing that many of the substructures clearly apparent in the higher angular resolution Chandra data are also visible at lower resolution in IXPE. The black X corresponds to the position of 3C 84. \textit{Right: }Cleaned 15 GHz Stokes I VLBA image of the parsec-scale jet of 3C 84, observed as part of the MOJAVE program on February 21st 2025, in units of $\mathrm{Jy} \thinspace \mathrm{beam}^{-1}$. Notice that the angular scale of this radio image is approximately three orders of magnitude smaller than the IXPE image. Overlaid in green is the X-ray polarization vector with the $1\sigma$ confidence interval of the polarization angle denoted by the two dashed lines. The polarization vector's direction is fully consistent with the direction of the radio jet.      }
    \label{fig:perseusimages}
\end{figure*}

The IXPE observations of 3C 84 were taken in five separate segments between 26 January 2025 and 26 March 2025. The total exposure before filtering for flares is 2.19 Ms (Fig. \ref{fig:perseusimages}). We perform good time interval (GTI) filtering of the light curve in the $8-10 \thinspace \mathrm{keV}$ band (where the mirrors have no effective area) to identify time intervals where solar flares may dominate the astrophysical X-ray signal. We also exclude 59 ks (see Appendix for more details). A small enhancement in the IXPE count rate is observed during this time period, and although the duration and the intensity of the flare correspond to only a small perturbation to the integrated signal, we conservatively exclude this time period out of an abundance of caution. After all filtering, the total exposure time is 2.13 Ms. For all subsequent results, all events identified as background using the algorithm of \cite{DiMarco2023} have been filtered out \textit{a priori}. All results presented, unless otherwise noted, are for the entire 2-8 $\thinspace \mathrm{keV}$ IXPE bandpass. 

Time-resolved spectroscopy of the IXPE data within a $1^{\prime}$ radius aperture centered on 3C 84 during the cleaned GTI's shows no evidence of additional variability in the overall flux or power-law photon index with time. As measured by IXPE, the overall flux of 3C 84 in this aperture is $\sim 1.5 \times 10^{-11} \thinspace \mathrm{erg} \thinspace \mathrm{cm^{-2}} \thinspace \mathrm{s^{-1}}$ and $\Gamma \sim 1.8$.   

\begin{figure}
    \centering
    \includegraphics[width=\linewidth]{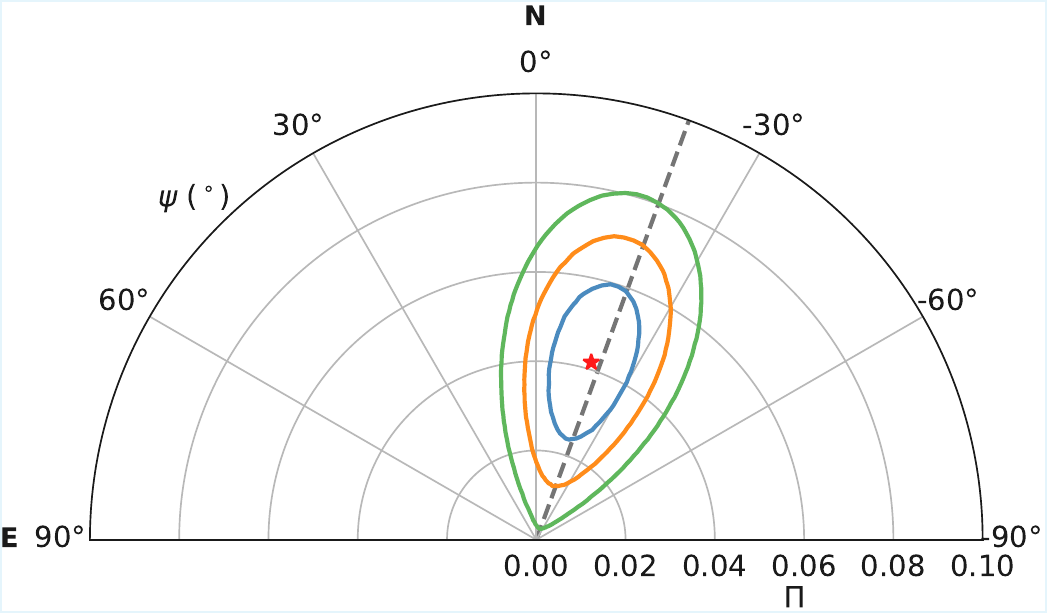}
    \caption{Polarization contours for 3C 84 using a spectropolarimetric model fit. The red star denotes the best-fit value, while the blue, orange, and green contours correspond to $1\sigma,2\sigma$ , and $3\sigma$ confidence intervals. The dashed line corresponds to the position angle of the radio jet.   }
    \label{fig:polplot}
\end{figure}

We measure the polarization of 3C 84 by performing a spectropolarimetric fit to the IXPE data in the 2-8 $\thinspace \mathrm{keV}$ band alongside Chandra, Swift XRT, and NuSTAR observations of the Perseus Cluster. This joint fit is required to separate the polarized power-law emission from the expected thermal emission from the ICM, and the details for how the spectra are extracted and modeled can be found in the Appendix. For the IXPE data, we use a source aperture of $30^{\prime \prime}$ in radius (see Appendix for details and justification). From this fit, we find a polarization degree of $\Pi_{X} = 4.18 \pm 1.31 \%$ and an angle of $\psi_{X} = 162.8^{\circ} \pm 9.2^{\circ} $, consistent with the simpler model-independent value. The corresponding 2--8 keV Minimum Detectable Polarization at 99\% (MDP$_{99}$), excluding the diffused background, is 2.39\%.
The polarization contours for this model fit are shown in Figure \ref{fig:polplot}. Extensive testing and analysis using data from Chandra, NuSTAR, and Swift alongside the extracted IXPE Stokes I/Q/U spectra have shown that these significance of the detection is robust to a large range of assumptions regarding the thermal emission of the Perseus Cluster and the non X-ray background in the IXPE data. An image of the Perseus Cluster as observed by IXPE as well as the X-ray polarization vector superimposed on the Very Long Baseline Array (VLBA, MOJAVE) observations can be found in Figure \ref{fig:perseusimages}. 

\subsection{Radio and optical polarization observations} \label{sec:ORpol}

\begin{figure*}
\centering
\includegraphics[scale=0.5]{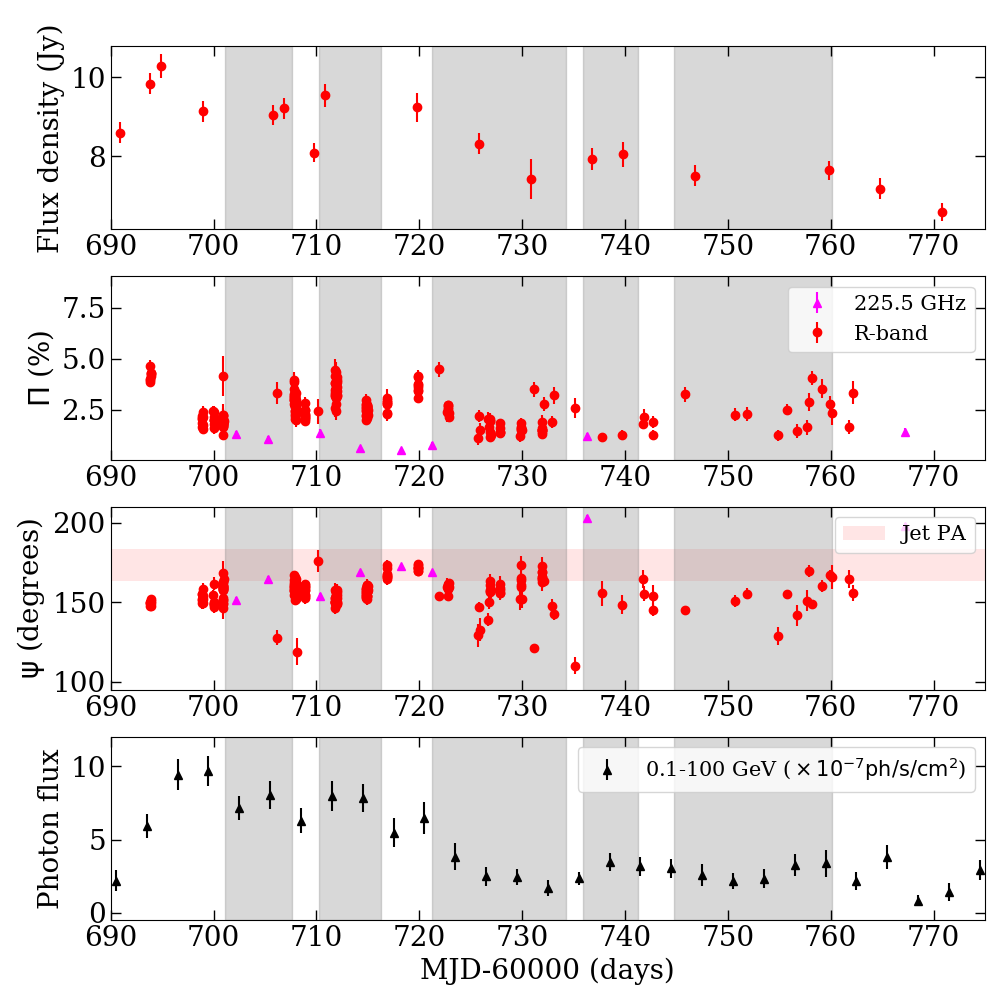}
 \caption{Multiwavelength brightness and polarization observation during the IXPE observations. The top panel shows the flux density in optical (R-band), the second panel from the top the host-corrected optical (R-band) and milli-meter radio (225.5~GHz) polarization degree (\%), the third panel from the top the optical and radio polarization angle (degrees), and the bottom panel the $\gamma$-ray photon flux 3-day binned light curves. The grey shaded areas mark the IXPE exposures over the entire observation. The horizontal red shaded area marks the projected direction of the jet. The error bars correspond to the 68\% (1$\sigma$) confidence interval. }
\label{plt:pol_all_gamma}
\end{figure*}

Contemporaneous to the IXPE observation, we initiated a multiwavelength campaign that included several ground-based optical and radio telescopes. We also conducted several Very Long Baseline Interferometry (VLBI) observations using the VLBA through the MOJAVE \citep{Lister2018}\footnote{\url{https://www.cv.nrao.edu/MOJAVE/index.html}} and BEAM-ME \citep{Jorstad2016}\footnote{https://www.bu.edu/blazars/BEAM-ME.html} programs that will be discussed separately in an upcoming publication (Paraschos et al., in preparation). The optical and radio observations are discussed in more detail in Appendix \ref{app_sec:mwl_obs}. We also take advantage of the  {\it Fermi} gamma-ray space telescope light curve repository \citep{repository2023} to extract the 3-day binned light curve. The R-band, millimeter radio (225.5~GHz), and $\gamma$-ray observations are shown in Fig. \ref{plt:pol_all_gamma}. We focus on the highest available radio frequency since that is the least affected from Faraday effects and is likely in the optical thin regime as the optical and X-rays. 

The optical, broadband, linear polarization shows rapid variability between $\sim$2 - 4\% about the jet axis (175$\pm$10 degrees, \citealp{Krichbaum1992,Nagai2010,Suzuki2012,Paraschos2022}). At the same time, the millimeter radio (225.5~GHz) shows milder variability at $\sim1\%$ also aligned with the jet axis. We find a strong correlation between the optical and $\gamma$-ray variability with a time lag about zero (see Appendix \ref{app_sec:cross-correlation}), consistent with typical behavior of blazars that indicates Compton scattering from relativistic electrons in the jet \citep{Liodakis2018,Liodakis2019,deJaeger2023}.

\section{Discussion \& Conclusions}\label{sec:disc_concl}

We presented novel multiwavelength polarization observations of 3C~84 during a bright $\gamma$-ray flare. The unprecedented sensitivity of the IXPE observation allowed us to detected an X-ray polarization degree of $\rm\Pi_X=4.2\pm1.3\%$  that is aligned with the jet axis. Previous IXPE observations of Seyfert 2 type galaxies such as 3C84, show the opposite polarization behavior, i.e., a high $\rm\Pi_X$ that is perpendicular to the jet axis \citep{Ursini2023,Marin2024}. There is in fact now an emerging X-ray polarization unification picture where both stellar and supermassive obscured black holes show the same behavior \citep{Saade2024}. The tendency of 3C~84 to behave more like BL Lac type objects has been already noted \citep{Veron1978} and was also found in our spectropolarimetric observations \citep{Marin2025}. This strongly rejects scattering on the dusty torus or on the polar winds as the origin of the X-ray polarization. 

We also find a low yet strongly variable optical ($\rm\Pi_O\sim$2-4\%) and radio ($\rm\Pi_R\sim$1\%) polarization degree that is also aligned with the jet. Given the orientation of the jet and the limited resolution of the optical and single-dish radio telescopes, the polarization degree is averaged over the entire jet with multiple emission regions with different magnetic field orientations likely depolarizing the observed emission. It is also likely that there is a minor contribution from the polar-scattered, inner thermal emission in the optical bands. However, the strong polarization variability, and the overall blazar-like behavior lead us to conclude that the optical and radio emission are dominated by the jet yet the intrinsic polarization is significantly depolarized in our limited resolution. This would suggest that the intrinsic radio and optical polarization degree (see also Paraschos et al., in preparation) is higher than the X-ray polarization degree.  For flaring jets as is the case for 3C~84, proton processes are expected to produce X-ray polarization at a similar level as the optical polarization \citep{Zhang2024} in tension with our observation. Finally, we find a strong correlation between optical and $\gamma$-rays with a time lag of about zero as is typically expected from Compton scattering from relativistic electrons. That is true for both the long-term behavior but also the flare in 2025 contemporaneous to our multiwavelength campaign (see Appendix \ref{app_sec:cross-correlation}).

All of the above strongly point to Compton scattering from relativistic electrons in the jet as the origin of the X-ray emission. The fact that we detect any polarization is contrary to the expectations from external Compton scattering where the target photon field is coming from structures external to the jet (e.g., accretion disk, broad-line region, torus, etc. \citealp{Zhang2013,Zhang2024}). Instead, it suggests synchrotron self-Compton likely taking place at parsec-scales downstream from the black hole. A similar conclusion was reached for Cen A through multiwavelength modeling \citep{Marin2023} suggesting that similar to blazars \citep{Liodakis2025}, X-rays in radio galaxies are dominated by Compton scattering of seed photons coming from the jet.

\begin{acknowledgments}
The Imaging X-ray Polarimetry Explorer (IXPE) is a joint US and Italian mission.  The US contribution is supported by the National Aeronautics and Space Administration (NASA) and led and managed by its Marshall Space Flight Center (MSFC), with industry partner Ball Aerospace (contract NNM15AA18C)---now, BAE Systems.  The Italian contribution is supported by the Italian Space Agency (Agenzia Spaziale Italiana, ASI) through contract ASI-OHBI-2022-13-I.0, agreements ASI-INAF-2022-19-HH.0 and ASI-INFN-2017.13-H0, and its Space Science Data Center (SSDC) with agreements ASI-INAF-2022-14-HH.0 and ASI-INFN 2021-43-HH.0, and by the Istituto Nazionale di Astrofisica (INAF) and the Istituto Nazionale di Fisica Nucleare (INFN) in Italy. This research used data products provided by the IXPE Team (MSFC, SSDC, INAF, and INFN) and distributed with additional software tools by the High-Energy Astrophysics Science Archive Research Center (HEASARC), at NASA Goddard Space Flight Center (GSFC). This research employs a list of Chandra datasets, obtained by the Chandra X-ray Observatory, contained in~\dataset[DOI: 10.25574/cdc.487]{https://doi.org/10.25574/cdc.487}. Some of the data are based on observations collected at the Observatorio de Sierra Nevada; which is owned and operated by the Instituto de Astrof\'isica de Andaluc\'ia (IAA-CSIC); and at the Centro Astron\'{o}mico Hispano en Andaluc\'ia (CAHA); which is operated jointly by Junta de Andaluc\'{i}a and Consejo Superior de Investigaciones Cient\'{i}ficas (IAA-CSIC). The Perkins Telescope Observatory, located in Flagstaff, AZ, USA, is owned and operated by Boston University. This research was partially supported by the Bulgarian National Science Fund of the Ministry of Education and Science under grants KP-06-H68/4 (2022) and KP-06-H88/4 (2024). The data in this study include observations made with the Nordic Optical Telescope, owned in collaboration by the University of Turku and Aarhus University, and operated jointly by Aarhus University, the University of Turku and the University of Oslo, representing Denmark, Finland and Norway, the University of Iceland and Stockholm University at the Observatorio del Roque de los Muchachos, La Palma, Spain, of the Instituto de Astrofisica de Canarias. The data presented here were obtained in part with ALFOSC, which is provided by the Instituto de Astrof\'{\i}sica de Andaluc\'{\i}a (IAA) under a joint agreement with the University of Copenhagen and NOT. The POLAMI observations reported here were carried out at the IRAM 30m Telescope. IRAM is supported by INSU/CNRS (France), MPG (Germany) and IGN (Spain). The Submillimeter Array (SMA) is a joint project between the Smithsonian Astrophysical Observatory and the Academia Sinica Institute of Astronomy and Astrophysics and is funded by the Smithsonian Institution and the Academia Sinica. Maunakea, the location of the SMA, is a culturally important site for the indigenous Hawaiian people; we are privileged to study the cosmos from its summit. The KVN is a facility operated by the Korea Astronomy and Space Science Institute. The KVN operations are supported by KREONET (Korea Research Environment Open NETwork) which is managed and operated by KISTI (Korea Institute of Science and Technology Information). S. Kang, S.-S. Lee, W. Y. Cheong, S.-H. Kim, and H.-W. Jeong were supported by the National Research Foundation of Korea (NRF) grant funded by the Korea government (MIST) (2020R1A2C2009003, RS-2025-00562700). Partly based on observations with the 100-m telescope of the MPIfR (Max-Planck-Institut f\"ur Radioastronomie) at Effelsberg. Observations with the 100-m radio telescope at Effelsberg have received funding from the European Union's Horizon 2020 research and innovation programme under grant agreement No 101004719 (ORP).  The IAA-CSIC co-authors acknowledge financial support from the Spanish "Ministerio de Ciencia e Innovaci\'{o}n" (MCIN/AEI/ 10.13039/501100011033) through the Center of Excellence Severo Ochoa award for the Instituto de Astrof\'{i}isica de Andaluc\'{i}a-CSIC (CEX2021-001131-S), and through grants PID2019-107847RB-C44 and PID2022-139117NB-C44.  I.L was funded by the European Union ERC-2022-STG - BOOTES - 101076343. Views and opinions expressed are however those of the author(s) only and do not necessarily reflect those of the European Union or the European Research Council Executive Agency. Neither the European Union nor the granting authority can be held responsible for them. J.O.-S. acknowledges financial support from the project ref. AST22\_00001\_9 with founding from the European Union - NextGenerationEU, the \textit{Ministerio de Ciencia, Innovaci\'on y Universidades, Plan de Recuperaci\'on, Transformaci\'on y Resiliencia}, the \textit{Consejer\'ia de Universidad, Investigaci\'on e Innovaci\'on} from the \textit{Junta de Andaluc\'ia} and the \textit{Consejo Superior de Investigaciones Cient\'ificas}, as well as from INFN Cap. U.1.01.01.01.009.
This work has been partially supported by the ASI-INAF program I/004/11/4. The research at Boston University was supported in part by National Science Foundation grant AST-2108622, NASA Fermi Guest Investigator grants 80NSSC23K1507 and 80NSSC23K1508, NASA NuSTAR Guest Investigator grant 80NSSC24K0547, and NASA Swift Guest Investigator grant 80NSSC23K1145.  E. L. was supported by Academy of Finland projects 317636 and 320045. We acknowledge funding to support our NOT observations from the Finnish Centre for Astronomy with ESO (FINCA), University of Turku, Finland (Academy of Finland grant nr 306531). This research has made use of data from the MOJAVE database that is maintained by the MOJAVE team \citep{Lister2018}. YYK was supported by the MuSES project, which has received funding from the European Union (ERC grant agreement No 101142396). Views and opinions expressed are however those of the author(s) only and do not necessarily reflect those of the European Union or ERCEA. Neither the European Union nor the granting authority can be held responsible for them. This work has made use of data from the Joan Oró Telescope (TJO) of the Montsec Observatory (OdM), which is owned by the Catalan Government and operated by the Institute for Space Studies of Catalonia (IEEC). C.C. and D.A. acknowledge support from the European Research Council (ERC) under the Horizon ERC Grants 2021 programme under grant agreement No.101040021. The University of New Hampshire group is supported in part by NASA Astrophysics Astrophysics Data Analysis Program grant 80NSSC24K0636. We acknowledge support from NASA grant number 80NSSC25K0002. 
\end{acknowledgments}

\facilities{IXPE, \textit{Swift}(XRT), Chandra, NuSTAR, \textit{Fermi},
Belogradchik Observatory, Calar Alto Observatory, IRAM-30m, Nordic Optical Telescope, LX-200, Perkins Telescope, VLBA, Sierra Nevada Observatory, SMA, Tuorla Observatory, KVN, Effelsberg 100-m Telescope.}

\appendix

\section{X-ray polarization data analysis}\label{app_sec:xpol}

Although the X-ray emission from 3C 84 itself is point-like, the thermal X-ray emission from the surrounding intracluster medium (ICM) is an astrophysical background that greatly complicates the analysis. Even in the case where the X-ray polarization of the ICM can be assumed to be exactly equal to zero \textit{a priori}, properly disentangling the flux of the thermal and non-thermal AGN components is crucial for properly measuring the polarization degree and its statistical significance. Unfortunately, IXPE has neither the spatial nor spectral resolution to decisively separate these two components. The polarization signal we measure from 3C 84 over the course of these 2.5 Ms could also be contaminated by variations in the background, the AGN flux, or even changes in its polarization state. In this appendix, we describe the results of an extensive battery of tests to which the 3C 84 data were subject to that quantify the potential contribution of these intervening signals. All analysis of the IXPE data uses a combination of \textit{ixpeobssim} version 31.1.0 \citep{Baldini_2022ixpeobssim} and the IXPE analysis tools included with HEASoft v6.35.1 \citep{Heasoft2014}. 

\subsection{Astrophysical and Non X-ray Backgrounds }\label{sec:nxb}
\begin{figure}
    \centering
    \includegraphics[width=0.8\linewidth]{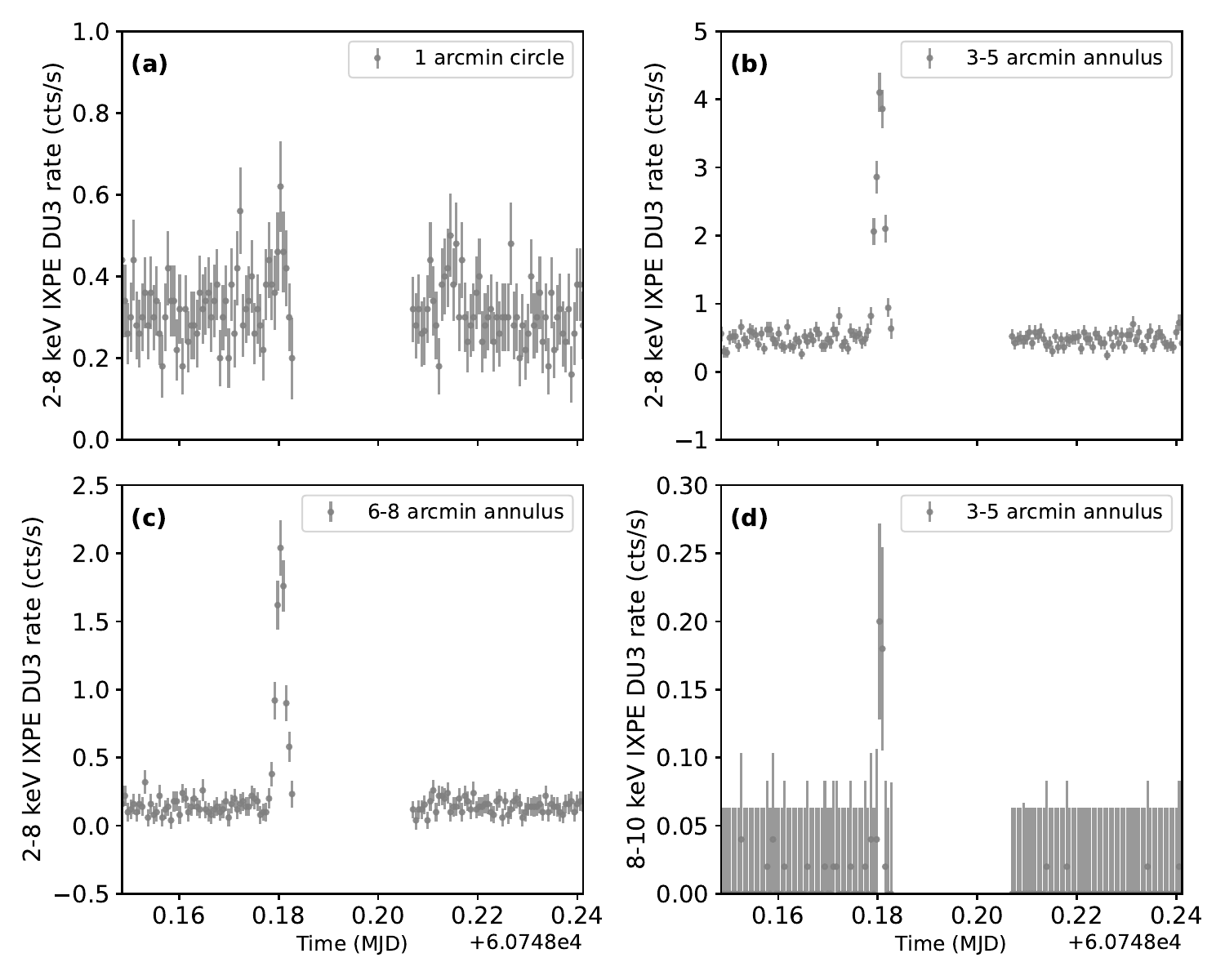}
        \caption{Example of a flare in the IXPE data, with DU3 light curve as a reference. Panel (a) shows the 2-8 keV light curve from a 1 arcmin circle centered around 3C 84. The flare is not obvious. Panels (b) and (c) show the light curve from the same time frame, but from the Perseus cluster from annular regions between 3-5 arcmin and 6-8 arcmin, respectively. The flare is much more prominent in the absence of 3C 84. Panel (d) the 3-5 arcmin light curve in 2-8 keV, outside the science window. The flare is still visible, indicating a likely nonastrophysical origin. In our subsequent analysis, we reject the flare times.}
    \label{fig:Xray_flares}
\end{figure}

\begin{figure}
    \centering
    \includegraphics[width=0.7\linewidth]{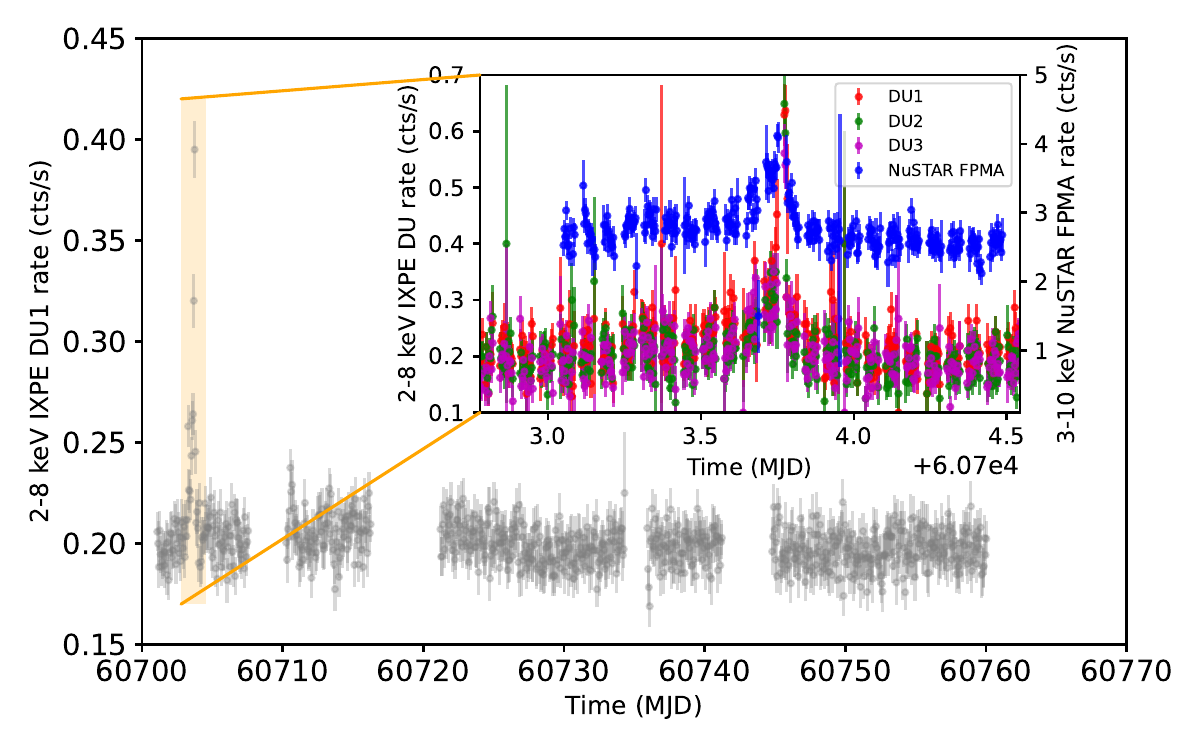}
    \caption{IXPE DU1 light curve of 3C 84 (in gray). The flare, shaded yellow, is observed simultaneously in all IXPE DUs (orange, green, and magenta light curves in the inset), as well as the simultaneous NuSTAR observation (blue light curve in the inset). This part of the light curve coincides with the period of enhanced activity from 3C 84.}
    \label{fig:XrayLC_NuSTAR}
\end{figure}

The long duration of this observation (approximately 2 months of wall clock time from beginning to end) and the fact that these observations occurred near the peak of the solar cycle demand extra scrutiny for periods of high non X-ray background (NXB) event rates. The NXB component is independent of the astrophysical background and can vary strongly with time.  This component is particularly important for IXPE data, for which the background has been shown the presence of spurious polarization signals \citep[][S. Silvestri et al in prep]{Bucciantini2025}. We perform additional good time interval (GTI) filtering of the IXPE observations using light curves measured in the 8-10 keV band (i.e. outside the effective area bandpass of the IXPE mirrors). From these light curves (see Fig. \ref{fig:Xray_flares}), we further filter out times of high activity. As this figure shows, the high rates observed in the 8-10 keV band correspond to similar flaring behavior in the 2-8 keV band. The first segment of the IXPE observation of Perseus also coincided with a period of enhanced $\gamma$-ray activity from 3C 84 (ATel 16998,17020). We detect a flare in the 2-8 keV IXPE light curve in the first segment of observation, concurrent with a similar flare in 3-8 keV NuSTAR data (see Fig. \ref{fig:XrayLC_NuSTAR}) and coincident with the period of renewed $\gamma$-ray activity. While there is little indication of significant spectral change during this flare and the total counts gathered during this flare are only a small fraction of the integrated signal, we nevertheless exclude it from our subsequent analysis.

\begin{figure}
    \centering
    \includegraphics[width=0.45\linewidth]{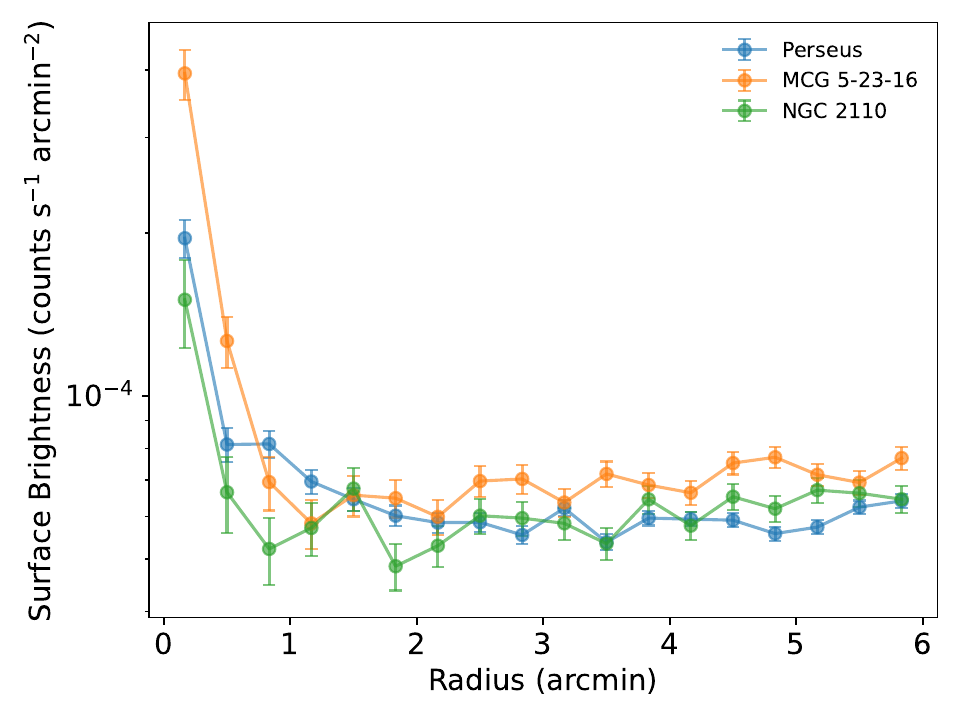}
    \includegraphics[width=0.45\linewidth]{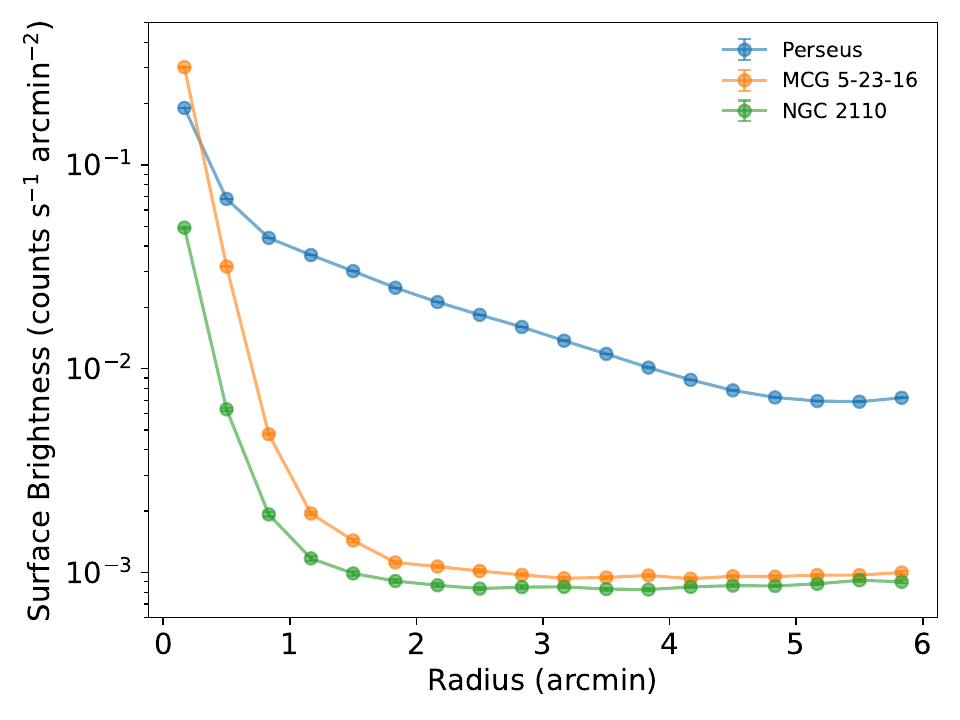}
    \caption{Surface brightness  profiles of Perseus (blue) and two example point source of different brightness, MCG 5-23-16 (orange) and NGC 2110 (green). Left panel: 8-10 keV surface brightness profiles. It demonstrates that the non-X-ray background levels are the same in all three IXPE observations. Right panel: the corresponding 2-8 keV surface brightness profiles. Even at distances as far away as 6 arcminutes, Perseus is at least a factor of $\sim$6 brighter than the point sources. Together, these two panels imply that the cluster is observed throughout and the non-X-ray background is much less than the astrophysical background.}
    \label{fig:sbprofs}
\end{figure}

In order to compare the relative contributions of the astrophysical background and NXB, we compare the surface brightness profile (see Fig. \ref{fig:sbprofs} of the Perseus Cluster with an archival IXPE observation of a brighter point source MCG-5-23-16 (Obs ID 02003299) as well as a fainter source NGC 2110 (Obs ID 03008799). 

In the 8-10 keV band (where the mirrors have no effective area to X-ray photons), we find that the average count rates of both the Perseus cluster and the point sources are similar outside of the central $\sim 1^{\prime}$, showing that their NXB count rates are consistent with each other. When we instead consider the 2-8 keV band, we find that Perseus's surface brightness is at least a factor of 2 larger than the point sources at a distance of $6^{\prime}$ from 3C 84. From these calculations we conclude that the astrophysical background dominates over the NXB throughout the entire usable field of view for the IXPE data. Near 3C 84 the astrophysical background is expected to be at least an order of magnitude higher than the inferred NXB. 

\begin{figure}[!h]
    \centering
    \includegraphics[width=0.5\linewidth]{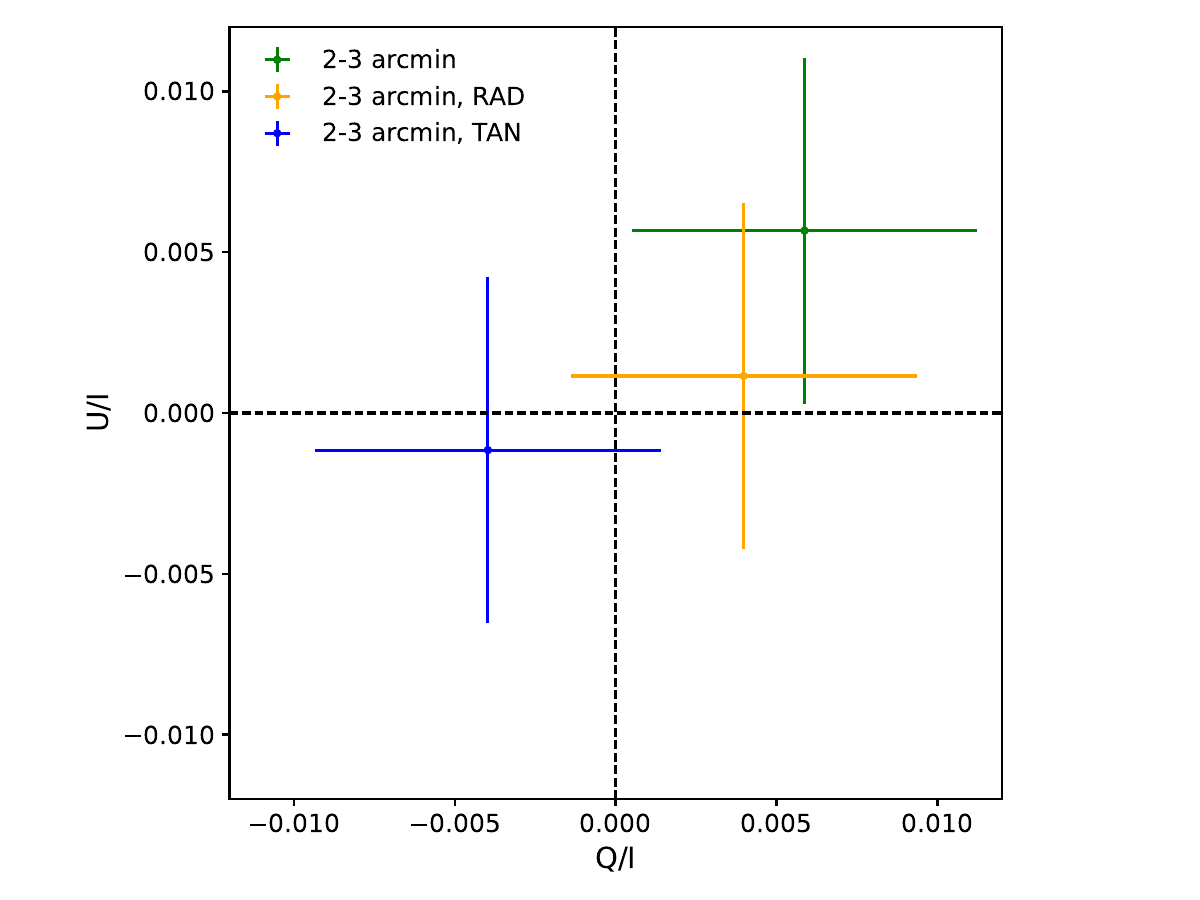}
    \caption{Q-U plot (green) for 2-3 arcmin region of Perseus cluster in 2-8 keV, along with the corresponding values for Stokes rotation along the radial (orange) and azimuthal (blue) directions. This annular region is dominated by thermal emission from the ICM. The error bars are at 1$\sigma$ level. For all three cases, there is no evidence of significant polarization of the ICM photons. 
    }
    \label{fig:q_u_contours}
\end{figure}

Before we further test the spectropolarimetric fit for 3C 84, we have also performed a simple model-independent test of the polarization in the $2^{\prime} - 3^{\prime}$ annular region just outside of 3C 84 using \textit{ixpeobssim} 31.1.0 \footnote{\url{https://ixpeobssim.readthedocs.io/en/latest/}} \citep{Baldini_2022ixpeobssim}. In this region, we find from Fig. \ref{fig:q_u_contours} that our polarization degree is $\sim 0.9\%$ and a value of $\rm MDP_{99} = 1.8 \%$. This provides strong and precise evidence that any polarization signal we measure for 3C 84 is not associated with the astrophysical background or NXB. A second model-independent test was performed after rotating the Stokes parameters for each event in this annulus with respect to its position relative to 3C 84. This rotation enables the possible detection of radial or azimuthal polarization structure that would otherwise average to zero net polarization geometrically. Similar to the first case, we find a polarization degree of $0.4\%$ and $\rm MDP_{99} = 1.6 \% $. When combined with the fact that the astrophysical background count rate in this region is more than an order of magnitude larger than the NXB count rate, we conclude that the NXB can be safely neglected in our spectropolarimetric fit. 

Performing a similar model-independent polarization cube analysis for the $1^{\prime}$ radius circular region around 3C 84 results provides the first evidence of a statistically significant detection of X-ray polarization for this AGN. The resultant polarization cube measures  a polarization degree of $\Pi_{X} = 2.42 \pm 0.79 \%$, corresponding to a $\sim 3\sigma$ detection. The polarization angle for this simple analysis is $\psi_{X} = 164.5^{\circ} \pm 9.3^{\circ}$, which is consistent with being parallel to the radio jet. 

\subsection{Spectral Modeling for 3C 84 and the ICM}

No single X-ray telescope can provide a complete and unambiguous spectral model for both 3C 84 and the surrounding ICM. In order to construct the most accurate model of the AGN plus ICM during our IXPE observations, we leverage existing observations of the Perseus Cluster with multiple X-ray telescopes to develop our joint model. After describing our rigorously derived spectral model for 3C 84+ICM as observed with IXPE, we will discuss the sensitivity of the resultant polarimetric measurement to possible systematic errors arising from modeling deficiencies by considering a few extreme cases when compared to our fiducial model. These will ultimately show that measured polarization signal is robust to systematic uncertainties in our joint model. 

\subsubsection{Chandra Observations}

Our best constraints on the thermal properties of the ICM surrounding 3C 84 come from Chandra observation ID \#4952 (carried out on 2004-10-14, with an exposure of 164.2 ks). This particular observation of the Perseus Cluster uses the ACIS-S detector, which minimizes systematics associated with the ACIS-I chip gaps that would be located in the regions of interest here. Unfortunately, Chandra observations show that 3C 84 itself suffers from pileup, making it impossible to extract an accurate spectrum of the AGN. We perform our analysis of the thermal spectrum using an annular region of $5^{\prime \prime}-30^{\prime \prime}$ centered on 3C 84, where we expect point-like emission from 3C 84 to be negligible even in the presence of pileup\footnote{See Figures 6.12 and 6.23 of the Chandra Proposer's Observatory Guide (\url{https://cxc.harvard.edu/proposer/POG/html/chap6.html}) for discussion for the Chandra on-axis PSF and how it varies with pileup.}. All of our thermal models assume the abundance model of Wilms \citep{Wilms_2000}, consistent with the study of the Perseus Cluster using Hitomi data \cite{Hitomi_2017_abundance}. Unless otherwise noted, all spectral fits are performed in the 0.5-7 keV band. The spectral and subsequent spectropolarimetric analyses are carried out with HEASoft v6.35.1, and errors are stated at 1$\sigma$ confidence level.

\begin{figure}
    \centering
    \includegraphics[width=0.5\linewidth]{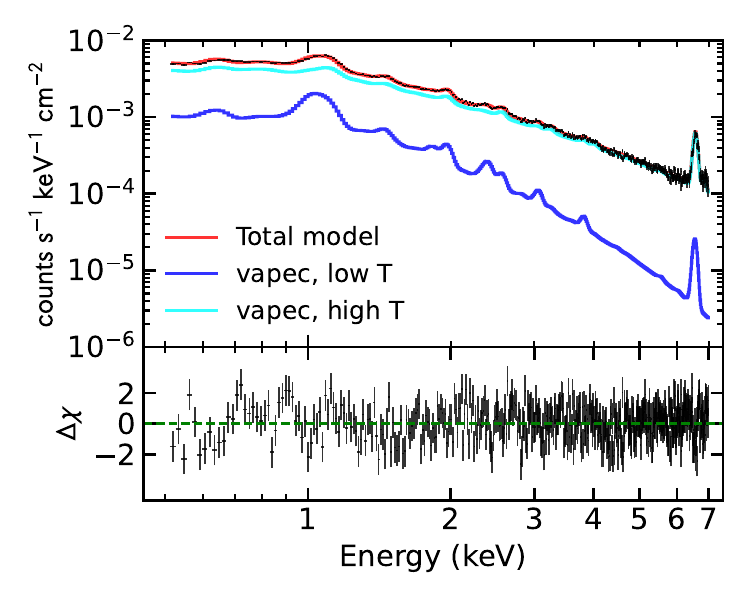}
    \caption{Chandra ACIS spectrum of the Perseus Cluster's thermal emission. We find a two-temperature model with low-temperature (1.65 keV) and high-temperature (4.39 keV) temperature \textsc{vapec} components to be the best description of the 0.5-7 keV spectrum.}
    \label{fig:chandra_spec}
\end{figure}

Because this annular region is significantly larger than the Chandra point spread function (PSF) and the annular binning used in past studies of the Perseus Cluster that study this same region \citep[e.g.][]{Schmidt2002,Fabian2006,Fabian2011,Fabian2017}, we fit its spectrum using several models of varying complexity to account for possible multiphase emission.  Past studies using Chandra observations have shown that subregions of our annular aperture are well described by a single thermal component with a temperature of $kT \sim  3 \thinspace \mathrm{keV}$ \citep{Schmidt2002,Fabian2006,Fabian2011,Fabian2017}, but the additional photon statistics gained by our larger region may provide evidence of additional components. We find that a simple absorbed thermal model (\texttt{tbabs * apec} in XSPEC nomenclature) does not result in a formally acceptable fit to the data ($\chi^{2} = 1038 $ for 437 degrees of freedom).  For this assumed emission model, clear residuals are seen around the emission lines at 1.2 and 2.3 keV. Motivated by the spectral model fit of a larger region of the Perseus Cluster with the Hitomi data \citep{Hitomi_2017_abundance}, we also consider a more complex two-component thermal model with free elemental abundances and intrinsic absorption ( \texttt{tbabs * (vapec + vapec})). We allow the abundances of Si, S, Ar, Ca, and Fe to vary freely with respect to the solar model, which are all free parameters in the Hitomi fit. Hitomi also allowed the Ni abundance as a free parameter, but given that these lines are predominantly above $8.0 $ keV we do not allow it to vary freely here. The abundance values are tied to equal values for the two temperature components. The best-fit model has temperatures of $kT = 1.65^{+0.04}_{-0.09} \thinspace \mathrm{keV}$ and $kT = 4.39^{+0.09}_{-0.20} \thinspace \mathrm{keV}$, respectively, along with supersolar abundances of Si, Ar, and Ca. These parameters are broadly consistent with the values measured in Hitomi and XMM-Newton, and we attribute the detailed differences between these parameters to the different spectral extraction regions and cross-calibration uncertainties associated with these two telescopes\footnote{For context, the Chandra region we have chosen is smaller than a single Hitomi SXS pixel ($1^{\prime} \times 1^{\prime}$), while the Hitomi result uses all of its 25 pixels for three overlapping but slightly offset pointings.}  The addition of a second temperature and free elemental abundances greatly improves the overall quality of the fit with a best-fit value of $\chi^{2} = 498$ and $434$ degrees of freedom (see Fig. \ref{fig:chandra_spec} for the Chandra spectrum, along with the best-fit model).  For completeness, we also consider a model where the thermal emission is a cooling flow from its upper temperature, again with free elemental abundances (\texttt{tbabs*(vapec + mkcflow)}) where the upper temperature of the cooling flow component is tied to the primary \texttt{apec} component, the metallicities of both components are tied to one another, and the normalization corresponds to a residual X-ray cooling flow. For this model, we find that the total cooling flow rate is $\dot{M} = 22 \pm 1.3 M_{\odot} \thinspace \mathrm{yr}^{-1}$ and the fit statistic is $\chi^{2} = 664$ for 438 degrees of freedom. We therefore conclude that this model fits the data significantly worse than our two-temperature model. Since this two-temperature model corresponds to a better overall fit and is further supported by the study of Hitomi, we choose the free elemental abundance model as our baseline for the spectropolarimetric fit. This model, whose parameters are described in Chandra column of Table \ref{table:ngc1275_params}, represents our fiducial thermal model.  

\subsubsection{NuSTAR and Swift Observations}
While the Chandra observations are able to provide a clear measurement of the ICM's thermal properties without contamination from 3C 84, the presence of pileup  makes it  unable to characterize the emission of the AGN itself. To model the AGN's spectrum, we instead use Swift XRT + NuSTAR calibration observations of the Perseus Cluster taken simultaneously with the IXPE observations to characterize the flux and spectrum of 3C 84. Although the Swift + NuSTAR observations do not suffer the same limitations as Chandra for modeling the spectrum of 3C 84, we do expect the observed spectrum in our source aperture to include photons from the ICM and the AGN. The use of data from all three telescopes is therefore crucial for characterizing the AGN spectrum but also the relative flux contributions of the two components.

We reduce the NuSTAR data using \textsc{NUSTARDAS} v2.1.4 using default parameters, in conjunction with CALDB v20250224. We create a GTI file by weeding out the previously mentioned flare using XSELECT. Then, we extract the source photons using \textsc{NUPRODUCTS} task from a 1 arcmin region centered on the AGN, and use a separate 1 arcmin region far away from the source as the background region. We bin the spectrum to a minimum of 25 counts per bin for further XSPEC analysis. The effective NuSTAR exposure time was 45.3 ks for FPMA and 45.0 ks for FPMB.

Swift/XRT observed Perseus multiple times during the IXPE observation. Since all but one of these observations were carried out in WT mode, we exclusively use all the WT mode data during the span of the IXPE observation. We follow the standard extraction method from \cite{Evans_2007,Evans_2009} and use the automated XRT product generation service\footnote{\url{https://www.swift.ac.uk/user_objects/}}. In this, the cleaned event lists and exposure maps per observation are using \textsc{xrtpipeline}, a box of 1 arcmin length is selected around the source, and the entire window excluding a 120-pixel (283 arcsec) wide box centered on the source is designated as the background region. For each of the observations an appropriate ancillary response file (ARF) is generated using \textsc{xrtmkarf} and the corresponding exposure map. The cleaned event lists are then combined, and the combined source and background spectra are generated using \textsc{XSELECT}. ARFs are added using the \textsc{ADDARF} tool, and BACKSCAL keyword is set appropriately. 3C 84 is not piled up in WT mode; hence we do not need any further correction. The combined XRT spectrum had an exposure time of 2.4 ks.

We fit the Swift + NuSTAR data to our fiducial thermal model (two temperatures with free elemental abundances), a power-law component, and a redshifted Gaussian Fe K$\alpha$ emission line. This emission line was found in \cite{Churazov2003} to be associated with 3C 84 itself, and while not visually obvious in our spectra, its inclusion nevertheless makes a significant improvement to the fit. 
Both the temperatures for our two thermal components are fixed to their Chandra-derived values, but their normalizations were allowed to be free parameters in the fit. This step is required to account for the fact that our NuSTAR + Swift data will inevitably sample different physical regions of the cluster given the variations in the PSF between these three telescopes. 

Due to the much lower angular resolution of Swift/XRT and NuSTAR, contributions from farther out gas could change the abundances relative to the Chandra values. That apart, the much lower spectral resolution in this case renders the fit relatively insensitive to the abundances except the Fe abundance. Therefore, for the Swift+NuSTAR fit, we freeze the other abundances to the solar value, keeping only the Fe abundance free.
We restrict our fit to $0.5-10$ keV band for the Swift data and $3-50$ keV for the NuSTAR data. Our best-fit power law model has $\Gamma = 1.64 \pm 0.04$, broadly similar to many values measured in the past. This model has a $\chi^{2}$ value of 1387 for 1357 degrees of freedom. 
We tried a variation of this model by including a second redshifted neutral hydrogen absorption component with zTBabs, but found that the second component attains values similar to $N_{\rm H}$ as the Galactic component with little improvement to the overall fit and leads to degeneracies with the Galactic TBabs component. Therefore, we chose a single TBabs as our preferred model. We find the best-fit $N_{\rm H}$ of $(1.88^{+0.12}_{-0.11}) \times 10^{21} \ {\rm cm^{-2}}$ to be consistent with the total Galactic hydrogen column density derived using the method of \cite{Willingale2013} \footnote{\url{https://www.swift.ac.uk/analysis/nhtot/}}. We also find a best-fit Fe abundance of $0.50 \pm 0.02$.
The fluxes of 3C 84 and the ICM in the 2-8 keV band are $6.6 \times 10^{-11}$ and $7.5 \times 10^{-11}$ $\rm{erg \ cm^{-2} \ s^{-1}}$ respectively, suggesting that the ICM should contribute $\sim 53\%$ of the photons observed in the IXPE aperture.  The equivalent width of the Gaussian line is found to be 77.7 eV.

\subsection{Spectropolarimetric fit with IXPE}

Based on the surface brightness profiles mentioned in Appendix \ref{sec:nxb}, we take an average average 2-8 keV  surface brightness profiles of MCG 5-23-16 and NGC 2110, scaled to match the on-axis count rate of Perseus, as a template for a point source at this flux level, and compare it with the 2-8 keV surface brightness profile of Perseus. This gives us an idea of the relative contribution between the power law component from 3C 84 and the thermal components from the cluster gas, as a function of the off-axis radius. We find that beyond $\sim40^{\prime\prime}$, the cluster dominates over the AGN contribution, thereby diluting the 3C 84 polarization and adding degeneracies. 

We therefore chose a source extraction region below this value, $30^{\prime\prime}$ in radius or about two times the in-orbit half-power diameter of IXPE to be precise. 
We keep the normalizations of the power-law and vapec components free in the subsequent spectropolarimetric fits to account for a different AGN/cluster flux ratio.

\begin{figure}[!h]
    \centering
    \includegraphics[width=0.45\linewidth]{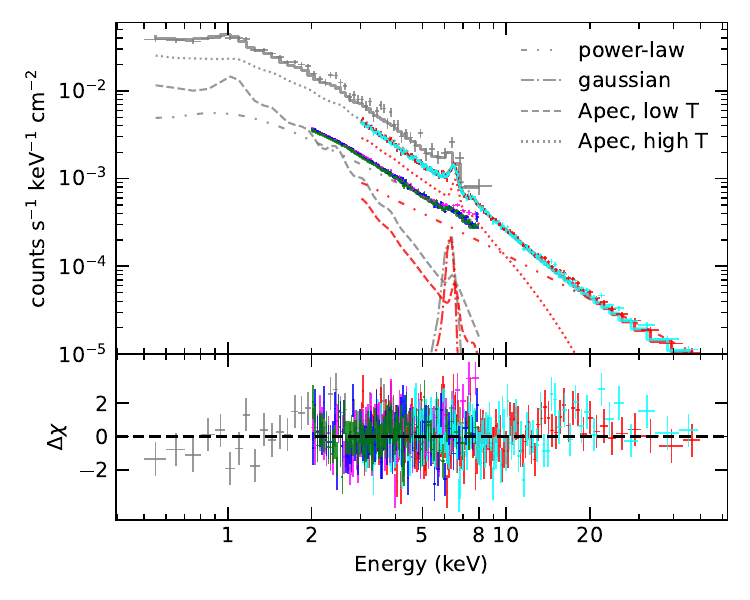}
    \includegraphics[width=0.45\linewidth]{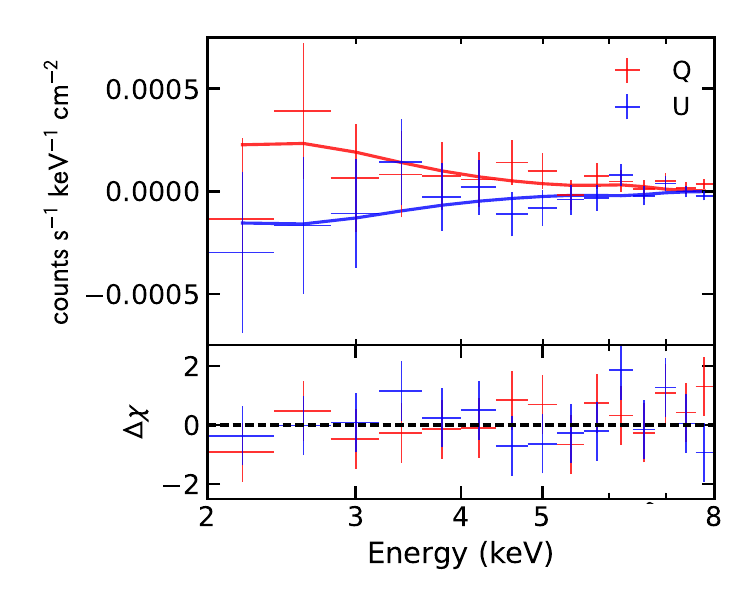}
    \caption{Spectropolarimetric fit between Swift/XRT, NuSTAR and IXPE. Left panel shows the 0.5-50 keV spectral fit, with XRT in gray, IXPE DUs in magenta, blue, and green, and NuSTAR FPMA and FPMB in red, and cyan, respectively. The best-fit model consists of an absorbed power law, a narrow gaussian at Fe K energies, as well as two-temperature thermal model based on our Chandra fit. The right panel shows the corresponding fit for the Q and U spectra. We have shown the fit for only the DU1 for greater clarity. We assumed a constant polarization model for the fit. }
    \label{fig:xrt_nu_ixpe_spectra}
\end{figure}

The combination of Chandra, Swift, and NuSTAR observations provides us with a complete spectral model that we will use to measure the polarization of 3C 84. The full spectral model is found in Table \ref{table:ngc1275_params}, includes both our fiducial thermal model from the Chandra observations and our power-law + Gaussian model from  the Swift + NuSTAR fit. We fit the IXPE, NUSTAR, and Swift data simultaneously in our spectropolarimetric fit (see Fig. \ref{fig:xrt_nu_ixpe_spectra}). To maximize polarization sensitivity, we use the $N_{\rm eff}$ weighting for the polarimetric data.  None of the key free parameters of this complex model fit differ significantly from their best-fit values to the Swift+ NuSTAR fit described above, with all parameter values within $1\sigma$ of one another. We find that the best-fit polarization degree is $\Pi_{X} = 4.18 \pm 1.31\%$, and the best fit angle is $\psi_{X} = 162.8 \pm 9.2^{\circ}$. This corresponds to a detection at the $\sim3.2\sigma$ level and a polarization angle approximately parallel to the jet. 

Further testing of this spectropolarimetric fit shows that this polarization degree is not sensitive to the particulars of the thermal background model. If we instead choose a  simpler single-temperature thermal model for the ICM, we get similar fit quality to the Swift + NuSTAR + IXPE spectra and a polarization degree of $\Pi_{X} = 3.51 \pm 1.12\% $ and $\psi_{X} = 162.4 \pm 9.3^{\circ}$. Even under the extreme and absurd assumption that the entire IXPE spectrum originates from a single power-law component, we measure a polarization degree of $\Pi_{X} = 3.1 \pm 0.88\% \ $ and $\psi_{X} = 165 \pm 7^{\circ}$. 
With these tests, we have shown that our measured polarization degree is robust to the full range of physically feasible thermal emission models.

The Gaussian line is found to be faint in IXPE, with an equivalent width of $\sim$7 eV. If we fix the equivalent width of the Gaussian line in IXPE to the NuSTAR+XRT equivalent width, the fit quality remains relatively unchanged, with a  $\chi^2$/d.o.f.=2741/2690, $\Pi_{\rm X}=4.2\pm1.3$\%, $\psi_{\rm X}=162.8\pm9.2^{\circ}$.
The origin of the Gaussian line, however, is debated. \cite{Aharonian_2018} argues that the Gaussian might be coming from scattering off of a torus. Irrespective of the origin, though, as the gaussian is rather weak in the IXPE data, assigning a polarization of 0 to it does not change our results. We still get a $\chi^2$/d.o.f.=2737/2690, $\Pi_{\rm X}=4.16\pm1.31$\%, $\psi_{\rm X}=162.8\pm9.2^{\circ}$. This further demonstrates the robustness of our polarization measurement to any systematic uncertainty in the emission line.

Although we find that our Swift + NuSTAR spectrum is well fit by a single power-law, we further test the possible presence of Compton scattering from a corona by fitting these spectral data to a non-thermal coronal Compton scattering \citep[specifically the \texttt{Nthcomp} model in \textsc{XSPEC, }][]{Lightman_1987}. We fix the seed photon spectrum to a disk blackbody spectrum with $ kT_{\rm bb}  = 100$ eV. We find that the best-fit \texttt{Nthcomp} model has a power-law index of $\Gamma = 1.67 \pm 0.03$, similar to what we measure for our simple power-law model. These data cannot directly measure the corona electron temperature, finding a lower limit of $kT_{e} > 455 keV$. Just as importantly, the overall fit statistic between the best-fit power-law and \texttt{Nthcomp} models practically indistinguishable ($\chi^2$/d.o.f. of 1387/1357 for the power-law vs 1386/1356 for \texttt{Nthcomp}). The spectral data are therefore unable to favor either the jet-dominated or corona-dominated origins for the X-rays. This is not surprising, as past studies have argued for both origins \citep[e.g.][]{Chitnis_2020,Reynolds_2021}. The nearly identical predictions between the two models further emphasize the power of X-ray polarimetry to break the long-standing degeneracy between these two different physical models for the X-rays.

\subsection{Tests for Polarization Variability}
With a $ \sim 3\sigma$ detection of X-ray polarization over such a long exposure, we further investigate the polarization data to see whether there is any evidence for time variability in the polarization signal. We calculate the average Stokes parameters for each of the five observation segments and perform the same analysis as in \cite{DiGesu2023,Ehlert2023} to test the null hypothesis that the average Stokes parameters in each observation segment are consistent with the time-integrated average. The $\chi^{2}$ value for the time-resolved Stokes parameters is $11.5$ with $2N-2 = 8 $ degrees of freedom. This $\chi^{2}$ value is not large enough to reject the null hypothesis of the Stokes parameters being constant throughout the entire exposure. We therefore conclude that the IXPE data are consistent with a constant polarization degree and angle throughout the exposure. 

\subsection{Test for energy-dependence of polarization}
We further investigated the energy dependence of the polarization using the energy bands of $2-4 \thinspace \mathrm{keV}$, $4-6 \thinspace \mathrm{keV}$, and $6-8 \thinspace \mathrm{keV}$. We carried out a methodology similar to the variability mentioned above by comparing the average polarization degree across the integrated $2-8 \thinspace \mathrm{keV}$ to those measured in these smaller energy bands. We find that the Stokes Q--U parameters for an assumed constant value with energy give a $\chi^2$ of 4.66 with $2N-2 = 4$ degrees of freedom.

As a second independent test, we have also fit the IXPE+NuSTAR+Swift/XRT spectra with \texttt{pollin} model in XSPEC, which assumes a constant + linear polarization degree and angle as a function of energy. We assume no change of the PA with energy, fixing that slope to be 0. Our best-fit value for the polarization degree slope is $A_{\Pi} = 0.01 \pm 0.01$, consistent with a constant polarization degree. The overall fit quality (\texttt{pollin} model, $\chi^2$/d.o.f.=2736/2689) is nearly identical to the fit quality of the model where we assume a constant polarization degree (\texttt{polconst} model,  $\chi^2$/d.o.f.=2737/2690). With these two results, we conclude that there is no statistically significant evidence that the polarization degree changes with energy.

\begin{table*}[t]
\centering
\caption{Best-fit spectral parameters for 3C 84 and the surrounding ICM using a combination of X-ray telescopes. Chandra observations are used to constrain the two thermal components of the ICM, while the NuSTAR + Swift data are included to constrain 3C 84's power-law emission in conjunction with the Chandra-derived thermal emission model. 
Model: \texttt{tbabs*(polconst*(powerlaw+zgauss) + polconst$_0$(vapec$_1$ + vapec$_2$)). }}
\begin{tabular}{ccccc }
\hline
\hline
\rule{0cm}{0.3cm}
 Model & Parameters & Chandra & Swift/XRT+NuSTAR & IXPE\\
 \hline
\textbf{tbabs} & N$_H$ ($\times$ 10$^{22}$ cm$^{-2}$) & $0.176_{-0.001}^{+0.001}$ & $0.188_{-0.011}^{+0.012}$ & $0.184_{-0.010}^{+0.011}$  \\
\hline
\textbf{polconst} & $\Pi$ (\%) & - & - & $4.18 \pm 1.31$  \\
& $\Psi \ (^{\circ})$ & - & - & $162.8_{-9.1}^{+9.2}$ \\
\hline
\textbf{powerlaw} & $\Gamma$ & - & 1.64$_{-0.04}^{+0.04}$ & 1.66$_{-0.03}^{+0.03}$  \\
& Norm & - & $9.2\times 10^{-3}$ & $3.6\times 10^{-2}$ \\
\hline
\textbf{zgauss} & Line energy (keV) & - & $6.4^{f}$ & $6.4^{f}$ \\
& Line width (keV) & - & $0^{f}$ & $0^{f}$ \\
& Norm & - & $1.6\times 10^{-4}$ & $2.2\times 10^{-5}$ \\
\hline
\textbf{polconst$_0$} & $\Pi$ (\%) & - & - & 0$^{f}$  \\
& $\Psi \ (^{\circ})$ & - & - & 0$^{f}$ \\
\hline
\textbf{vapec$_1$} & $kT$ (keV) & $1.65^{+0.04}_{-0.10}$ & $1.65^{f}$ & $1.65^{f}$ \\
& Norm  & $4.15\times 10^{-3}$ & $5.56\times 10^{-2}$ & $6.55\times 10^{-2}$ \\
\hline
\textbf{vapec$_2$} & $kT$ (keV) & $4.39^{+0.09}_{-0.20}$ & $4.39^{f}$ & $4.39^{f}$ \\
& Si & $1.67^{+0.06}_{-0.06}$ & $1.0^{f}$ & $1.0^{f}$  \\
&  S & $0.99^{+0.06}_{-0.06}$ & $1.0^{f}$ & $1.0^{f}$  \\
& Ar & $1.16^{+0.17}_{-0.17}$ & $1.0^{f}$ & $1.0^{f}$  \\
& Ca & $1.30^{+0.18}_{-0.18}$ & $1.0^{f}$ & $1.0^{f}$  \\
& Fe & $0.87^{+0.02}_{-0.02}$ & $0.50^{+0.02}_{-0.02}$ & $0.50^{f}$  \\
& Norm  & $1.93\times 10^{-2}$ & $1.29\times 10^{-1}$ & $1.29\times 10^{-1}$ \\
\hline
& \textbf{$\chi^2$/d.o.f} & 498/434 & 1387/1357 & 2737/2690 \\
\hline
\end{tabular}
\label{table:ngc1275_params}
\end{table*}

\section{Radio \& optical observations} \label{app_sec:mwl_obs}
Here we further discuss the contemporaneous radio, optical, and archival optical observations. 

\subsection{Polarization observations}

\begin{figure}
\centering
\includegraphics[scale=0.5]{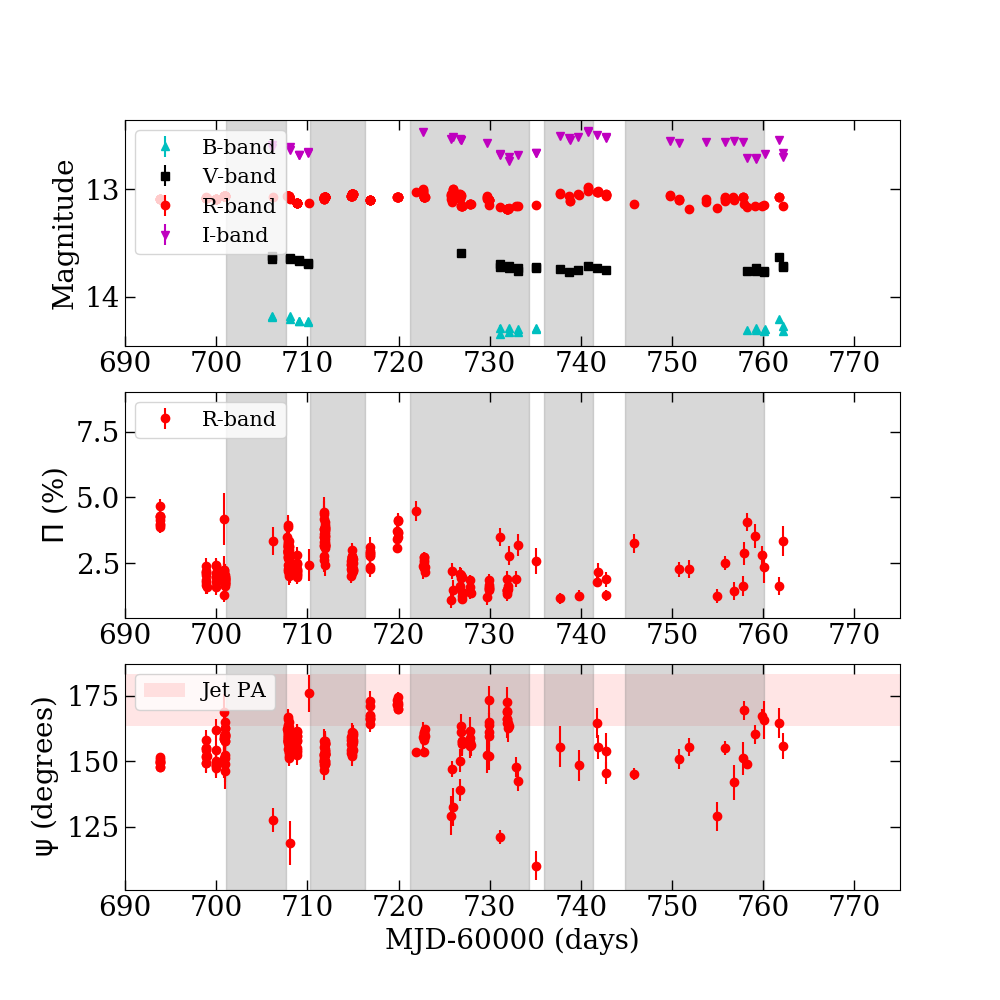}
 \caption{Optical brightness and polarization measurements during the IXPE observations. The top panel shows the brightness in magnitudes for different optical bands (BVRI), the middle panel the host-corrected polarization degree (\%, R-band), and the bottom panel the polarization angle (degrees, R-band). The grey shaded areas mark the IXPE exposures over the entire observation. The different optical bands are marked with different symbols and colors as shown in the legend. The horizontal red shaded area marks the projected direction of the jet. The error bars correspond to the 68\%  (1$\sigma$) confidence interval. }
\label{plt:optical_pol_nogamma}
\end{figure}

\begin{figure}
\centering
\includegraphics[scale=0.5]{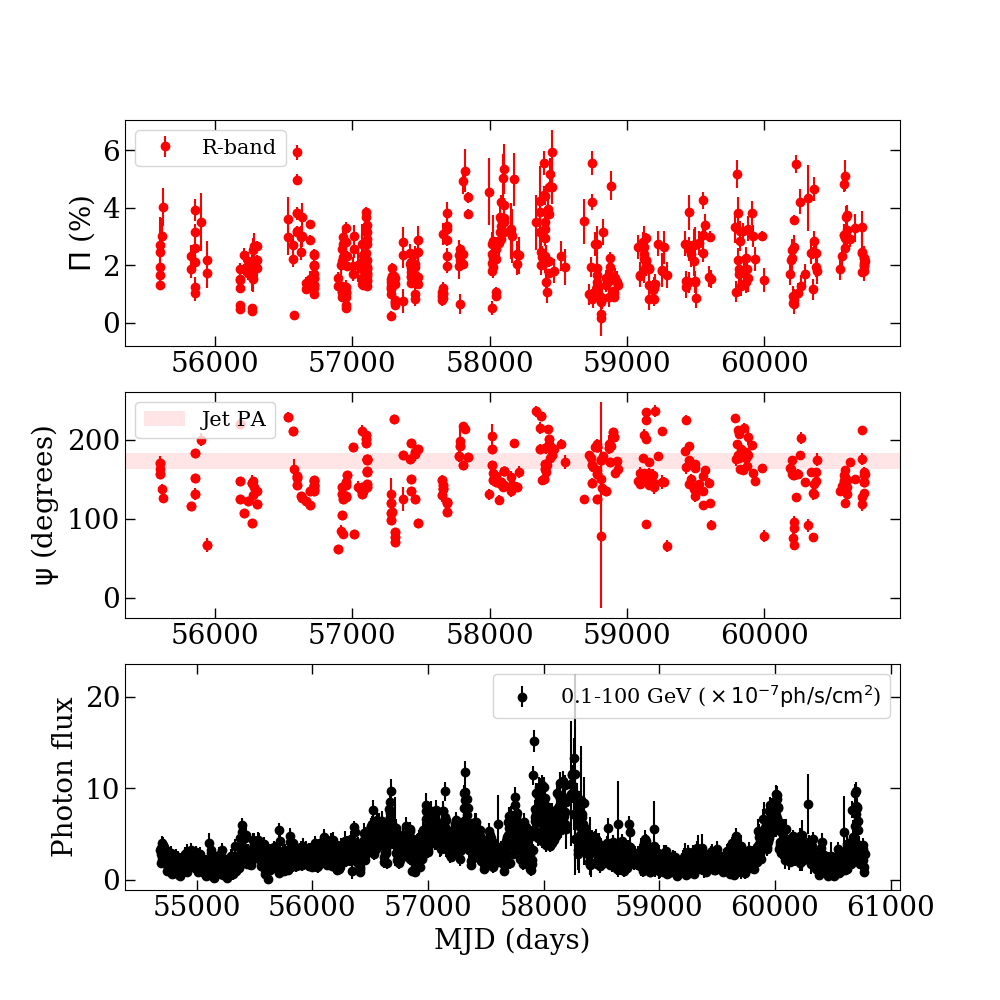}
 \caption{Archival optical and $\gamma$-ray polarization observations. The top panel shows the host-corrected polarization degree (\%, R-band), the middle panel the polarization angle (degrees, R-band) and the bottom panel the 3-day binned {\it Fermi} photon flux light curve in the 0.1-100~GeV band.  The horizontal red shaded area marks the projected direction of the jet. The error bars correspond to the 68\%  (1$\sigma$) confidence interval.}
\label{plt:optical_pol_archival}
\end{figure}

\begin{figure}
\centering
\includegraphics[scale=0.5]{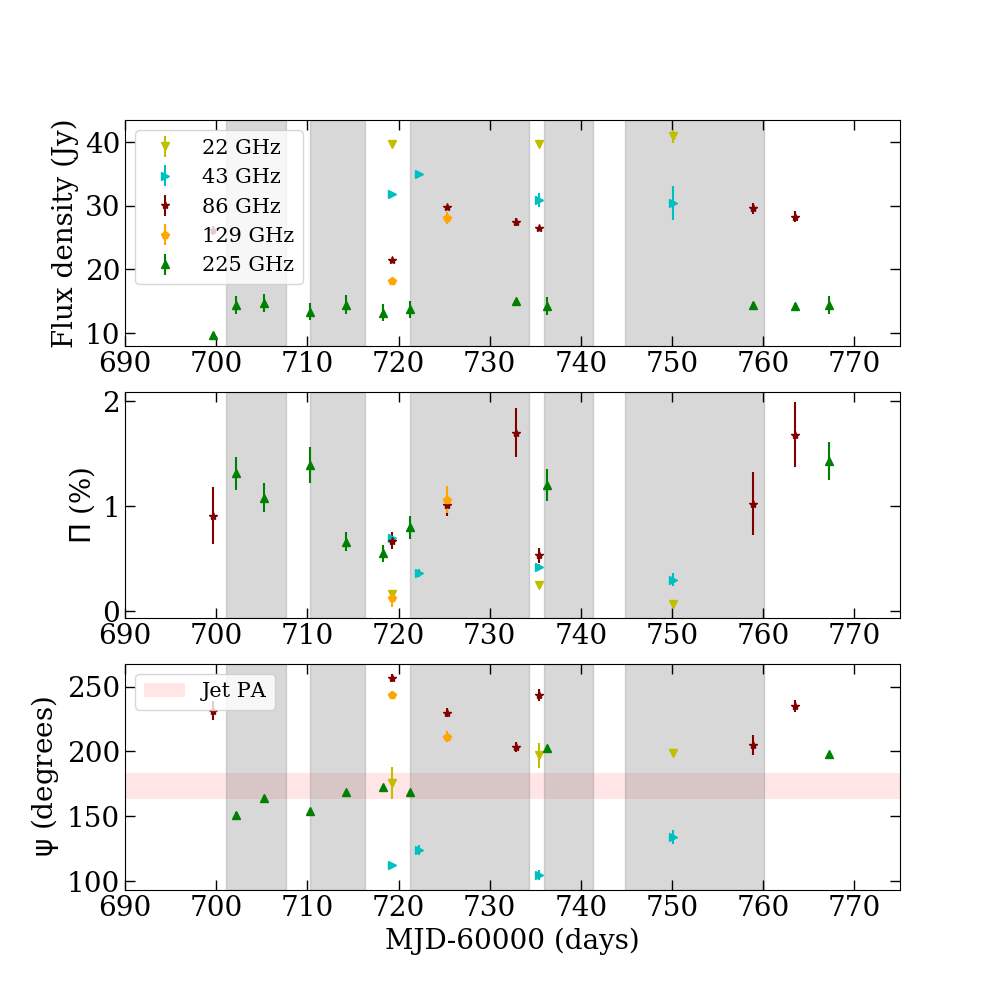}
 \caption{Radio flux density and polarization measurements during the IXPE observations. The top panel shows the flux density in Jansky, the middle panel the polarization degree (\%) and the bottom panel the polarization angle (degrees). The grey shaded areas mark the IXPE exposures over the entire observation. The different radio frequencies are marked with different symbols and colors as shown in the legend. The horizontal red shaded area marks the projected direction of the jet. The error bars correspond to the 68\%  (1$\sigma$) confidence interval.}
\label{plt:radio_pol_nogamma}
\end{figure}

Optical observations covering the IXPE observation were taken using multiple telescopes across Europe and the United States. Those were namely the Belogradchik Observatory in Bulgaria \citep{Bachev2023}, the Calar Alto Observatory \citep{juan_escudero:2023,escudero2024} and Sierra Nevada Observatory \citep{otero2024} in continental Spain, the Nordic Optical Telescope \citep{Nilsson2018,MAGIC2018} in La Palma, Spain, the LX-200 in Russia, and the Perkins Telescope \citep{Jorstad2010} in the United States. All the observations were reduced following standard polarimetric procedures and corrected for instrumental polarization using unpolarized and polarized standard stars. More details on the observations, data reduction, and observing strategies of the individual observatories can be found in previous IXPE-collaboration-led publications \cite[e.g.,][]{Liodakis2022,DiGesu2022,Peirson2023,Kouch2024}.

3C~84 has a prominent host galaxy that contributes a significant fraction of unpolarized light. This depolarizing effect is taken into account in broadband measurements by making a model of the host galaxy light distribution (as in \citealp{Nilsson2007}) and estimating the flux density contribution of the host and $H_\alpha$ line (see \citealp{Aleksic2014}) to the aperture used for the individual observatories. That contribution is then subtracted from the total light, and the observed polarization degree is corrected following \cite{Hovatta2016}. The intrinsic R-band polarization degree is plotted in Figure \ref{plt:optical_pol_nogamma}. Spectropolarimetry taken with the Nordic Optical Telescope are presented in an accompanying publication \citep{Marin2025}. In addition, we obtained archival optical polarization data from the Boston University monitoring program (now called BEAM-ME) to compare with the measurements obtained during the IXPE observation. We find a consistent behavior where the polarization degree varies from 2\% to 4\%, aligned with the jet direction (Fig. \ref{plt:optical_pol_archival}).

In radio,  3C~84 was observed in nine frequencies from 2.5 to 225.5~GHz using the Effelsberg 100-m telescope as part of the Monitoring the Stokes $Q$, $U$, $I$ and $V$ Emission of AGN jets in Radio (QUIVER) program \cite[2.5, 4.8, 8.3, 10.4~GHz, ][]{Krauss2003,Myserlis2018,Myserlis2025}, the Korean VLBI Network \citep{Kang2015} in single dish mode (22, 43, 86, 129~GHz), the IRAM-30m telescope through the Polarimetric Monitoring of AGN at Millimeter Wavelengths \citep{agudo2018} program (86, 225~GHz) and by the SMA Monitoring of AGNs with Polarization (SMAPOL) program (225~GHz, \citealp{Myserlis2025}) using the Submillimeter Array. The radio observations are shown in Figure \ref{plt:radio_pol_nogamma}.

\subsection{Cross-correlation analysis}\label{app_sec:cross-correlation}

\begin{figure}
\centering
\includegraphics[scale=0.5]{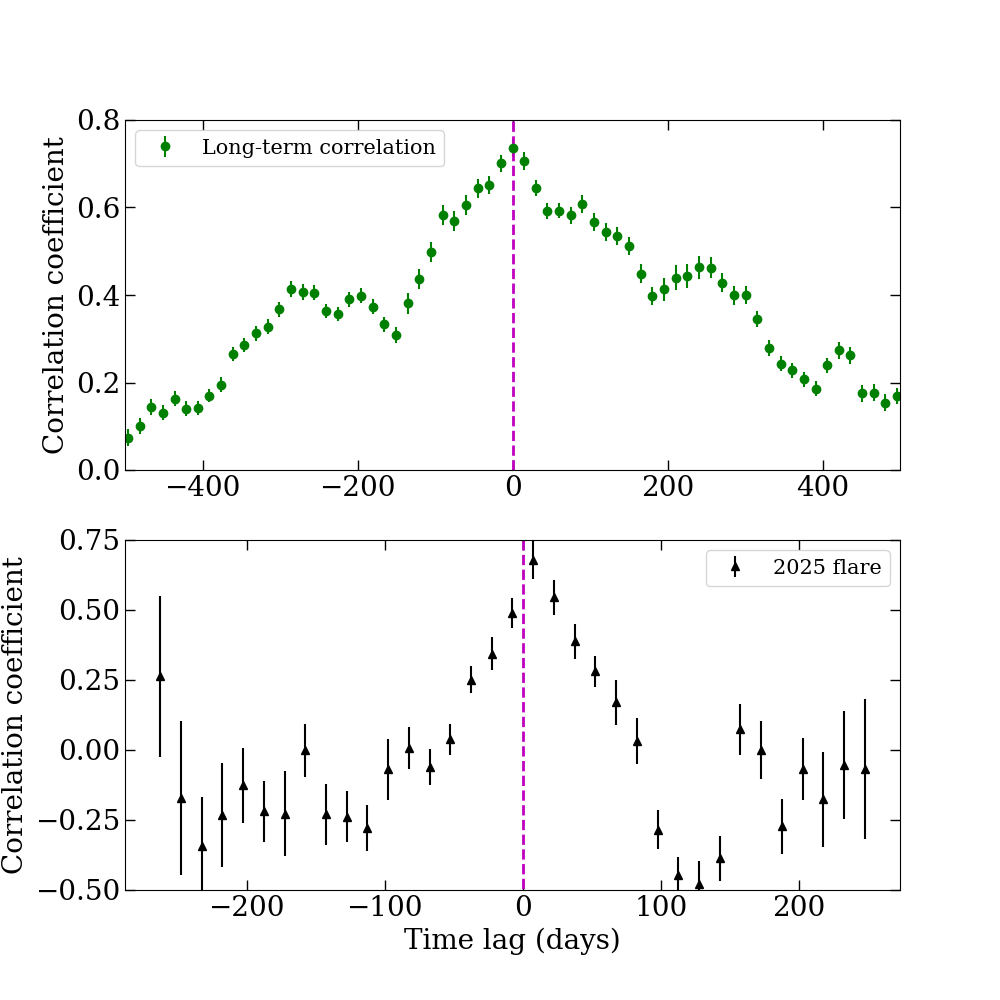}
 \caption{DCF between optical and the 3-day binned $\gamma$-ray light curves. The top panel shows the DCF considering all the available observations, and the bottom panel shows the DCF for the 2025 flare that coincided with the IXPE observations. In both panel the vertical magenta line marks a time lag of zero.}
\label{plt:dcf}
\end{figure}

To probe the optical-$\gamma$-ray connection, we use contemporaneous and archival data from the Tuorla Monitoring program \citep{Nilsson2018} \footnote{\url{https://tuorlablazar.utu.fi/}}. The observations are taken in the R-band and are corrected for the host galaxy contribution revealing a blazar-like light curve with several flares that appear to be well aligned with the $\gamma$-ray flares observed by {\it Fermi}. To better quantify the optical-$\gamma$-ray connection, we use the Discrete Correlation Function \cite[DCF, ][]{Edelson1988}. We find a strong correlation between the long-term optical and $\gamma$-ray light curves with a time lag that is consistent with zero (8$\pm$7 days, Fig. \ref{plt:dcf} - top panel). We repeated the analysis considering only the observations of the flare in 2025 that was contemporaneous to the IXPE observation. Similar to the long-term behavior, we also find the time lag to be consistent with zero (10$\pm$6 days, Fig. \ref{plt:dcf} - top panel), suggesting that optical and $\gamma$-rays are emitted cospatially in both cases as expected from inverse Compton scattering.

\bibliography{sample7}{}

@INPROCEEDINGS{Myserlis2025,
       author = {{Myserlis}, I. and {Agudo}, I. and {Thum}, C. and {Rao}, R. and {Homan}, D.~C. and {Jorstad}, S.~G. and {Marscher}, A.~P. and {Kraus}, A. and {Angelakis}, E.},
        title = "{The physics of AGN jets revealed by full-Stokes monitoring at mm and cm wavelengths}",
    booktitle = {Highlights of Spanish Astrophysics XII},
         year = 2025,
       editor = {{Manteiga}, M. and {Gonz{\'a}lez-Galindo}, F. and {Labiano-Ortega}, A. and {Mart{\'\i}nez-Gonz{\'a}lez}, M.~J. and {Rea}, N. and {Romero-G{\'o}mez}, M. and {Ulla-Miguel}, A. and {Yepes}, G. and {Rodr{\'\i}guez-L{\'o}pez}, C. and {G{\'o}mez-Garc{\'\i}a}, A. and {Dafonte}, C.},
        month = may,
        pages = {121},
       adsurl = {https://ui.adsabs.harvard.edu/abs/2025hsa..conf..121M}
}

@ARTICLE{Marin2025,
       author = {{Marin}, F. and {Pursimo}, T. and {Liodakis}, I. and {Lindfors}, E. and {Biedermann}, J. and {Hutsem{\'e}kers}, D. and {Turkki}, M.},
        title = "{Spectropolarimetry of NGC 1275 reveals a narrow-line radio galaxy with polarization parallel to its radio jet axis}",
      journal = {arXiv e-prints},
         year = 2025,
        month = oct,
          eid = {arXiv:2510.06713},
        pages = {arXiv:2510.06713},
archivePrefix = {arXiv},
       eprint = {2510.06713},
 primaryClass = {astro-ph.GA},
       adsurl = {https://ui.adsabs.harvard.edu/abs/2025arXiv251006713M}
}

@ARTICLE{Tagliacozzo2023,
       author = {{Tagliacozzo}, D. and {Marinucci}, A. and {Ursini}, F. and {Matt}, G. and {Bianchi}, S. and {Baldini}, L. and {Barnouin}, T. and {Cavero Rodriguez}, N. and {De Rosa}, A. and {Di Gesu}, L. and {Dov{\v{c}}iak}, M. and {Harper}, D. and {Ingram}, A. and {Karas}, V. and {Kim}, D.~E. and {Krawczynski}, H. and {Madejski}, G. and {Marin}, F. and {Middei}, R. and {Marshall}, H.~L. and {Muleri}, F. and {Panagiotou}, C. and {Petrucci}, P. -O. and {Podgorny}, J. and {Poutanen}, J. and {Puccetti}, S. and {Soffitta}, P. and {Tombesi}, F. and {Veledina}, A. and {Zhang}, W. and {Agudo}, I. and {Antonelli}, L.~A. and {Bachetti}, M. and {Baumgartner}, W.~H. and {Bellazzini}, R. and {Bongiorno}, S.~D. and {Bonino}, R. and {Brez}, A. and {Bucciantini}, N. and {Capitanio}, F. and {Castellano}, S. and {Cavazzuti}, E. and {Chen}, C. -T. and {Ciprini}, S. and {Costa}, E. and {Del Monte}, E. and {Di Lalla}, N. and {Di Marco}, A. and {Donnarumma}, I. and {Doroshenko}, V. and {Ehlert}, S.~R. and {Enoto}, T. and {Evangelista}, Y. and {Fabiani}, S. and {Ferrazzoli}, R. and {Garcia}, J.~A. and {Gunji}, S. and {Heyl}, J. and {Iwakiri}, W. and {Jorstad}, S.~G. and {Kaaret}, P. and {Kislat}, F. and {Kitaguchi}, T. and {Kolodziejczak}, J.~J. and {La Monaca}, F. and {Latronico}, L. and {Liodakis}, I. and {Maldera}, S. and {Manfreda}, A. and {Marscher}, A.~P. and {Massaro}, F. and {Mitsuishi}, I. and {Mizuno}, T. and {Negro}, M. and {Ng}, C. -Y. and {O'Dell}, S.~L. and {Omodei}, N. and {Oppedisano}, C. and {Papitto}, A. and {Pavlov}, G.~G. and {Peirson}, A.~L. and {Perri}, M. and {Pesce-Rollins}, M. and {Pilia}, M. and {Possenti}, A. and {Ramsey}, B.~D. and {Rankin}, J. and {Ratheesh}, A. and {Roberts}, O.~J. and {Romani}, R.~W. and {Sgr{\`o}}, C. and {Slane}, P. and {Spandre}, G. and {Swartz}, D.~A. and {Tamagawa}, T. and {Tavecchio}, F. and {Taverna}, R. and {Tawara}, Y. and {Tennant}, A.~F. and {Thomas}, N.~E. and {Trois}, A. and {Tsygankov}, S.~S. and {Turolla}, R. and {Vink}, J. and {Weisskopf}, M.~C. and {Wu}, K. and {Xie}, F. and {Zane}, S.},
        title = "{The geometry of the hot corona in MCG-05-23-16 constrained by X-ray polarimetry}",
      journal = {\mnras},
         year = 2023,
        month = nov,
       volume = {525},
       number = {3},
        pages = {4735-4743},
          doi = {10.1093/mnras/stad2627},
archivePrefix = {arXiv},
       eprint = {2305.10213},
 primaryClass = {astro-ph.HE},
       adsurl = {https://ui.adsabs.harvard.edu/abs/2023MNRAS.525.4735T}
}

@ARTICLE{Gianolli2023,
       author = {{Gianolli}, V.~E. and {Kim}, D.~E. and {Bianchi}, S. and {Ag{\'\i}s-Gonz{\'a}lez}, B. and {Madejski}, G. and {Marin}, F. and {Marinucci}, A. and {Matt}, G. and {Middei}, R. and {Petrucci}, P. -O. and {Soffitta}, P. and {Tagliacozzo}, D. and {Tombesi}, F. and {Ursini}, F. and {Barnouin}, T. and {De Rosa}, A. and {Di Gesu}, L. and {Ingram}, A. and {Loktev}, V. and {Panagiotou}, C. and {Podgorny}, J. and {Poutanen}, J. and {Puccetti}, S. and {Ratheesh}, A. and {Veledina}, A. and {Zhang}, W. and {Agudo}, I. and {Antonelli}, L.~A. and {Bachetti}, M. and {Baldini}, L. and {Baumgartner}, W.~H. and {Bellazzini}, R. and {Bongiorno}, S.~D. and {Bonino}, R. and {Brez}, A. and {Bucciantini}, N. and {Capitanio}, F. and {Castellano}, S. and {Cavazzuti}, E. and {Chen}, C. -T. and {Ciprini}, S. and {Costa}, E. and {Del Monte}, E. and {Di Lalla}, N. and {Di Marco}, A. and {Donnarumma}, I. and {Doroshenko}, V. and {Dov{\v{c}}iak}, M. and {Ehlert}, S.~R. and {Enoto}, T. and {Evangelista}, Y. and {Fabiani}, S. and {Ferrazzoli}, R. and {Garc{\'\i}a}, J.~A. and {Gunji}, S. and {Heyl}, J. and {Iwakiri}, W. and {Jorstad}, S.~G. and {Kaaret}, P. and {Karas}, V. and {Kislat}, F. and {Kitaguchi}, T. and {Kolodziejczak}, J.~J. and {Krawczynski}, H. and {La Monaca}, F. and {Latronico}, L. and {Liodakis}, I. and {Maldera}, S. and {Manfreda}, A. and {Marscher}, A.~P. and {Marshall}, H.~L. and {Massaro}, F. and {Mitsuishi}, I. and {Mizuno}, T. and {Muleri}, F. and {Negro}, M. and {Ng}, C. -Y. and {O'Dell}, S.~L. and {Omodei}, N. and {Oppedisano}, C. and {Papitto}, A. and {Pavlov}, G.~G. and {Peirson}, A.~L. and {Perri}, M. and {Pesce-Rollins}, M. and {Pilia}, M. and {Possenti}, A. and {Ramsey}, B.~D. and {Rankin}, J. and {Roberts}, O.~J. and {Romani}, R.~W. and {Sgr{\`o}}, C. and {Slane}, P. and {Spandre}, G. and {Swartz}, D.~A. and {Tamagawa}, T. and {Tavecchio}, F. and {Taverna}, R. and {Tawara}, Y. and {Tennant}, A.~F. and {Thomas}, N.~E. and {Trois}, A. and {Tsygankov}, S.~S. and {Turolla}, R. and {Vink}, J. and {Weisskopf}, M.~C. and {Wu}, K. and {Xie}, F. and {Zane}, S.},
        title = "{Uncovering the geometry of the hot X-ray corona in the Seyfert galaxy NGC 4151 with IXPE}",
      journal = {\mnras},
         year = 2023,
        month = aug,
       volume = {523},
       number = {3},
        pages = {4468-4476},
          doi = {10.1093/mnras/stad1697},
archivePrefix = {arXiv},
       eprint = {2303.12541},
 primaryClass = {astro-ph.GA},
       adsurl = {https://ui.adsabs.harvard.edu/abs/2023MNRAS.523.4468G}
}

@ARTICLE{Ingram2023,
       author = {{Ingram}, A. and {Ewing}, M. and {Marinucci}, A. and {Tagliacozzo}, D. and {Rosario}, D.~J. and {Veledina}, A. and {Kim}, D.~E. and {Marin}, F. and {Bianchi}, S. and {Poutanen}, J. and {Matt}, G. and {Marshall}, H.~L. and {Ursini}, F. and {De Rosa}, A. and {Petrucci}, P. -O. and {Madejski}, G. and {Barnouin}, T. and {Gesu}, L. Di and {Dov{\v{c}}iak}, M. and {Gianolli}, V.~E. and {Krawczynski}, H. and {Loktev}, V. and {Middei}, R. and {Podgorny}, J. and {Puccetti}, S. and {Ratheesh}, A. and {Soffitta}, P. and {Tombesi}, F. and {Ehlert}, S.~R. and {Massaro}, F. and {Agudo}, I. and {Antonelli}, L.~A. and {Bachetti}, M. and {Baldini}, L. and {Baumgartner}, W.~H. and {Bellazzini}, R. and {Bongiorno}, S.~D. and {Bonino}, R. and {Brez}, A. and {Bucciantini}, N. and {Capitanio}, F. and {Castellano}, S. and {Cavazzuti}, E. and {Chen}, C. -T. and {Ciprini}, S. and {Costa}, E. and {Del Monte}, E. and {Lalla}, N. Di and {Marco}, A. Di and {Donnarumma}, I. and {Doroshenko}, V. and {Enoto}, T. and {Evangelista}, Y. and {Fabiani}, S. and {Ferrazzoli}, R. and {Garc{\'\i}a}, J.~A. and {Gunji}, S. and {Heyl}, J. and {Iwakiri}, W. and {Jorstad}, S.~G. and {Kaaret}, P. and {Karas}, V. and {Kislat}, F. and {Kitaguchi}, T. and {Kolodziejczak}, J.~J. and {Monaca}, F. La and {Latronico}, L. and {Liodakis}, I. and {Maldera}, S. and {Manfreda}, A. and {Marscher}, A.~P. and {Mitsuishi}, I. and {Mizuno}, T. and {Muleri}, F. and {Negro}, M. and {Ng}, C. -Y. and {O'Dell}, S.~L. and {Omodei}, N. and {Oppedisano}, C. and {Papitto}, A. and {Pavlov}, G.~G. and {Peirson}, A.~L. and {Perri}, M. and {Pesce-Rollins}, M. and {Pilia}, M. and {Possenti}, A. and {Ramsey}, B.~D. and {Rankin}, J. and {Roberts}, O.~J. and {Romani}, R.~W. and {Sgr{\`o}}, C. and {Slane}, P. and {Spandre}, G. and {Swartz}, D.~A. and {Tamagawa}, T. and {Tavecchio}, F. and {Taverna}, R. and {Tawara}, Y. and {Tennant}, A.~F. and {Thomas}, N.~E. and {Trois}, A. and {Tsygankov}, S.~S. and {Turolla}, R. and {Vink}, J. and {Weisskopf}, M.~C. and {Wu}, K. and {Xie}, F. and {Zane}, S.},
        title = "{The X-ray polarization of the Seyfert 1 galaxy IC 4329A}",
      journal = {\mnras},
         year = 2023,
        month = nov,
       volume = {525},
       number = {4},
        pages = {5437-5449},
          doi = {10.1093/mnras/stad2625},
archivePrefix = {arXiv},
       eprint = {2305.13028},
 primaryClass = {astro-ph.HE},
       adsurl = {https://ui.adsabs.harvard.edu/abs/2023MNRAS.525.5437I}
}

@ARTICLE{Chakraborty2025,
       author = {{Chakraborty}, Sudip and {Ratheesh}, Ajay and {Tagliacozzo}, Daniele and {Kaaret}, Philip and {Podgorn{\'y}}, Jakub and {Marin}, Fr{\'e}d{\'e}ric and {Tombesi}, Francesco and {Ehlert}, Steven R. and {Chen}, Chien-Ting J. and {Kim}, Dawoon E. and {Liodakis}, Ioannis and {Ursini}, Francesco and {Middei}, Riccardo and {Di Marco}, Alessandro and {La Monaca}, Fabio and {Banerjee}, Srimanta and {Fukumura}, Keigo and {Maksym}, W. Peter and {Miku{\v{s}}incov{\'a}}, Romana and {Nemmen}, Rodrigo and {Petrucci}, Pierre-Olivier and {Soffitta}, Paolo and {Svoboda}, Ji{\v{r}}{\'\i} and {Zhang}, Wenda},
        title = "{First X-Ray Polarimetric View of a Low-luminosity Active Galactic Nucleus: The Case of NGC 2110}",
      journal = {\apj},
         year = 2025,
        month = sep,
       volume = {990},
       number = {1},
          eid = {89},
        pages = {89},
          doi = {10.3847/1538-4357/ade87d},
archivePrefix = {arXiv},
       eprint = {2503.01071},
 primaryClass = {astro-ph.HE},
       adsurl = {https://ui.adsabs.harvard.edu/abs/2025ApJ...990...89C}
}

@ARTICLE{Ursini2023,
       author = {{Ursini}, F. and {Marinucci}, A. and {Matt}, G. and {Bianchi}, S. and {Marin}, F. and {Marshall}, H.~L. and {Middei}, R. and {Poutanen}, J. and {Rogantini}, D. and {De Rosa}, A. and {Di Gesu}, L. and {Garc{\'\i}a}, J.~A. and {Ingram}, A. and {Kim}, D.~E. and {Krawczynski}, H. and {Puccetti}, S. and {Soffitta}, P. and {Svoboda}, J. and {Tombesi}, F. and {Weisskopf}, M.~C. and {Barnouin}, T. and {Perri}, M. and {Podgorny}, J. and {Ratheesh}, A. and {Zaino}, A. and {Agudo}, I. and {Antonelli}, L.~A. and {Bachetti}, M. and {Baldini}, L. and {Baumgartner}, W.~H. and {Bellazzini}, R. and {Bongiorno}, S.~D. and {Bonino}, R. and {Brez}, A. and {Bucciantini}, N. and {Capitanio}, F. and {Castellano}, S. and {Cavazzuti}, E. and {Ciprini}, S. and {Costa}, E. and {Del Monte}, E. and {Di Lalla}, N. and {Di Marco}, A. and {Donnarumma}, I. and {Doroshenko}, V. and {Dovciak}, M. and {Ehlert}, S.~R. and {Enoto}, T. and {Evangelista}, Y. and {Fabiani}, S. and {Ferrazzoli}, R. and {Gunji}, S. and {Heyl}, J. and {Iwakiri}, W. and {Jorstad}, S.~G. and {Karas}, V. and {Kitaguchi}, T. and {Kolodziejczak}, J.~J. and {La Monaca}, F. and {Latronico}, L. and {Liodakis}, I. and {Maldera}, S. and {Manfreda}, A. and {Marscher}, A.~P. and {Mitsuishi}, I. and {Mizuno}, T. and {Muleri}, F. and {Ng}, C.~Y. and {O'Dell}, S.~L. and {Omodei}, N. and {Oppedisano}, C. and {Papitto}, A. and {Pavlov}, G.~G. and {Peirson}, A.~L. and {Pesce-Rollins}, M. and {Petrucci}, P. -O. and {Pilia}, M. and {Possenti}, A. and {Ramsey}, B.~D. and {Rankin}, J. and {Romani}, R.~W. and {Sgr{\`o}}, C. and {Slane}, P. and {Spandre}, G. and {Tamagawa}, T. and {Tavecchio}, F. and {Taverna}, R. and {Tawara}, Y. and {Tennant}, A.~F. and {Thomas}, N.~E. and {Trois}, A. and {Tsygankov}, S.~S. and {Turolla}, R. and {Vink}, J. and {Wu}, K. and {Xie}, F. and {Zane}, S.},
        title = "{Mapping the circumnuclear regions of the Circinus galaxy with the Imaging X-ray Polarimetry Explorer}",
      journal = {\mnras},
         year = 2023,
        month = feb,
       volume = {519},
       number = {1},
        pages = {50-58},
          doi = {10.1093/mnras/stac3189},
archivePrefix = {arXiv},
       eprint = {2211.01697},
 primaryClass = {astro-ph.HE},
       adsurl = {https://ui.adsabs.harvard.edu/abs/2023MNRAS.519...50U}
}

@ARTICLE{Singh2011,
       author = {{Singh}, V. and {Shastri}, P. and {Risaliti}, G.},
        title = "{X-ray spectral properties of Seyfert galaxies and the unification scheme}",
      journal = {\aap},
         year = 2011,
        month = aug,
       volume = {532},
          eid = {A84},
        pages = {A84},
          doi = {10.1051/0004-6361/201016387},
archivePrefix = {arXiv},
       eprint = {1101.0252},
 primaryClass = {astro-ph.HE},
       adsurl = {https://ui.adsabs.harvard.edu/abs/2011A&A...532A..84S}
}

@ARTICLE{Kara2016,
       author = {{Kara}, E. and {Alston}, W.~N. and {Fabian}, A.~C. and {Cackett}, E.~M. and {Uttley}, P. and {Reynolds}, C.~S. and {Zoghbi}, A.},
        title = "{A global look at X-ray time lags in Seyfert galaxies}",
      journal = {\mnras},
         year = 2016,
        month = oct,
       volume = {462},
       number = {1},
        pages = {511-531},
          doi = {10.1093/mnras/stw1695},
archivePrefix = {arXiv},
       eprint = {1605.02631},
 primaryClass = {astro-ph.HE},
       adsurl = {https://ui.adsabs.harvard.edu/abs/2016MNRAS.462..511K}
}

@ARTICLE{Haardt1991,
       author = {{Haardt}, F. and {Maraschi}, L.},
        title = "{A Two-Phase Model for the X-Ray Emission from Seyfert Galaxies}",
      journal = {\apjl},
         year = 1991,
        month = oct,
       volume = {380},
        pages = {L51},
          doi = {10.1086/186171},
       adsurl = {https://ui.adsabs.harvard.edu/abs/1991ApJ...380L..51H}
}

@ARTICLE{Mastichiadis2021,
       author = {{Mastichiadis}, Apostolos and {Petropoulou}, Maria},
        title = "{Hadronic X-Ray Flares from Blazars}",
      journal = {\apj},
         year = 2021,
        month = jan,
       volume = {906},
       number = {2},
          eid = {131},
        pages = {131},
          doi = {10.3847/1538-4357/abc952},
archivePrefix = {arXiv},
       eprint = {2009.12158},
 primaryClass = {astro-ph.HE},
       adsurl = {https://ui.adsabs.harvard.edu/abs/2021ApJ...906..131M}
}

@ARTICLE{Mastichiadis1996,
       author = {{Mastichiadis}, Apostolos},
        title = "{The Hadronic Model of Active Galactic Nuclei}",
      journal = {\ssr},
         year = 1996,
        month = jan,
       volume = {75},
       number = {1-2},
        pages = {317-329},
          doi = {10.1007/BF00195042},
       adsurl = {https://ui.adsabs.harvard.edu/abs/1996SSRv...75..317M}
}

@ARTICLE{Mannheim1993-II,
       author = {{Mannheim}, K.},
        title = "{The proton blazar.}",
      journal = {\aap},
         year = 1993,
        month = mar,
       volume = {269},
        pages = {67-76},
          doi = {10.48550/arXiv.astro-ph/9302006},
archivePrefix = {arXiv},
       eprint = {astro-ph/9302006},
 primaryClass = {astro-ph},
       adsurl = {https://ui.adsabs.harvard.edu/abs/1993A&A...269...67M}
}

@ARTICLE{Mastichiadis1997,
       author = {{Mastichiadis}, A. and {Kirk}, J.~G.},
        title = "{Variability in the synchrotron self-Compton model of blazar emission.}",
      journal = {\aap},
         year = 1997,
        month = apr,
       volume = {320},
        pages = {19-25},
          doi = {10.48550/arXiv.astro-ph/9610058},
archivePrefix = {arXiv},
       eprint = {astro-ph/9610058},
 primaryClass = {astro-ph},
       adsurl = {https://ui.adsabs.harvard.edu/abs/1997A&A...320...19M}
}

@software{Heasoft2014,
       author = {{Nasa High Energy Astrophysics Science Archive Research Center (Heasarc)}},
        title = "{HEAsoft: Unified Release of FTOOLS and XANADU}",
 howpublished = {Astrophysics Source Code Library, record ascl:1408.004},
         year = 2014,
        month = aug,
          eid = {ascl:1408.004},
archivePrefix = {ascl},
       eprint = {1408.004},
       adsurl = {https://ui.adsabs.harvard.edu/abs/2014ascl.soft08004N}
}

@ARTICLE{Capecchiacci2025,
       author = {{Capecchiacci}, Sara and {Liodakis}, Ioannis and {Middei}, Riccardo and {Kim}, Dawoon E. and {Di Gesu}, Laura and {Agudo}, Ivan and {Agis-Gonzalez}, Beatriz and {Arbet-Engels}, Axel and {Blinov}, Dmitry and {Chen}, Chien-Ting and {Ehlert}, Steven R. and {Gau}, Ephraim and {Heckmann}, Lea and {Hu}, Kun and {Jorstad}, Svetlana G. and {Kaaret}, Philip and {Kouch}, Pouya M. and {Krawczynski}, Henric and {Lindfors}, Elina and {Marin}, Frederic and {Marscher}, Alan P. and {Myserlis}, Ioannis and {O'Dell}, Stephen L. and {Pacciani}, Luigi and {Paneque}, David and {Perri}, Matteo and {Puccetti}, Simonetta and {Saade}, M. Lynne and {Tavecchio}, Fabrizio and {Tennant}, Allyn F. and {Traianou}, Efthalia and {Weisskopf}, Martin C. and {Wu}, Kinwah and {Aceituno}, Francisco Jose and {Bonnoli}, Giacomo and {Casanova}, Victor and {Emery}, Gabriel and {Escudero}, Juan and {Morcuende}, Daniel and {Otero-Santos}, Jorge and {Sota}, Alfredo and {Piirola}, Vilppu and {Borman}, George A. and {Kopatskaya}, Evgenia N. and {Larionova}, Elena G. and {Morozova}, Daria A. and {Shishkina}, Ekaterina V. and {Savchenko}, Sergey S. and {Vasilyev}, Andrey A. and {Grishina}, Tatiana S. and {Troitskiy}, Ivan S. and {Zhovtan}, Alexey V. and {McCall}, Callum and {Jermak}, Helen E. and {Steele}, Iain A. and {Bachev}, Rumen and {Strigachev}, Anton and {Imazawa}, Ryo and {Sasada}, Mahito and {Fukazawa}, Yasushi and {Kawabata}, Koji S. and {Uemura}, Makoto and {Mizuno}, Tsunefumi and {Nakaoka}, Tatsuya and {Tochihara}, Sumie and {Akai}, Takahiro and {Akitaya}, Hiroshi and {Berdyugin}, Andrei V. and {Kagitani}, Masato and {Kravtsov}, Vadim and {Poutanen}, Juri and {Sakanoi}, Takeshi and {Alvarez-Ortega}, Diego and {Casadio}, Carolina and {Kang}, Sincheol and {Lee}, Sang-Sung and {Kim}, Sanghyun and {Cheong}, Whee Yeon and {Jeong}, Hyeon-Woo and {Song}, Chanwoo and {Li}, Shan and {Nam}, Myeong-Seok and {Gurwell}, Mark and {Keating}, Garrett and {Rao}, Ramprasad and {Angelakis}, Emmanouil and {Kraus}, Alexander and {Benke}, Petra and {Debbrecht}, Lena and {Eich}, Julia and {Eppel}, Florian and {Gokus}, Andrea and {Hammerich}, Steven and {Hessdorfer}, Jonas and {Kadler}, Matthias and {Kirchner}, Dana and {Paraschos}, Georgios Filippos and {Rosch}, Florian and {Schulga}, Wladislaw},
        title = "{Unveiling blazar synchrotron emission: a multiwavelength polarimetric study of HSP and LSP populations}",
      journal = {arXiv e-prints},
         year = 2025,
        month = aug,
          eid = {arXiv:2508.14168},
        pages = {arXiv:2508.14168},
archivePrefix = {arXiv},
       eprint = {2508.14168},
 primaryClass = {astro-ph.HE},
       adsurl = {https://ui.adsabs.harvard.edu/abs/2025arXiv250814168C}
}

@ARTICLE{MAGIC2018,
       author = {{MAGIC Collaboration} and {Ansoldi}, S. and {Antonelli}, L.~A. and {Arcaro}, C. and {Baack}, D. and {Babi{\'c}} and {}, A. and {Banerjee}, B. and {Bangale}, P. and {Barres de Almeida}, U. and {Barrio}, J.~A. and {Becerra Gonz{\'a}lez}, J. and {Bednarek}, W. and {Bernardini}, E. and {Berse}, R. Ch and {Berti}, A. and {Besenrieder}, J. and {Bhattacharyya}, W. and {Bigongiari}, C. and {Biland}, A. and {Blanch}, O. and {Bonnoli}, G. and {Carosi}, R. and {Ceribella}, G. and {Chatterjee}, A. and {Colak}, S.~M. and {Colin}, P. and {Colombo}, E. and {Contreras}, J.~L. and {Cortina}, J. and {Covino}, S. and {Cumani}, P. and {D'Elia}, V. and {da Vela}, P. and {Dazzi}, F. and {de Angelis}, A. and {de Lotto}, B. and {Delfino}, M. and {Delgado}, J. and {di Pierro}, F. and {Dom{\'\i}nguez}, A. and {Dominis Prester}, D. and {Dorner}, D. and {Doro}, M. and {Einecke}, S. and {Elsaesser}, D. and {Fallah Ramazani}, V. and {Fattorini}, A. and {Fern{\'a}ndez-Barral}, A. and {Ferrara}, G. and {Fidalgo}, D. and {Foffano}, L. and {Fonseca}, M.~V. and {Font}, L. and {Fruck}, C. and {Gallozzi}, S. and {Garc{\'\i}a L{\'o}pez}, R.~J. and {Garczarczyk}, M. and {Gaug}, M. and {Giammaria}, P. and {Godinovi{\'c}} and {}, N. and {Guberman}, D. and {Hadasch}, D. and {Hahn}, A. and {Hassan}, T. and {Hayashida}, M. and {Herrera}, J. and {Hoang}, J. and {Hrupec}, D. and {Inoue}, S. and {Ishio}, K. and {Iwamura}, Y. and {Konno}, Y. and {Kubo}, H. and {Kushida}, J. and {Lamastra}, A. and {Lelas}, D. and {Leone}, F. and {Lindfors}, E. and {Lombardi}, S. and {Longo}, F. and {L{\'o}pez}, M. and {Maggio}, C. and {Majumdar}, P. and {Makariev}, M. and {Maneva}, G. and {Manganaro}, M. and {Mannheim}, K. and {Maraschi}, L. and {Mariotti}, M. and {Mart{\'\i}nez}, M. and {Masuda}, S. and {Mazin}, D. and {Mielke}, K. and {Minev}, M. and {Miranda}, J.~M. and {Mirzoyan}, R. and {Moralejo}, A. and {Moreno}, V. and {Moretti}, E. and {Neustroev}, V. and {Niedzwiecki}, A. and {Nievas Rosillo}, M. and {Nigro}, C. and {Nilsson}, K. and {Ninci}, D. and {Nishijima}, K. and {Noda}, K. and {Nogu{\'e}s}, L. and {Paiano}, S. and {Palacio}, J. and {Paneque}, D. and {Paoletti}, R. and {Paredes}, J.~M. and {Pedaletti}, G. and {Pe{\~n}il}, P. and {Peresano}, M. and {Persic}, M. and {Pfrang}, K. and {Prada Moroni}, P.~G. and {Prandini}, E. and {Puljak}, I. and {Garcia}, J.~R. and {Rhode}, W. and {Rib{\'o}}, M. and {Rico}, J. and {Righi}, C. and {Rugliancich}, A. and {Saha}, L. and {Saito}, T. and {Satalecka}, K. and {Schweizer}, T. and {Sitarek}, J. and {{\v{S}}nidari{\'c}} and {}, I. and {Sobczynska}, D. and {Stamerra}, A. and {Strzys}, M. and {Suri{\'c}} and {}, T. and {Tavecchio}, F. and {Temnikov}, P. and {Terzi{\'c}} and {}, T. and {Teshima}, M. and {Torres-Alb{\`a}}, N. and {Tsujimoto}, S. and {Vanzo}, G. and {Vazquez Acosta}, M. and {Vovk}, I. and {Ward}, J.~E. and {Will}, M. and {Zari{\'c}} and {}, D. and {Ciprini S.} and {Fermi-Lat Collaboration} and {Desiante}, R. and {Barcewicz}, S. and {Hovatta}, T. and {Jormanainen}, J. and {Takalo}, L. and {Reinthal}, R. and {Mankuzhiyil}, N. and {Wierda}, F. and {L{\"a}hteenm{\"a}ki}, A. and {Tammi}, J. and {Tornikoski}, M. and {Vera}, R.~J.~C. and {Kiehlmann}, S. and {Max-Moerbeck}, W. and {Readhead}, A.~C.~S.},
        title = "{The broad-band properties of the intermediate synchrotron peaked BL Lac S2 0109+22 from radio to VHE gamma-rays}",
      journal = {\mnras},
         year = 2018,
        month = oct,
       volume = {480},
       number = {1},
        pages = {879-892},
          doi = {10.1093/mnras/sty1753},
archivePrefix = {arXiv},
       eprint = {1807.02095},
 primaryClass = {astro-ph.HE},
       adsurl = {https://ui.adsabs.harvard.edu/abs/2018MNRAS.480..879M}
}

@ARTICLE{Paraschos2021,
       author = {{Paraschos}, G.~F. and {Kim}, J. -Y. and {Krichbaum}, T.~P. and {Zensus}, J.~A.},
        title = "{Pinpointing the jet apex of 3C 84}",
      journal = {\aap},
         year = 2021,
        month = jun,
       volume = {650},
          eid = {L18},
        pages = {L18},
          doi = {10.1051/0004-6361/202140776},
archivePrefix = {arXiv},
       eprint = {2106.04918},
 primaryClass = {astro-ph.GA},
       adsurl = {https://ui.adsabs.harvard.edu/abs/2021A&A...650L..18P}
}

@ARTICLE{Tugliani2025,
       author = {{Tugliani}, Stefano and {Massaro}, Francesco and {Negro}, Michela and {Bonino}, Raffaella and {Cibrario}, Nicol{\`o} and {Latronico}, Luca and {Maldera}, Simone and {Paggi}, Alessandro and {Tramacere}, Andrea and {Meyer}, Eileen T. and {Liodakis}, Ioannis and {Marshall}, Herman L. and {Perlman}, Eric S.},
        title = "{X-Ray Polarimetric Observations of the Western Hotspot of Pictor A}",
      journal = {\apjl},
         year = 2025,
        month = jun,
       volume = {985},
       number = {2},
          eid = {L32},
        pages = {L32},
          doi = {10.3847/2041-8213/adcdfe},
       adsurl = {https://ui.adsabs.harvard.edu/abs/2025ApJ...985L..32T}
}

@ARTICLE{Liodakis2025,
       author = {{Liodakis}, Ioannis and {Zhang}, Haocheng and {Boula}, Stella and {Middei}, Riccardo and {Otero-Santos}, Jorge and {Blinov}, Dmitry and {Agudo}, Iv{\'a}n and {B{\"o}ttcher}, Markus and {Chen}, Chien-Ting and {Ehlert}, Steven R. and {Jorstad}, Svetlana G. and {Kaaret}, Philip and {Krawczynski}, Henric and {Peirson}, Abel L. and {Romani}, Roger W. and {Tavecchio}, Fabrizio and {Weisskopf}, Martin C. and {Kouch}, Pouya M. and {Lindfors}, Elina and {Nilsson}, Kari and {McCall}, Callum and {Jermak}, Helen E. and {Steele}, Iain A. and {Myserlis}, Ioannis and {Gurwell}, Mark and {Keating}, Garrett K. and {Rao}, Ramprasad and {Kang}, Sincheol and {Lee}, Sang-Sung and {Kim}, Sanghyun and {Cheong}, Whee Yeon and {Jeong}, Hyeon-Woo and {Angelakis}, Emmanouil and {Kraus}, Alexander and {Jos{\'e} Aceituno}, Francisco and {Bonnoli}, Giacomo and {Casanova}, V{\'\i}ctor and {Escudero}, Juan and {Ag{\'\i}s-Gonz{\'a}lez}, Beatriz and {Morcuende}, Daniel and {Sota}, Alfredo and {Bachev}, Rumen and {Grishina}, Tatiana S. and {Kopatskaya}, Evgenia N. and {Larionova}, Elena G. and {Morozova}, Daria A. and {Savchenko}, Sergey S. and {Shishkina}, Ekaterina V. and {Troitskiy}, Ivan S. and {Troitskaya}, Yulia V. and {Vasilyev}, Andrey A.},
        title = "{Origin of the X-ray emission in blazars through multiwavelength polarization}",
      journal = {arXiv e-prints},
         year = 2025,
        month = may,
          eid = {arXiv:2505.13603},
        pages = {arXiv:2505.13603},
          doi = {10.48550/arXiv.2505.13603},
archivePrefix = {arXiv},
       eprint = {2505.13603},
 primaryClass = {astro-ph.HE},
       adsurl = {https://ui.adsabs.harvard.edu/abs/2025arXiv250513603L}
}

@ARTICLE{Marin2023,
       author = {{Marin}, Fr{\'e}d{\'e}ric and {Barnouin}, Thibault and {Ehlert}, Steven R. and {Peirson}, Abel Lawrence and {Lopez-Rodriguez}, Enrique and {Petropoulou}, Maria and {Wu}, Kinwah and {Mart{\'\i}-Vidal}, Iv{\'a}n},
        title = "{An X-rays-to-radio investigation of the nuclear polarization from the radio-galaxy Centaurus A}",
      journal = {\mnras},
         year = 2023,
        month = dec,
       volume = {526},
       number = {4},
        pages = {6321-6329},
          doi = {10.1093/mnras/stad3059},
archivePrefix = {arXiv},
       eprint = {2310.04260},
 primaryClass = {astro-ph.HE},
       adsurl = {https://ui.adsabs.harvard.edu/abs/2023MNRAS.526.6321M}
}

@ARTICLE{Veron1978,
       author = {{Veron}, P.},
        title = "{NGC1275: a BL Lacertae object?}",
      journal = {\nat},
         year = 1978,
        month = mar,
       volume = {272},
       number = {5652},
        pages = {430-431},
          doi = {10.1038/272430a0},
       adsurl = {https://ui.adsabs.harvard.edu/abs/1978Natur.272..430V}
}

@ARTICLE{Khachikian1974,
       author = {{Khachikian}, E.~Y. and {Weedman}, D.~W.},
        title = "{An atlas of Seyfert galaxies.}",
      journal = {\apj},
         year = 1974,
        month = sep,
       volume = {192},
        pages = {581-589},
          doi = {10.1086/153093},
       adsurl = {https://ui.adsabs.harvard.edu/abs/1974ApJ...192..581K}
}

@ARTICLE{Sinitsyna2025,
       author = {{Sinitsyna}, Vera G. and {Sinitsyna}, Vera Y.},
        title = "{Multiwavelength Long-term Studies of Radio Galaxy NGC 1275}",
      journal = {\apj},
         year = 2025,
        month = may,
       volume = {985},
       number = {1},
          eid = {39},
        pages = {39},
          doi = {10.3847/1538-4357/adc112},
       adsurl = {https://ui.adsabs.harvard.edu/abs/2025ApJ...985...39S}
}

@ARTICLE{Saade2024,
       author = {{Saade}, M. Lynne and {Kaaret}, Philip and {Liodakis}, Ioannis and {Ehlert}, Steven R.},
        title = "{A Comparison of the X-Ray Polarimetric Properties of Stellar and Supermassive Black Holes}",
      journal = {\apj},
         year = 2024,
        month = oct,
       volume = {974},
       number = {1},
          eid = {101},
        pages = {101},
          doi = {10.3847/1538-4357/ad73a3},
archivePrefix = {arXiv},
       eprint = {2408.12746},
 primaryClass = {astro-ph.HE},
       adsurl = {https://ui.adsabs.harvard.edu/abs/2024ApJ...974..101S}
}

@ARTICLE{Marin2024,
       author = {{Marin}, F. and {Marinucci}, A. and {Laurenti}, M. and {Kim}, D.~E. and {Barnouin}, T. and {Di Marco}, A. and {Ursini}, F. and {Bianchi}, S. and {Ravi}, S. and {Marshall}, H.~L. and {Matt}, G. and {Chen}, C. -T. and {Gianolli}, V.~E. and {Ingram}, A. and {Middei}, R. and {Maksym}, W.~P. and {Panagiotou}, C. and {Podgorny}, J. and {Puccetti}, S. and {Ratheesh}, A. and {Tombesi}, F. and {Agudo}, I. and {Antonelli}, L.~A. and {Bachetti}, M. and {Baldini}, L. and {Baumgartner}, W. and {Bellazzini}, R. and {Bongiorno}, S. and {Bonino}, R. and {Brez}, A. and {Bucciantini}, N. and {Capitanio}, F. and {Castellano}, S. and {Cavazzuti}, E. and {Ciprini}, S. and {Costa}, E. and {De Rosa}, A. and {Del Monte}, E. and {Di Gesu}, L. and {Di Lalla}, N. and {Donnarumma}, I. and {Doroshenko}, V. and {Dov{\v{c}}iak}, M. and {Ehlert}, S. and {Enoto}, T. and {Evangelista}, Y. and {Fabiani}, S. and {Ferrazzoli}, R. and {Garcia}, J. and {Gunji}, S. and {Heyl}, J. and {Iwakiri}, W. and {Jorstad}, S. and {Kaaret}, P. and {Karas}, V. and {Kislat}, F. and {Kitaguchi}, T. and {Kolodziejczak}, J. and {Krawczynski}, H. and {La Monaca}, F. and {Latronico}, L. and {Liodakis}, I. and {Madejski}, G. and {Maldera}, S. and {Manfreda}, A. and {Marscher}, A. and {Massaro}, F. and {Mitsuishi}, I. and {Mizuno}, T. and {Muleri}, F. and {Negro}, M. and {Ng}, S. and {O'Dell}, S. and {Omodei}, N. and {Oppedisano}, C. and {Papitto}, A. and {Pavlov}, G. and {Perri}, M. and {Pesce-Rollins}, M. and {Petrucci}, P. -O. and {Pilia}, M. and {Possenti}, A. and {Poutanen}, J. and {Ramsey}, B. and {Rankin}, J. and {Roberts}, O. and {Romani}, R. and {Sgr{\`o}}, C. and {Slane}, P. and {Soffitta}, P. and {Spandre}, G. and {Swartz}, D. and {Tamagawa}, T. and {Tavecchio}, F. and {Taverna}, R. and {Tawara}, Y. and {Tennant}, A. and {Thomas}, N. and {Trois}, A. and {Tsygankov}, S. and {Turolla}, R. and {Vink}, J. and {Weisskopf}, M. and {Wu}, K. and {Xie}, F. and {Zane}, S.},
        title = "{X-ray polarization measurement of the gold standard of radio-quiet active galactic nuclei: NGC 1068}",
      journal = {\aap},
         year = 2024,
        month = sep,
       volume = {689},
          eid = {A238},
        pages = {A238},
          doi = {10.1051/0004-6361/202449760},
archivePrefix = {arXiv},
       eprint = {2403.02061},
 primaryClass = {astro-ph.HE},
       adsurl = {https://ui.adsabs.harvard.edu/abs/2024A&A...689A.238M}
}

@ARTICLE{Hodgson2018,
       author = {{Hodgson}, Jeffrey A. and {Rani}, Bindu and {Lee}, Sang-Sung and {Algaba}, Juan Carlos and {Kino}, Motoki and {Trippe}, Sascha and {Park}, Jong-Ho and {Zhao}, Guang-Yao and {Byun}, Do-Young and {Kang}, Sincheol and {Kim}, Jae-Young and {Kim}, Jeong-Sook and {Kim}, Soon-Wook and {Miyazaki}, Atsushi and {Wajima}, Kiyoaki and {Oh}, Junghwan and {Kim}, Dae-won and {Gurwell}, Mark},
        title = "{KVN observations reveal multiple {\ensuremath{\gamma}}-ray emission regions in 3C 84?}",
      journal = {\mnras},
         year = 2018,
        month = mar,
       volume = {475},
       number = {1},
        pages = {368-378},
          doi = {10.1093/mnras/stx3041},
archivePrefix = {arXiv},
       eprint = {1802.02763},
 primaryClass = {astro-ph.HE},
       adsurl = {https://ui.adsabs.harvard.edu/abs/2018MNRAS.475..368H}
}

@ARTICLE{Kim2019,
       author = {{Kim}, J. -Y. and {Krichbaum}, T.~P. and {Marscher}, A.~P. and {Jorstad}, S.~G. and {Agudo}, I. and {Thum}, C. and {Hodgson}, J.~A. and {MacDonald}, N.~R. and {Ros}, E. and {Lu}, R. -S. and {Bremer}, M. and {de Vicente}, P. and {Lindqvist}, M. and {Trippe}, S. and {Zensus}, J.~A.},
        title = "{Spatially resolved origin of millimeter-wave linear polarization in the nuclear region of 3C 84}",
      journal = {\aap},
         year = 2019,
        month = feb,
       volume = {622},
          eid = {A196},
        pages = {A196},
          doi = {10.1051/0004-6361/201832920},
archivePrefix = {arXiv},
       eprint = {1811.07815},
 primaryClass = {astro-ph.GA},
       adsurl = {https://ui.adsabs.harvard.edu/abs/2019A&A...622A.196K}
}

@ARTICLE{Giovannini2018,
       author = {{Giovannini}, G. and {Savolainen}, T. and {Orienti}, M. and {Nakamura}, M. and {Nagai}, H. and {Kino}, M. and {Giroletti}, M. and {Hada}, K. and {Bruni}, G. and {Kovalev}, Y.~Y. and {Anderson}, J.~M. and {D'Ammando}, F. and {Hodgson}, J. and {Honma}, M. and {Krichbaum}, T.~P. and {Lee}, S. -S. and {Lico}, R. and {Lisakov}, M.~M. and {Lobanov}, A.~P. and {Petrov}, L. and {Sohn}, B.~W. and {Sokolovsky}, K.~V. and {Voitsik}, P.~A. and {Zensus}, J.~A. and {Tingay}, S.},
        title = "{A wide and collimated radio jet in 3C84 on the scale of a few hundred gravitational radii}",
      journal = {Nature Astronomy},
         year = 2018,
        month = apr,
       volume = {2},
        pages = {472-477},
          doi = {10.1038/s41550-018-0431-2},
archivePrefix = {arXiv},
       eprint = {1804.02198},
 primaryClass = {astro-ph.GA},
       adsurl = {https://ui.adsabs.harvard.edu/abs/2018NatAs...2..472G}
}

@ARTICLE{Nagai2014,
       author = {{Nagai}, H. and {Haga}, T. and {Giovannini}, G. and {Doi}, A. and {Orienti}, M. and {D'Ammando}, F. and {Kino}, M. and {Nakamura}, M. and {Asada}, K. and {Hada}, K. and {Giroletti}, M.},
        title = "{Limb-brightened Jet of 3C 84 Revealed by the 43 GHz Very-Long-Baseline-Array Observation}",
      journal = {\apj},
         year = 2014,
        month = apr,
       volume = {785},
       number = {1},
          eid = {53},
        pages = {53},
          doi = {10.1088/0004-637X/785/1/53},
archivePrefix = {arXiv},
       eprint = {1402.5930},
 primaryClass = {astro-ph.HE},
       adsurl = {https://ui.adsabs.harvard.edu/abs/2014ApJ...785...53N}
}

@ARTICLE{Suzuki2012,
       author = {{Suzuki}, Kenta and {Nagai}, Hiroshi and {Kino}, Motoki and {Kataoka}, Jun and {Asada}, Keiichi and {Doi}, Akihiro and {Inoue}, Makoto and {Orienti}, Monica and {Giovannini}, Gabriele and {Giroletti}, Marcello and {L{\"a}hteenm{\"a}ki}, Anne and {Tornikoski}, Merja and {Le{\'o}n-Tavares}, Jonathan and {Bach}, Uwe and {Kameno}, Seiji and {Kobayashi}, Hideyuki},
        title = "{Exploring the Central Sub-parsec Region of the {\ensuremath{\gamma}}-Ray Bright Radio Galaxy 3C 84 with VLBA at 43 GHz in the Period of 2002-2008}",
      journal = {\apj},
         year = 2012,
        month = feb,
       volume = {746},
       number = {2},
          eid = {140},
        pages = {140},
          doi = {10.1088/0004-637X/746/2/140},
archivePrefix = {arXiv},
       eprint = {1112.0756},
 primaryClass = {astro-ph.HE},
       adsurl = {https://ui.adsabs.harvard.edu/abs/2012ApJ...746..140S}
}

@ARTICLE{Nagai2010,
       author = {{Nagai}, Hiroshi and {Suzuki}, Kenta and {Asada}, Keiichi and {Kino}, Motoki and {Kameno}, Seiji and {Doi}, Akihiro and {Inoue}, Makoto and {Kataoka}, Jun and {Bach}, Uwe and {Hirota}, Tomoya and {Matsumoto}, Naoko and {Honma}, Mareki and {Kobayashi}, Hideyuki and {Fujisawa}, Kenta},
        title = "{VLBI Monitoring of 3C 84 (NGC 1275) in Early Phase of the 2005 Outburst}",
      journal = {\pasj},
         year = 2010,
        month = apr,
       volume = {62},
        pages = {L11},
          doi = {10.1093/pasj/62.2.L11},
archivePrefix = {arXiv},
       eprint = {1001.3852},
 primaryClass = {astro-ph.HE},
       adsurl = {https://ui.adsabs.harvard.edu/abs/2010PASJ...62L..11N}
}

@ARTICLE{Krichbaum1992,
       author = {{Krichbaum}, T.~P. and {Witzel}, A. and {Graham}, D.~A. and {Alef}, W. and {Pauliny-Toth}, I.~I.~K. and {Hummel}, C.~A. and {Quirrenbach}, A. and {Inoue}, M. and {Hirabayashi}, H. and {Morimoto}, M. and {Rogers}, A.~E.~E. and {Zensus}, J.~A. and {Lawrence}, C.~R. and {Readhead}, A.~C.~S. and {Booth}, R.~S. and {Ronnang}, B.~O. and {Kus}, A.~J. and {Johnston}, K.~J. and {Spencer}, J.~H. and {Burke}, B.~F. and {Dhawan}, V. and {Bartel}, N. and {Shapiro}, I.~I. and {Alberdi}, A. and {Marcaide}, J.~M.},
        title = "{The evolution of the sub-parsec structure of 3C 84 at 43 GHz.}",
      journal = {\aap},
         year = 1992,
        month = jul,
       volume = {260},
        pages = {33-48},
       adsurl = {https://ui.adsabs.harvard.edu/abs/1992A&A...260...33K}
}

@ARTICLE{Ehlert2023,
       author = {{Ehlert}, Steven R. and {Liodakis}, Ioannis and {Middei}, Riccardo and {Marscher}, Alan P. and {Tavecchio}, Fabrizio and {Agudo}, Iv{\'a}n and {Kouch}, Pouya M. and {Lindfors}, Elina and {Nilsson}, Kari and {Myserlis}, Ioannis and {Gurwell}, Mark and {Rao}, Ramprasad and {Aceituno}, Francisco Jos{\'e} and {Bonnoli}, Giacomo and {Casanova}, V{\'\i}ctor and {Ag{\'\i}s-Gonz{\'a}lez}, Beatriz and {Escudero}, Juan and {Husillos}, C{\'e}sar and {Otero Santos}, Jorge and {Sota}, Alfredo and {Angelakis}, Emmanouil and {Kraus}, Alexander and {Keating}, Garrett K. and {Antonelli}, Lucio A. and {Bachetti}, Matteo and {Baldini}, Luca and {Baumgartner}, Wayne H. and {Bellazzini}, Ronaldo and {Bianchi}, Stefano and {Bongiorno}, Stephen D. and {Bonino}, Raffaella and {Brez}, Alessandro and {Bucciantini}, Niccol{\'o} and {Capitanio}, Fiamma and {Castellano}, Simone and {Cavazzuti}, Elisabetta and {Chen}, Chien-Ting and {Ciprini}, Stefano and {Costa}, Enrico and {De Rosa}, Alessandra and {Del Monte}, Ettore and {Di Gesu}, Laura and {Di Lalla}, Niccol{\'o} and {Di Marco}, Alessandro and {Donnarumma}, Immacolata and {Doroshenko}, Victor and {Dov{\v{c}}iak}, Michal and {Enoto}, Teruaki and {Evangelista}, Yuri and {Fabiani}, Sergio and {Ferrazzoli}, Riccardo and {Garcia}, Javier A. and {Gunji}, Shuichi and {Hayashida}, Kiyoshi and {Heyl}, Jeremy and {Iwakiri}, Wataru and {Jorstad}, Svetlana G. and {Kaaret}, Philip and {Karas}, Vladimir and {Kislat}, Fabian and {Kitaguchi}, Takao and {Kolodziejczak}, Jeffery J. and {Krawczynski}, Henric and {La Monaca}, Fabio and {Latronico}, Luca and {Maldera}, Simone and {Manfreda}, Alberto and {Marin}, Fr{\'e}d{\'e}ric and {Marinucci}, Andrea and {Marshall}, Herman L. and {Massaro}, Francesco and {Matt}, Giorgio and {Mitsuishi}, Ikuyuki and {Mizuno}, Tsunefumi and {Muleri}, Fabio and {Negro}, Michela and {Ng}, C. -Y. and {O'Dell}, Stephen L. and {Omodei}, Nicola and {Oppedisano}, Chiara and {Papitto}, Alessandro and {Pavlov}, George G. and {Peirson}, Abel L. and {Perri}, Matteo and {Pesce-Rollins}, Melissa and {Petrucci}, Pierre-Olivier and {Pilia}, Maura and {Possenti}, Andrea and {Poutanen}, Juri and {Puccetti}, Simonetta and {Ramsey}, Brian D. and {Rankin}, John and {Ratheesh}, Ajay and {Roberts}, Oliver J. and {Romani}, Roger W. and {Sgr{\'o}}, Carmelo and {Slane}, Patrick and {Soffitta}, Paolo and {Spandre}, Gloria and {Swartz}, Douglas A. and {Tamagawa}, Toru and {Taverna}, Roberto and {Tawara}, Yuzuru and {Tennant}, Allyn F. and {Thomas}, Nicholas E. and {Tombesi}, Francesco and {Trois}, Alessio and {Tsygankov}, Sergey S. and {Turolla}, Roberto and {Vink}, Jacco and {Weisskopf}, Martin C. and {Wu}, Kinwah and {Xie}, Fei and {Zane}, Silvia},
        title = "{X-Ray Polarization of the BL Lacertae Type Blazar 1ES 0229+200}",
      journal = {\apj},
         year = 2023,
        month = dec,
       volume = {959},
       number = {1},
          eid = {61},
        pages = {61},
          doi = {10.3847/1538-4357/ad05c4},
archivePrefix = {arXiv},
       eprint = {2310.01635},
 primaryClass = {astro-ph.HE},
       adsurl = {https://ui.adsabs.harvard.edu/abs/2023ApJ...959...61E}
}

@ARTICLE{Ehlert2022,
       author = {{Ehlert}, Steven R. and {Ferrazzoli}, Riccardo and {Marinucci}, Andrea and {Marshall}, Herman L. and {Middei}, Riccardo and {Pacciani}, Luigi and {Perri}, Matteo and {Petrucci}, Pierre-Olivier and {Puccetti}, Simonetta and {Barnouin}, Thibault and {Bianchi}, Stefano and {Liodakis}, Ioannis and {Madejski}, Grzegorz and {Marin}, Fr{\'e}d{\'e}ric and {Marscher}, Alan P. and {Matt}, Giorgio and {Poutanen}, Juri and {Wu}, Kinwah and {Agudo}, Iv{\'a}n and {Antonelli}, Lucio A. and {Bachetti}, Matteo and {Baldini}, Luca and {Baumgartner}, Wayne H. and {Bellazzini}, Ronaldo and {Bongiorno}, Stephen D. and {Bonino}, Raffaella and {Brez}, Alessandro and {Bucciantini}, Niccol{\'o} and {Capitanio}, Fiamma and {Castellano}, Simone and {Cavazzuti}, Elisabetta and {Ciprini}, Stefano and {Costa}, Enrico and {De Rosa}, Alessandra and {Del Monte}, Ettore and {Di Gesu}, Laura and {Di Lalla}, Niccol{\'o} and {Di Marco}, Alessandro and {Donnarumma}, Immacolata and {Doroshenko}, Victor and {Dov{\v{c}}iak}, Michal and {Enoto}, Teruaki and {Evangelista}, Yuri and {Fabiani}, Sergio and {Garcia}, Javier A. and {Gunji}, Shuichi and {Hayashida}, Kiyoshi and {Heyl}, Jeremy and {Iwakiri}, Wataru and {Jorstad}, Svetlana G. and {Karas}, Vladimir and {Kitaguchi}, Takao and {Kolodziejczak}, Jeffery J. and {Krawczynski}, Henric and {La Monaca}, Fabio and {Latronico}, Luca and {Maldera}, Simone and {Manfreda}, Alberto and {Massaro}, Francesco and {Mitsuishi}, Ikuyuki and {Mizuno}, Tsunefumi and {Muleri}, Fabio and {Negro}, Michela and {Ng}, C. -Y. and {O'Dell}, Stephen L. and {Omodei}, Nicola and {Oppedisano}, Chiara and {Papitto}, Alessandro and {Pavlov}, George G. and {Peirson}, Abel L. and {Pesce-Rollins}, Melissa and {Pilia}, Maura and {Possenti}, Andrea and {Ramsey}, Brian D. and {Rankin}, John and {Ratheesh}, Ajay and {Romani}, Roger W. and {Sgr{\`o}}, Carmelo and {Slane}, Patrick and {Soffitta}, Paolo and {Spandre}, Gloria and {Tamagawa}, Toru and {Tavecchio}, Fabrizio and {Taverna}, Roberto and {Tawara}, Yuzuru and {Tennant}, Allyn F. and {Thomas}, Nicholas E. and {Tombesi}, Francesco and {Trois}, Alessio and {Tsygankov}, Sergey and {Turolla}, Roberto and {Vink}, Jacco and {Weisskopf}, Martin C. and {Xie}, Fei and {Zane}, Silvia and {IXPE Collaboration} and {Rodi}, James and {Jourdain}, Elisabeth and {Roques}, Jean-Pierre},
        title = "{Limits on X-Ray Polarization at the Core of Centaurus A as Observed with the Imaging X-Ray Polarimetry Explorer}",
      journal = {\apj},
         year = 2022,
        month = aug,
       volume = {935},
       number = {2},
          eid = {116},
        pages = {116},
          doi = {10.3847/1538-4357/ac8056},
archivePrefix = {arXiv},
       eprint = {2207.06625},
 primaryClass = {astro-ph.HE},
       adsurl = {https://ui.adsabs.harvard.edu/abs/2022ApJ...935..116E}
}

@ARTICLE{Middei2023-II,
       author = {{Middei}, Riccardo and {Perri}, Matteo and {Puccetti}, Simonetta and {Liodakis}, Ioannis and {Di Gesu}, Laura and {Marscher}, Alan P. and {Rodriguez Cavero}, Nicole and {Tavecchio}, Fabrizio and {Donnarumma}, Immacolata and {Laurenti}, Marco and {Jorstad}, Svetlana G. and {Agudo}, Iv{\'a}n and {Marshall}, Herman L. and {Pacciani}, Luigi and {Kim}, Dawoon E. and {Aceituno}, Francisco Jos{\'e} and {Bonnoli}, Giacomo and {Casanova}, V{\'\i}ctor and {Ag{\'\i}s-Gonz{\'a}lez}, Beatriz and {Sota}, Alfredo and {Casadio}, Carolina and {Escudero}, Juan and {Myserlis}, Ioannis and {Sievers}, Albrecht and {Kouch}, Pouya M. and {Lindfors}, Elina and {Gurwell}, Mark and {Keating}, Garrett K. and {Rao}, Ramprasad and {Kang}, Sincheol and {Lee}, Sang-Sung and {Kim}, Sang-Hyun and {Cheong}, Whee Yeon and {Jeong}, Hyeon-Woo and {Angelakis}, Emmanouil and {Kraus}, Alexander and {Antonelli}, Lucio A. and {Bachetti}, Matteo and {Baldini}, Luca and {Baumgartner}, Wayne H. and {Bellazzini}, Ronaldo and {Bianchi}, Stefano and {Bongiorno}, Stephen D. and {Bonino}, Raffaella and {Brez}, Alessandro and {Bucciantini}, Niccol{\`o} and {Capitanio}, Fiamma and {Castellano}, Simone and {Cavazzuti}, Elisabetta and {Chen}, Chien-Ting and {Ciprini}, Stefano and {Costa}, Enrico and {De Rosa}, Alessandra and {Del Monte}, Ettore and {Di Lalla}, Niccol{\`o} and {Di Marco}, Alessandro and {Doroshenko}, Victor and {Dov{\v{c}}iak}, Michal and {Ehlert}, Steven R. and {Enoto}, Teruaki and {Evangelista}, Yuri and {Fabiani}, Sergio and {Ferrazzoli}, Riccardo and {Garc{\'\i}a}, Javier A. and {Gunji}, Shuichi and {Hayashida}, Kiyoshi and {Heyl}, Jeremy and {Iwakiri}, Wataru and {Kaaret}, Philip and {Karas}, Vladimir and {Kislat}, Fabian and {Kitaguchi}, Takao and {Kolodziejczak}, Jeffery J. and {Krawczynski}, Henric and {La Monaca}, Fabio and {Latronico}, Luca and {Maldera}, Simone and {Manfreda}, Alberto and {Marin}, Fr{\'e}d{\'e}ric and {Marinucci}, Andrea and {Massaro}, Francesco and {Matt}, Giorgio and {Mitsuishi}, Ikuyuki and {Mizuno}, Tsunefumi and {Muleri}, Fabio and {Negro}, Michela and {Ng}, Chi-Yung and {O'Dell}, Stephen L. and {Omodei}, Nicola and {Oppedisano}, Chiara and {Papitto}, Alessandro and {Pavlov}, George G. and {Peirson}, Abel L. and {Pesce-Rollins}, Melissa and {Petrucci}, Pierre-Olivier and {Pilia}, Maura and {Possenti}, Andrea and {Poutanen}, Juri and {Ramsey}, Brian D. and {Rankin}, John and {Ratheesh}, Ajay and {Roberts}, Oliver J. and {Romani}, Roger W. and {Sgr{\`o}}, Carmelo and {Slane}, Patrick and {Soffitta}, Paolo and {Spandre}, Gloria and {Swartz}, Douglas A. and {Tamagawa}, Toru and {Taverna}, Roberto and {Tawara}, Yuzuru and {Tennant}, Allyn F. and {Thomas}, Nicholas E. and {Tombesi}, Francesco and {Trois}, Alessio and {Tsygankov}, Sergey S. and {Turolla}, Roberto and {Vink}, Jacco and {Weisskopf}, Martin C. and {Wu}, Kinwah and {Xie}, Fei and {Zane}, Silvia},
        title = "{IXPE and Multiwavelength Observations of Blazar PG 1553+113 Reveal an Orphan Optical Polarization Swing}",
      journal = {\apjl},
         year = 2023,
        month = aug,
       volume = {953},
       number = {2},
          eid = {L28},
        pages = {L28},
          doi = {10.3847/2041-8213/acec3e},
archivePrefix = {arXiv},
       eprint = {2308.00039},
 primaryClass = {astro-ph.HE},
       adsurl = {https://ui.adsabs.harvard.edu/abs/2023ApJ...953L..28M}
}

@ARTICLE{Chen2024,
       author = {{Chen}, Chien-Ting J. and {Liodakis}, Ioannis and {Middei}, Riccardo and {Kim}, Dawoon E. and {Di Gesu}, Laura and {Di Marco}, Alessandro and {Ehlert}, Steven R. and {Errando}, Manel and {Negro}, Michela and {Jorstad}, Svetlana G. and {Marscher}, Alan P. and {Wu}, Kinwah and {Agudo}, Iv{\'a}n and {Poutanen}, Juri and {Mizuno}, Tsunefumi and {Kouch}, Pouya M. and {Lindfors}, Elina and {Borman}, George A. and {Grishina}, Tatiana S. and {Kopatskaya}, Evgenia N. and {Larionova}, Elena G. and {Morozova}, Daria A. and {Savchenko}, Sergey S. and {Troitsky}, Ivan S. and {Troitskaya}, Yulia V. and {Vasilyev}, Andrey A. and {Zhovtan}, Alexey V. and {Aceituno}, Francisco Jos{\'e} and {Bonnoli}, Giacomo and {Casanova}, V{\'\i}ctor and {Escudero}, Juan and {Ag{\'\i}s-Gonz{\'a}lez}, Beatriz and {Husillos}, C{\'e}sar and {Otero Santos}, Jorge and {Sota}, Alfredo and {Piirola}, Vilppu and {Myserlis}, Ioannis and {Angelakis}, Emmanouil and {Kraus}, Alexander and {Gurwell}, Mark and {Keating}, Garrett and {Rao}, Ramprasad and {Kang}, Sincheol and {Lee}, Sang-Sung and {Kim}, Sang-Hyun and {Cheong}, Whee Yeon and {Jeong}, Hyeon-Woo and {Song}, Chanwoo and {Berdyugin}, Andrei V. and {Kagitani}, Masato and {Kravtsov}, Vadim and {Nitindala}, Anagha P. and {Sakanoi}, Takeshi and {Imazawa}, Ryo and {Sasada}, Mahito and {Fukazawa}, Yasushi and {Kawabata}, Koji S. and {Uemura}, Makoto and {Nakaoka}, Tatsuya and {Akitaya}, Hiroshi and {Casadio}, Carolina and {Sievers}, Albrecht and {Antonelli}, Lucio Angelo and {Bachetti}, Matteo and {Baldini}, Luca and {Baumgartner}, Wayne H. and {Bellazzini}, Ronaldo and {Bianchi}, Stefano and {Bongiorno}, Stephen D. and {Bonino}, Raffaella and {Brez}, Alessandro and {Bucciantini}, Niccol{\'o} and {Capitanio}, Fiamma and {Castellano}, Simone and {Cavazzuti}, Elisabetta and {Ciprini}, Stefano and {Costa}, Enrico and {De Rosa}, Alessandra and {Del Monte}, Ettore and {Di Lalla}, Niccol{\'o} and {Donnarumma}, Immacolata and {Doroshenko}, Victor and {Dov{\v{c}}iak}, Michal and {Enoto}, Teruaki and {Evangelista}, Yuri and {Fabiani}, Sergio and {Ferrazzoli}, Riccardo and {Garcia}, Javier A. and {Gunji}, Shuichi and {Hayashida}, Kiyoshi and {Heyl}, Jeremy and {Iwakiri}, Wataru and {Kaaret}, Philip and {Karas}, Vladimir and {Kislat}, Fabian and {Kitaguchi}, Takao and {Kolodziejczak}, Jeffery J. and {Krawczynski}, Henric and {La Monaca}, Fabio and {Latronico}, Luca and {Maldera}, Simone and {Manfreda}, Alberto and {Marin}, Fr{\'e}d{\'e}ric and {Marinucci}, Andrea and {Marshall}, Herman L. and {Massaro}, Francesco and {Matt}, Giorgio and {Mitsuishi}, Ikuyuki and {Muleri}, Fabio and {Ng}, C. -Y. and {O'Dell}, Stephen L. and {Omodei}, Nicola and {Oppedisano}, Chiara and {Papitto}, Alessandro and {Pavlov}, George G. and {Peirson}, Abel Lawrence and {Perri}, Matteo and {Pesce-Rollins}, Melissa and {Petrucci}, Pierre-Olivier and {Pilia}, Maura and {Possenti}, Andrea and {Puccetti}, Simonetta and {Ramsey}, Brian D. and {Rankin}, John and {Ratheesh}, Ajay and {Roberts}, Oliver J. and {Romani}, Roger W. and {Sgr{\'o}}, Carmelo and {Slane}, Patrick and {Soffitta}, Paolo and {Spandre}, Gloria and {Swartz}, Douglas A. and {Tamagawa}, Toru and {Tavecchio}, Fabrizio and {Taverna}, Roberto and {Tawara}, Yuzuru and {Tennant}, Allyn F. and {Thomas}, Nicholas E. and {Tombesi}, Francesco and {Trois}, Alessio and {Tsygankov}, Sergey S. and {Turolla}, Roberto and {Vink}, Jacco and {Weisskopf}, Martin C. and {Xie}, Fei and {Zane}, Silvia},
        title = "{X-Ray and Multiwavelength Polarization of Mrk 501 from 2022 to 2023}",
      journal = {\apj},
         year = 2024,
        month = oct,
       volume = {974},
       number = {1},
          eid = {50},
        pages = {50},
          doi = {10.3847/1538-4357/ad63a1},
archivePrefix = {arXiv},
       eprint = {2407.11128},
 primaryClass = {astro-ph.HE},
       adsurl = {https://ui.adsabs.harvard.edu/abs/2024ApJ...974...50C}
}

@ARTICLE{Hodgson2021,
       author = {{Hodgson}, Jeffrey A. and {Rani}, Bindu and {Oh}, Junghwan and {Marscher}, Alan and {Jorstad}, Svetlana and {Mizuno}, Yosuke and {Park}, Jongho and {Lee}, S.~S. and {Trippe}, Sascha and {Mertens}, Florent},
        title = "{A Detailed Kinematic Study of 3C 84 and Its Connection to {\ensuremath{\gamma}}-Rays}",
      journal = {\apj},
         year = 2021,
        month = jun,
       volume = {914},
       number = {1},
          eid = {43},
        pages = {43},
          doi = {10.3847/1538-4357/abf6dd},
archivePrefix = {arXiv},
       eprint = {2104.03081},
 primaryClass = {astro-ph.HE},
       adsurl = {https://ui.adsabs.harvard.edu/abs/2021ApJ...914...43H}
}

@INPROCEEDINGS{Soffitta2023,
       author = {{Soffitta}, Paolo and {Baldini}, Luca and {Baumgartner}, Wayne and {Bellazzini}, Ronaldo and {Bongiorno}, Stephen D. and {Bucciantini}, Niccol{\`o} and {Costa}, Enrico and {Dov{\v{c}}iak}, Michal and {Ehlert}, Steven and {Kaaret}, Philip E. and {Kolodziejczak}, Jeffery J. and {Latronico}, Luca and {Marin}, Fr{\'e}d{\'e}ric and {Marscher}, Alan P. and {Marshall}, Herman L. and {Matt}, Giorgio and {Muleri}, Fabio and {O'Dell}, Stephen L. and {Poutanen}, Juri and {Ramsey}, Brian and {Romani}, Roger W. and {Slane}, Patrick and {Tennant}, Allyn F. and {Turolla}, Roberto and {Weisskopf}, Martin C. and {Agudo}, Iv{\'a}n. and {Antonelli}, Lucio Angelo and {Bachetti}, Matteo and {Bianchi}, Stefano and {Bonino}, Raffaella and {Brez}, Alessandro and {Capitanio}, Fiamma and {Castellano}, Simone and {Cavazzuti}, Elisabetta and {Chen}, Chiel-Ting and {Ciprini}, Stefano and {De Rosa}, Alessandra and {Del Monte}, Ettore and {Di Gesu}, Laura and {Di Lalla}, Niccol{\`o} and {Di Marco}, Alessandro and {Donnarumma}, Immacolata and {Doroshenko}, Victor and {Enoto}, Teruaki and {Evangelista}, Yuri and {Fabiani}, Sergio and {Ferrazzoli}, Riccardo and {Garcia}, Javier A. and {Gunji}, Shuichi and {Hayashida}, Kiyoshi and {Heyl}, Jeremy and {Iwakiri}, Wataru and {Jorstad}, Svetlana G. and {Karas}, Vladimir and {Kislat}, Fabian and {Kitaguchi}, Takao and {Krawczynski}, Henric and {La Monaca}, Fabio and {Liodakis}, Ioannis and {Maldera}, Simone and {Manfreda}, Alberto and {Marinucci}, Andrea and {Massaro}, Francesco and {Mitsuishi}, Ikuyuki and {Mizuno}, Tsunefumi and {Negro}, Michela and {Ng}, C. -Y. and {Omodei}, Nicola and {Oppedisano}, Chiara and {Papitto}, Alessandro and {Pavlov}, George G. and {Peirson}, Abel Lawrence and {Perri}, Matteo and {Pesce-Rollins}, Melissa and {Petrucci}, Pierre-Olivier and {Pilia}, Maura and {Possenti}, Andrea and {Puccetti}, Simonetta and {Rankin}, John and {Ratheesh}, Ajay and {Roberts}, Oliver J. and {Sgr{\`o}}, Carmelo and {Spandre}, Gloria and {Swartz}, Douglas A. and {Tamagawa}, Toru and {Tavecchio}, Fabrizio and {Taverna}, Roberto and {Tawara}, Yuzuru and {Thomas}, Nicholas E. and {Tombesi}, Francesco and {Trois}, Alessio and {Tsygankov}, Sergey S. and {Vink}, Jacco and {Wu}, Kinwah and {Xie}, Fei and {Zane}, Silvia},
        title = "{The Imaging x-ray polarimetry explorer (IXPE) at last!}",
    booktitle = {UV, X-Ray, and Gamma-Ray Space Instrumentation for Astronomy XXIII},
         year = 2023,
       editor = {{Siegmund}, Oswald H. and {Hoadley}, Keri},
       series = {Society of Photo-Optical Instrumentation Engineers (SPIE) Conference Series},
       volume = {12678},
        month = oct,
          eid = {1267803},
        pages = {1267803},
          doi = {10.1117/12.2677296},
       adsurl = {https://ui.adsabs.harvard.edu/abs/2023SPIE12678E..03S}
}

@ARTICLE{Agudo2025,
       author = {{Agudo}, Ivan and {Liodakis}, Ioannis and {Otero-Santos}, Jorge and {Middei}, Riccardo and {Marscher}, Alan and {Jorstad}, Svetlana and {Zhang}, Haocheng and {Li}, Hui and {Di Gesu}, Laura and {Romani}, Roger W. and {Kim}, Dawoon E. and {Fenu}, Francesco and {Marshall}, Herman L. and {Pacciani}, Luigi and {Escudero Pedrosa}, Juan and {Aceituno}, Francisco Jose and {Agis-Gonzalez}, Beatriz and {Bonnoli}, Giacomo and {Casanova}, Victor and {Morcuende}, Daniel and {Piirola}, Vilppu and {Sota}, Alfredo and {Kouch}, Pouya M. and {Lindfors}, Elina and {McCall}, Callum and {Jermak}, Helen E. and {Steele}, Iain A. and {Borman}, George A. and {Grishina}, Tatiana S. and {Hagen-Thorn}, Vladimir A. and {Kopatskaya}, Evgenia N. and {Larionova}, Elena G. and {Morozova}, Daria A. and {Savchenko}, Sergey S. and {Shishkina}, Ekaterina V. and {Troitskiy}, Ivan S. and {Troitskaya}, Yulia V. and {Vasilyev}, Andrey A. and {Zhovtan}, Alexey V. and {Myserlis}, Ioannis and {Gurwell}, Mark and {Keating}, Garrett and {Rao}, Ramprasad and {Kang}, Sincheol and {Lee}, Sang-Sung and {Kim}, Sanghyun and {Cheong}, Whee Yeon and {Jeong}, Hyeon-Woo and {Angelakis}, Emmanouil and {Kraus}, Alexander and {Blinov}, Dmitry and {Maharana}, Siddharth and {Bachev}, Rumen and {Jormanainen}, Jenni and {Nilsson}, Kari and {Fallah Ramazani}, Vandad and {Casadio}, Carolina and {Fuentes}, Antonio and {Traianou}, Efthalia and {Thum}, Clemens and {Gomez}, Jose L. and {Antonelli}, Lucio Angelo and {Bachetti}, Matteo and {Baldini}, Luca and {Baumgartner}, Wayne H. and {Bellazzini}, Ronaldo and {Bianchi}, Stefano and {Bongiorno}, Stephen D. and {Bonino}, Raffaella and {Brez}, Alessandro and {Bucciantini}, Niccolo and {Capitanio}, Fiamma and {Castellano}, Simone and {Cavazzuti}, Elisabetta and {Chen}, Chien-Ting and {Ciprini}, Stefano and {Costa}, Enrico and {De Rosa}, Alessandra and {Del Monte}, Ettore and {Di Lalla}, Niccolo and {Di Marco}, Alessandro and {Donnarumma}, Immacolata and {Doroshenko}, Victor and {Dovciak}, Michal and {Ehlert}, Steven R. and {Enoto}, Teruaki and {Evangelista}, Yuri and {Fabiani}, Sergio and {Ferrazzoli}, Riccardo and {Garcia}, Javier A. and {Gunji}, Shuichi and {Hayashida}, Kiyoshi and {Heyl}, Jeremy and {Iwakiri}, Wataru and {Kaaret}, Philip and {Karas}, Vladimir and {Kislat}, Fabian and {Kitaguchi}, Takao and {Kolodziejczak}, Jeffery J. and {Krawczynski}, Henric and {La Monaca}, Fabio and {Latronico}, Luca and {Maldera}, Simone and {Manfreda}, Alberto and {Marin}, Frederic and {Marinucci}, Andrea and {Massaro}, Francesco and {Matt}, Giorgio and {Mitsuishi}, Ikuyuki and {Mizuno}, Tsunefumi and {Muleri}, Fabio and {Negro}, Michela and {Ng}, Chi-Yung and {O'Dell}, Stephen L. and {Omodei}, Nicola and {Oppedisano}, Chiara and {Papitto}, Alessandro and {Pavlov}, George G. and {Peirson}, Abel L. and {Perri}, Matteo and {Pesce-Rollins}, Melissa and {Petrucci}, Pierre-Olivier and {Pilia}, Maura and {Possenti}, Andrea and {Poutanen}, Juri and {Puccetti}, Simonetta and {Ramsey}, Brian D. and {Rankin}, John and {Ratheesh}, Ajay and {Roberts}, Oliver J. and {Sgro}, Carmelo and {Slane}, Patrick and {Soffitta}, Paolo and {Spandre}, Gloria and {Swartz}, Douglas A. and {Tamagawa}, Toru and {Tavecchio}, Fabrizio and {Taverna}, Roberto and {Tawara}, Yuzuru and {Tennant}, Allyn F. and {Thomas}, Nicholas E. and {Tombesi}, Francesco and {Trois}, Alessio and {Tsygankov}, Sergey S. and {Turolla}, Roberto and {Vink}, Jacco and {Weisskopf}, Martin C. and {Wu}, Kinwah and {Xie}, Fei and {Zane}, Silvia},
        title = "{High optical to X-ray polarization ratio reveals Compton scattering in BL Lacertae's jet}",
      journal = {arXiv e-prints},
         year = 2025,
        month = may,
          eid = {arXiv:2505.01832},
        pages = {arXiv:2505.01832},
          doi = {10.48550/arXiv.2505.01832},
archivePrefix = {arXiv},
       eprint = {2505.01832},
 primaryClass = {astro-ph.HE},
       adsurl = {https://ui.adsabs.harvard.edu/abs/2025arXiv250501832A}
}

@ARTICLE{Paraschos2025,
       author = {{Paraschos}, G.~F. and {Mpisketzis}, V.},
        title = "{Unravelling the dynamics of cosmic vortices: Probing a Kelvin-Helmholtz instability in the jet of 3C 84}",
      journal = {\aap},
         year = 2025,
        month = apr,
       volume = {696},
          eid = {L7},
        pages = {L7},
          doi = {10.1051/0004-6361/202554201},
archivePrefix = {arXiv},
       eprint = {2503.16647},
 primaryClass = {astro-ph.HE},
       adsurl = {https://ui.adsabs.harvard.edu/abs/2025A&A...696L...7P}
}

@ARTICLE{Paraschos2023,
       author = {{Paraschos}, G.~F. and {Mpisketzis}, V. and {Kim}, J. -Y. and {Witzel}, G. and {Krichbaum}, T.~P. and {Zensus}, J.~A. and {Gurwell}, M.~A. and {L{\"a}hteenm{\"a}ki}, A. and {Tornikoski}, M. and {Kiehlmann}, S. and {Readhead}, A.~C.~S.},
        title = "{A multi-band study and exploration of the radio wave-{\ensuremath{\gamma}}-ray connection in 3C 84}",
      journal = {\aap},
         year = 2023,
        month = jan,
       volume = {669},
          eid = {A32},
        pages = {A32},
          doi = {10.1051/0004-6361/202244814},
archivePrefix = {arXiv},
       eprint = {2210.09795},
 primaryClass = {astro-ph.HE},
       adsurl = {https://ui.adsabs.harvard.edu/abs/2023A&A...669A..32P}
}

@ARTICLE{Paraschos2024-II,
       author = {{Paraschos}, G.~F. and {Kim}, J. -Y. and {Wielgus}, M. and {R{\"o}der}, J. and {Krichbaum}, T.~P. and {Ros}, E. and {Agudo}, I. and {Myserlis}, I. and {Moscibrodzka}, M. and {Traianou}, E. and {Zensus}, J.~A. and {Blackburn}, L. and {Chan}, C. -K. and {Issaoun}, S. and {Janssen}, M. and {Johnson}, M.~D. and {Fish}, V.~L. and {Akiyama}, K. and {Alberdi}, A. and {Alef}, W. and {Algaba}, J.~C. and {Anantua}, R. and {Asada}, K. and {Azulay}, R. and {Bach}, U. and {Baczko}, A. -K. and {Ball}, D. and {Balokovi{\'c}}, M. and {Barrett}, J. and {Baub{\"o}ck}, M. and {Benson}, B.~A. and {Bintley}, D. and {Blundell}, R. and {Bouman}, K.~L. and {Bower}, G.~C. and {Boyce}, H. and {Bremer}, M. and {Brinkerink}, C.~D. and {Brissenden}, R. and {Britzen}, S. and {Broderick}, A.~E. and {Broguiere}, D. and {Bronzwaer}, T. and {Bustamante}, S. and {Byun}, D. -Y. and {Carlstrom}, J.~E. and {Ceccobello}, C. and {Chael}, A. and {Chang}, D.~O. and {Chatterjee}, K. and {Chatterjee}, S. and {Chen}, M.~T. and {Chen}, Y. and {Cheng}, X. and {Cho}, I. and {Christian}, P. and {Conroy}, N.~S. and {Conway}, J.~E. and {Cordes}, J.~M. and {Crawford}, T.~M. and {Crew}, G.~B. and {Cruz-Osorio}, A. and {Cui}, Y. and {Dahale}, R. and {Davelaar}, J. and {De Laurentis}, M. and {Deane}, R. and {Dempsey}, J. and {Desvignes}, G. and {Dexter}, J. and {Dhruv}, V. and {Doeleman}, S.~S. and {Dougal}, S. and {Dzib}, S.~A. and {Eatough}, R.~P. and {Emami}, R. and {Falcke}, H. and {Farah}, J. and {Fomalont}, E. and {Ford}, H.~A. and {Foschi}, M. and {Fraga-Encinas}, R. and {Freeman}, W.~T. and {Friberg}, P. and {Fromm}, C.~M. and {Fuentes}, A. and {Galison}, P. and {Gammie}, C.~F. and {Garc{\'\i}a}, R. and {Gentaz}, O. and {Georgiev}, B. and {Goddi}, C. and {Gold}, R. and {G{\'o}mez-Ruiz}, A.~I. and {G{\'o}mez}, J.~L. and {Gu}, M. and {Gurwell}, M. and {Hada}, K. and {Haggard}, D. and {Haworth}, K. and {Hecht}, M.~H. and {Hesper}, R. and {Heumann}, D. and {Ho}, L.~C. and {Ho}, P. and {Honma}, M. and {Huang}, C.~L. and {Huang}, L. and {Hughes}, D.~H. and {Ikeda}, S. and {Impellizzeri}, C.~M.~V. and {Inoue}, M. and {James}, D.~J. and {Jannuzi}, B.~T. and {Jeter}, B. and {Jaing}, W. and {Jim{\'e}nez-Rosales}, A. and {Jorstad}, S. and {Joshi}, A.~V. and {Jung}, T. and {Karami}, M. and {Karuppusamy}, R. and {Kawashima}, T. and {Keating}, G.~K. and {Kettenis}, M. and {Kim}, D. -J. and {Kim}, J. and {Kim}, J. and {Kino}, M. and {Koay}, J.~Y. and {Kocherlakota}, P. and {Kofuji}, Y. and {Koch}, P.~M. and {Koyama}, S. and {Kramer}, C. and {Kramer}, J.~A. and {Kramer}, M. and {Kuo}, C. -Y. and {La Bella}, N. and {Lauer}, T.~R. and {Lee}, D. and {Lee}, S. -S. and {Leung}, P.~K. and {Levis}, A. and {Li}, Z. and {Lico}, R. and {Lindahl}, G. and {Lindqvist}, M. and {Lisakov}, M. and {Liu}, J. and {Liu}, K. and {Liuzzo}, E. and {Lo}, W. -P. and {Lobanov}, A.~P. and {Loinard}, L. and {Lonsdale}, C.~J. and {Lowitz}, A.~E. and {Lu}, R. -S. and {MacDonald}, N.~R. and {Mao}, J. and {Marchili}, N. and {Markoff}, S. and {Marrone}, D.~P. and {Marscher}, A.~P. and {Mart{\'\i}-Vidal}, I. and {Matsushita}, S. and {Matthews}, L.~D. and {Medeiros}, L. and {Menten}, K.~M. and {Michalik}, D. and {Mizuno}, I. and {Mizuno}, Y. and {Moran}, J.~M. and {Moriyama}, K. and {Mulaudzi}, W. and {M{\"u}ller}, C. and {M{\"u}ller}, H. and {Mus}, A. and {Musoke}, G. and {Nadolski}, A. and {Nagai}, H. and {Nagar}, N.~M. and {Nakamura}, M. and {Narayanan}, G. and {Natarajan}, I. and {Nathanail}, A. and {Navarro Fuentes}, S. and {Neilsen}, J. and {Neri}, R. and {Ni}, C. and {Noutsos}, A. and {Nowak}, M.~A. and {Oh}, J. and {Okino}, H. and {Olivares}, H. and {Ortiz-Le{\'o}n}, G.~N. and {Oyama}, T. and {{\"O}zel}, F. and {Palumbo}, D.~C.~M. and {Park}, J.},
        title = "{Ordered magnetic fields around the 3C 84 central black hole}",
      journal = {\aap},
         year = 2024,
        month = feb,
       volume = {682},
          eid = {L3},
        pages = {L3},
          doi = {10.1051/0004-6361/202348308},
archivePrefix = {arXiv},
       eprint = {2402.00927},
 primaryClass = {astro-ph.HE},
       adsurl = {https://ui.adsabs.harvard.edu/abs/2024A&A...682L...3P}
}

@ARTICLE{Paraschos2024,
       author = {{Paraschos}, G.~F. and {Debbrecht}, L.~C. and {Kramer}, J.~A. and {Traianou}, E. and {Liodakis}, I. and {Krichbaum}, T.~P. and {Kim}, J. -Y. and {Janssen}, M. and {Nair}, D.~G. and {Savolainen}, T. and {Ros}, E. and {Bach}, U. and {Hodgson}, J.~A. and {Lisakov}, M. and {MacDonald}, N.~R. and {Zensus}, J.~A.},
        title = "{Evidence of a toroidal magnetic field in the core of 3C 84}",
      journal = {\aap},
         year = 2024,
        month = jun,
       volume = {686},
          eid = {L5},
        pages = {L5},
          doi = {10.1051/0004-6361/202450218},
archivePrefix = {arXiv},
       eprint = {2405.00097},
 primaryClass = {astro-ph.HE},
}

@ARTICLE{Paraschos2022,
       author = {{Paraschos}, G.~F. and {Krichbaum}, T.~P. and {Kim}, J. -Y. and {Hodgson}, J.~A. and {Oh}, J. and {Ros}, E. and {Zensus}, J.~A. and {Marscher}, A.~P. and {Jorstad}, S.~G. and {Gurwell}, M.~A. and {L{\"a}hteenm{\"a}ki}, A. and {Tornikoski}, M. and {Kiehlmann}, S. and {Readhead}, A.~C.~S.},
        title = "{Jet kinematics in the transversely stratified jet of 3C 84. A two-decade overview}",
      journal = {\aap},
         year = 2022,
        month = sep,
       volume = {665},
          eid = {A1},
        pages = {A1},
          doi = {10.1051/0004-6361/202243343},
archivePrefix = {arXiv},
       eprint = {2205.10281},
 primaryClass = {astro-ph.HE},
       adsurl = {https://ui.adsabs.harvard.edu/abs/2022A&A...665A...1P}
}

@ARTICLE{Antonucci1993,
       author = {{Antonucci}, Robert},
        title = "{Unified models for active galactic nuclei and quasars.}",
      journal = {\araa},
         year = 1993,
        month = jan,
       volume = {31},
        pages = {473-521},
          doi = {10.1146/annurev.aa.31.090193.002353},
       adsurl = {https://ui.adsabs.harvard.edu/abs/1993ARA&A..31..473A}
}

@ARTICLE{Jorstad2016,
       author = {{Jorstad}, Svetlana and {Marscher}, Alan},
        title = "{The VLBA-BU-BLAZAR Multi-Wavelength Monitoring Program}",
      journal = {Galaxies},
         year = 2016,
        month = oct,
       volume = {4},
       number = {4},
          eid = {47},
        pages = {47},
          doi = {10.3390/galaxies4040047},
       adsurl = {https://ui.adsabs.harvard.edu/abs/2016Galax...4...47J}
}

@ARTICLE{Lister2018,
       author = {{Lister}, M.~L. and {Aller}, M.~F. and {Aller}, H.~D. and {Hodge}, M.~A. and {Homan}, D.~C. and {Kovalev}, Y.~Y. and {Pushkarev}, A.~B. and {Savolainen}, T.},
        title = "{MOJAVE. XV. VLBA 15 GHz Total Intensity and Polarization Maps of 437 Parsec-scale AGN Jets from 1996 to 2017}",
      journal = {\apjs},
         year = 2018,
        month = jan,
       volume = {234},
       number = {1},
          eid = {12},
        pages = {12},
          doi = {10.3847/1538-4365/aa9c44},
archivePrefix = {arXiv},
       eprint = {1711.07802},
 primaryClass = {astro-ph.GA},
       adsurl = {https://ui.adsabs.harvard.edu/abs/2018ApJS..234...12L}
}

@ARTICLE{Edelson1988,
       author = {{Edelson}, R.~A. and {Krolik}, J.~H.},
        title = "{The Discrete Correlation Function: A New Method for Analyzing Unevenly Sampled Variability Data}",
      journal = {\apj},
         year = 1988,
        month = oct,
       volume = {333},
        pages = {646},
          doi = {10.1086/166773},
       adsurl = {https://ui.adsabs.harvard.edu/abs/1988ApJ...333..646E}
}

@ARTICLE{Abdo2009,
       author = {{Abdo}, A.~A. and {Ackermann}, M. and {Ajello}, M. and {Asano}, K. and {Baldini}, L. and {Ballet}, J. and {Barbiellini}, G. and {Bastieri}, D. and {Baughman}, B.~M. and {Bechtol}, K. and {Bellazzini}, R. and {Blandford}, R.~D. and {Bloom}, E.~D. and {Bonamente}, E. and {Borgland}, A.~W. and {Bregeon}, J. and {Brez}, A. and {Brigida}, M. and {Bruel}, P. and {Burnett}, T.~H. and {Caliandro}, G.~A. and {Cameron}, R.~A. and {Caraveo}, P.~A. and {Casandjian}, J.~M. and {Cavazzuti}, E. and {Cecchi}, C. and {Celotti}, A. and {Chekhtman}, A. and {Cheung}, C.~C. and {Chiang}, J. and {Ciprini}, S. and {Claus}, R. and {Cohen-Tanugi}, J. and {Colafrancesco}, S. and {Cominsky}, L.~R. and {Conrad}, J. and {Costamante}, L. and {Dermer}, C.~D. and {de Angelis}, A. and {de Palma}, F. and {Digel}, S.~W. and {Donato}, D. and {do Couto e Silva}, E. and {Drell}, P.~S. and {Dubois}, R. and {Dumora}, D. and {Farnier}, C. and {Favuzzi}, C. and {Finke}, J. and {Focke}, W.~B. and {Frailis}, M. and {Fukazawa}, Y. and {Funk}, S. and {Fusco}, P. and {Gargano}, F. and {Georganopoulos}, M. and {Germani}, S. and {Giebels}, B. and {Giglietto}, N. and {Giordano}, F. and {Glanzman}, T. and {Grenier}, I.~A. and {Grondin}, M. -H. and {Grove}, J.~E. and {Guillemot}, L. and {Guiriec}, S. and {Hanabata}, Y. and {Harding}, A.~K. and {Hartman}, R.~C. and {Hayashida}, M. and {Hays}, E. and {Hughes}, R.~E. and {J{\'o}hannesson}, G. and {Johnson}, A.~S. and {Johnson}, R.~P. and {Johnson}, W.~N. and {Kadler}, M. and {Kamae}, T. and {Kanai}, Y. and {Katagiri}, H. and {Kataoka}, J. and {Kawai}, N. and {Kerr}, M. and {Kn{\"o}dlseder}, J. and {Kuehn}, F. and {Kuss}, M. and {Latronico}, L. and {Lemoine-Goumard}, M. and {Longo}, F. and {Loparco}, F. and {Lott}, B. and {Lovellette}, M.~N. and {Lubrano}, P. and {Madejski}, G.~M. and {Makeev}, A. and {Mazziotta}, M.~N. and {McEnery}, J.~E. and {Meurer}, C. and {Michelson}, P.~F. and {Mitthumsiri}, W. and {Mizuno}, T. and {Moiseev}, A.~A. and {Monte}, C. and {Monzani}, M.~E. and {Morselli}, A. and {Moskalenko}, I.~V. and {Murgia}, S. and {Nakamori}, T. and {Nolan}, P.~L. and {Norris}, J.~P. and {Nuss}, E. and {Ohsugi}, T. and {Omodei}, N. and {Orlando}, E. and {Ormes}, J.~F. and {Paneque}, D. and {Panetta}, J.~H. and {Parent}, D. and {Pepe}, M. and {Pesce-Rollins}, M. and {Piron}, F. and {Porter}, T.~A. and {Rain{\`o}}, S. and {Razzano}, M. and {Reimer}, A. and {Reimer}, O. and {Reposeur}, T. and {Ritz}, S. and {Rodriguez}, A.~Y. and {Romani}, R.~W. and {Ryde}, F. and {Sadrozinski}, H.~F. -W. and {Sambruna}, R. and {Sanchez}, D. and {Sander}, A. and {Sato}, R. and {Parkinson}, P.~M. Saz and {Sgr{\`o}}, C. and {Smith}, D.~A. and {Smith}, P.~D. and {Spandre}, G. and {Spinelli}, P. and {Starck}, J. -L. and {Strickman}, M.~S. and {Strong}, A.~W. and {Suson}, D.~J. and {Tajima}, H. and {Takahashi}, H. and {Takahashi}, T. and {Tanaka}, T. and {Taylor}, G.~B. and {Thayer}, J.~G. and {Thompson}, D.~J. and {Torres}, D.~F. and {Tosti}, G. and {Uchiyama}, Y. and {Usher}, T.~L. and {Vilchez}, N. and {Vitale}, V. and {Waite}, A.~P. and {Wood}, K.~S. and {Ylinen}, T. and {Ziegler}, M. and {Aller}, H.~D. and {Aller}, M.~F. and {Kellermann}, K.~I. and {Kovalev}, Y.~Y. and {Kovalev}, Yu. A. and {Lister}, M.~L. and {Pushkarev}, A.~B.},
        title = "{Fermi Discovery of Gamma-ray Emission from NGC 1275}",
      journal = {\apj},
         year = 2009,
        month = jul,
       volume = {699},
       number = {1},
        pages = {31-39},
          doi = {10.1088/0004-637X/699/1/31},
archivePrefix = {arXiv},
       eprint = {0904.1904},
 primaryClass = {astro-ph.HE},
       adsurl = {https://ui.adsabs.harvard.edu/abs/2009ApJ...699...31A}
}

@ARTICLE{Cao2024,
       author = {{Cao}, Zhen and {Aharonian}, F. and {Axikegu} and {Bai}, Y.~X. and {Bao}, Y.~W. and {Bastieri}, D. and {Bi}, X.~J. and {Bi}, Y.~J. and {Bian}, W. and {Bukevich}, A.~V. and {Cao}, Q. and {Cao}, W.~Y. and {Cao}, Zhe and {Chang}, J. and {Chang}, J.~F. and {Chen}, A.~M. and {Chen}, E.~S. and {Chen}, H.~X. and {Chen}, Liang and {Chen}, Lin and {Chen}, Long and {Chen}, M.~J. and {Chen}, M.~L. and {Chen}, Q.~H. and {Chen}, S. and {Chen}, S.~H. and {Chen}, S.~Z. and {Chen}, T.~L. and {Chen}, Y. and {Cheng}, N. and {Cheng}, Y.~D. and {Chu}, M.~C. and {Cui}, M.~Y. and {Cui}, S.~W. and {Cui}, X.~H. and {Cui}, Y.~D. and {Dai}, B.~Z. and {Dai}, H.~L. and {Dai}, Z.~G. and {Danzengluobu} and {Dong}, X.~Q. and {Duan}, K.~K. and {Fan}, J.~H. and {Fan}, Y.~Z. and {Fang}, J. and {Fang}, J.~H. and {Fang}, K. and {Feng}, C.~F. and {Feng}, H. and {Feng}, L. and {Feng}, S.~H. and {Feng}, X.~T. and {Feng}, Y. and {Feng}, Y.~L. and {Gabici}, S. and {Gao}, B. and {Gao}, C.~D. and {Gao}, Q. and {Gao}, W. and {Gao}, W.~K. and {Ge}, M.~M. and {Ge}, T.~T. and {Geng}, L.~S. and {Giacinti}, G. and {Gong}, G.~H. and {Gou}, Q.~B. and {Gu}, M.~H. and {Guo}, F.~L. and {Guo}, J. and {Guo}, X.~L. and {Guo}, Y.~Q. and {Guo}, Y.~Y. and {Han}, Y.~A. and {Hannuksela}, O.~A. and {Hasan}, M. and {He}, H.~H. and {He}, H.~N. and {He}, J.~Y. and {He}, Y. and {Hor}, Y.~K. and {Hou}, B.~W. and {Hou}, C. and {Hu}, H.~B. and {Hu}, Q. and {Hu}, S.~C. and {Huang}, C. and {Huang}, D.~H. and {Huang}, T.~Q. and {Huang}, W.~J. and {Huang}, X.~T. and {Huang}, X.~Y. and {Huang}, Y. and {Huang}, Y.~Y. and {Ji}, X.~L. and {Jia}, H.~Y. and {Jia}, K. and {Jiang}, H.~B. and {Jiang}, K. and {Jiang}, X.~W. and {Jiang}, Z.~J. and {Jin}, M. and {Kang}, M.~M. and {Karpikov}, I. and {Khangulyan}, D. and {Kuleshov}, D. and {Kurinov}, K. and {Li}, B.~B. and {Li}, C.~M. and {Li}, Cheng and {Li}, Cong and {Li}, D. and {Li}, F. and {Li}, H.~B. and {Li}, H.~C. and {Li}, Jian and {Li}, Jie and {Li}, K. and {Li}, S.~D. and {Li}, W.~L. and {Li}, W.~L. and {Li}, X.~R. and {Li}, Xin and {Li}, Y.~Z. and {Li}, Zhe and {Li}, Zhuo and {Liang}, E.~W. and {Liang}, Y.~F. and {Lin}, S.~J. and {Liu}, B. and {Liu}, C. and {Liu}, D. and {Liu}, D.~B. and {Liu}, H. and {Liu}, H.~D. and {Liu}, J. and {Liu}, J.~L. and {Liu}, M.~Y. and {Liu}, R.~Y. and {Liu}, S.~M. and {Liu}, W. and {Liu}, Y. and {Liu}, Y.~N. and {Luo}, Q. and {Luo}, Y. and {Lv}, H.~K. and {Ma}, B.~Q. and {Ma}, L.~L. and {Ma}, X.~H. and {Mao}, J.~R. and {Min}, Z. and {Mitthumsiri}, W. and {Mu}, H.~J. and {Nan}, Y.~C. and {Neronov}, A. and {Ng}, K.~C.~Y. and {Ou}, L.~J. and {Pattarakijwanich}, P. and {Pei}, Z.~Y. and {Qi}, J.~C. and {Qi}, M.~Y. and {Qiao}, B.~Q. and {Qin}, J.~J. and {Raza}, A. and {Ruffolo}, D. and {S{\'a}iz}, A. and {Saeed}, M. and {Semikoz}, D. and {Shao}, L. and {Shchegolev}, O. and {Sheng}, X.~D. and {Shu}, F.~W. and {Song}, H.~C. and {Stenkin}, Yu V. and {Stepanov}, V. and {Su}, Y. and {Sun}, D.~X. and {Sun}, Q.~N. and {Sun}, X.~N. and {Sun}, Z.~B. and {Takata}, J. and {Tam}, P.~H.~T. and {Tang}, Q.~W. and {Tang}, R. and {Tang}, Z.~B. and {Tian}, W.~W. and {Wan}, L.~H. and {Wang}, C. and {Wang}, C.~B. and {Wang}, G.~W. and {Wang}, H.~G. and {Wang}, H.~H. and {Wang}, J.~C. and {Wang}, Kai and {Wang}, Kai and {Wang}, L.~P. and {Wang}, L.~Y. and {Wang}, P.~H. and {Wang}, R. and {Wang}, W. and {Wang}, X.~G.},
        title = "{Detection of two TeV gamma-ray outbursts from NGC 1275 by LHAASO}",
      journal = {\mnras},
         year = 2024,
        month = nov,
          doi = {10.1093/mnras/stae2512},
archivePrefix = {arXiv},
       eprint = {2411.01215},
 primaryClass = {astro-ph.HE},
       adsurl = {https://ui.adsabs.harvard.edu/abs/2024MNRAS.tmp.2493C}
}

@ARTICLE{Fukazawa2018,
       author = {{Fukazawa}, Yasushi and {Shiki}, Kensei and {Tanaka}, Yasuyuki and {Itoh}, Ryosuke and {Takahashi}, Hiromitsu and {Imazato}, Fumiya and {D'Ammando}, Filippo and {Ojha}, Roopesh and {Nagai}, Hiroshi},
        title = "{X-Ray and GeV Gamma-Ray Variability of the Radio Galaxy NGC 1275}",
      journal = {\apj},
         year = 2018,
        month = mar,
       volume = {855},
       number = {2},
          eid = {93},
        pages = {93},
          doi = {10.3847/1538-4357/aaabc0},
       adsurl = {https://ui.adsabs.harvard.edu/abs/2018ApJ...855...93F}
}

@ARTICLE{Tanada2018,
       author = {{Tanada}, K. and {Kataoka}, J. and {Arimoto}, M. and {Akita}, M. and {Cheung}, C.~C. and {Digel}, S.~W. and {Fukazawa}, Y.},
        title = "{The Origins of the Gamma-Ray Flux Variations of NGC 1275 Based on Eight Years of Fermi-LAT Observations}",
      journal = {\apj},
         year = 2018,
        month = jun,
       volume = {860},
       number = {1},
          eid = {74},
        pages = {74},
          doi = {10.3847/1538-4357/aac26b},
archivePrefix = {arXiv},
       eprint = {1805.02361},
 primaryClass = {astro-ph.HE},
       adsurl = {https://ui.adsabs.harvard.edu/abs/2018ApJ...860...74T}
}

@ARTICLE{Rani2018,
       author = {{Rani}, B. and {Madejski}, G.~M. and {Mushotzky}, R.~F. and {Reynolds}, C. and {Hodgson}, J.~A.},
        title = "{NuStar View of the Central Region of the Perseus Cluster}",
      journal = {\apjl},
         year = 2018,
        month = oct,
       volume = {866},
       number = {1},
          eid = {L13},
        pages = {L13},
          doi = {10.3847/2041-8213/aae48f},
archivePrefix = {arXiv},
       eprint = {1809.10163},
 primaryClass = {astro-ph.HE},
       adsurl = {https://ui.adsabs.harvard.edu/abs/2018ApJ...866L..13R}
}

@ARTICLE{Imazato2021,
       author = {{Imazato}, Fumiya and {Fukazawa}, Yasushi and {Sasada}, Mahito and {Sakamoto}, Takanori},
        title = "{Origin of the UV to X-Ray Emission of Radio Galaxy NGC 1275 Explored by Analyzing Its Variability}",
      journal = {\apj},
         year = 2021,
        month = jan,
       volume = {906},
       number = {1},
          eid = {30},
        pages = {30},
          doi = {10.3847/1538-4357/abc7bc},
archivePrefix = {arXiv},
       eprint = {2011.10299},
 primaryClass = {astro-ph.HE},
       adsurl = {https://ui.adsabs.harvard.edu/abs/2021ApJ...906...30I}
}

@software{juan_escudero:2023,
       author = {{Escudero Pedrosa}, Juan and {Morcuende Parrilla}, Daniel and {Otero-Santos}, Jorge},
        title = "{IOP4}",
         year = 2024,
        month = jun,
          eid = {10.5281/zenodo.10222722},
          doi = {10.5281/zenodo.10222722},
      version = {v1.2.0},
    publisher = {Zenodo},
       adsurl = {https://ui.adsabs.harvard.edu/abs/2024zndo..10222722E}
}
\bibliographystyle{aasjournalv7}



\end{document}